\documentclass[nofootinbib,aps,pre,twocolumn,citeautoscript,superscriptaddress,floatfix,notitlepage]{revtex4-1}
\usepackage[utf8]{inputenc}
\usepackage[colorlinks=true]{hyperref}

\DeclareMathAlphabet{\mathcald}{U}{dutchcal}{m}{n}
\SetMathAlphabet{\mathcald}{bold}{U}{dutchcal}{b}{n}
\DeclareMathAlphabet{\mathalt}{U}{dutchcal}{b}{n}

\usepackage{amssymb,bm,xspace,soul} 
\usepackage{color} 
\usepackage[papersize={8.5in,11in}]{geometry}
\usepackage{natbib}
\usepackage{graphicx}
\usepackage{amsmath}
\usepackage{braket}
\usepackage{enumitem}
\usepackage{feynmp}
\usepackage{math tools} 
\usepackage{dsfont} 
\usepackage{mathrsfs} 
\usepackage{slashed}

\usepackage{pifont}
\newcommand{\xmark}{\ding{55}}%
\newcommand{\cmark}{\ding{51}}%

\newcommand{\M}{\mathcal{M}}

\newcommand{\T}{\mathcal{T}}

\newcommand{\vk}{{\boldsymbol{k}}}
\renewcommand{\vr}{{\boldsymbol{r}}}
\newcommand{\vQ}{{\boldsymbol{Q}}}
\newcommand{\vK}{{\boldsymbol{K}}}
\newcommand{\vq}{{\boldsymbol{q}}}
\newcommand{\vp}{{\boldsymbol{p}}}

\newcommand{\vphi}{{\varphi}}

\newcommand{\tr}{\text{tr}}

\renewcommand{\o}{\over}
\newcommand{\eq}[1]{\begin{align}#1\end{align}}

\renewcommand{\(}{\left(}
\renewcommand{\)}{\right)}
\renewcommand{\[}{\left[}
\renewcommand{\]}{\right]}
\newcommand{\abs}[1]{\left| #1 \right|}

\newcommand{\nt}{\notag\\}

\let\v\boldsymbol

\newcommand{\ph}{\phantom}

\newcommand{\ep}{\epsilon}

\renewcommand{\a}{\alpha}
\renewcommand{\b}{\beta}
\renewcommand{\d}{\delta}
\newcommand{\g}{\gamma}
\newcommand{\n}{\nu}
\newcommand{\m}{\mu}
\renewcommand{\t}{\tau}
\newcommand{\s}{\sigma}

\renewcommand{\th}{\theta}

\newcommand{\lam}{\lambda}
\renewcommand{\r}{\rho}

\renewcommand{\k}{\kappa}

\let\ptl\partial
\renewcommand{\dag}{\dagger}

\newcommand{\id}{\mathds{1}}
\newcommand{\Zt}{\mathds{Z}_2}

\newcommand{\ua}{\uparrow}
\newcommand{\da}{\downarrow}

\DeclareMathOperator{\sech}{sech}

\usepackage{tocloft} 
\addtocontents{toc}{\cftpagenumbersoff{chapter}}
\addtocontents{toc}{\cftpagenumbersoff{section}}

\setlist[enumerate]{leftmargin=*}

\usepackage{anyfontsize} 
\usepackage[table,dvipsnames]{xcolor}
\usepackage{wasysym}
\usepackage{units}
\usepackage{times}
\usepackage{nicefrac}

\usepackage{setspace}

\usepackage{titlesec}

\titleformat{\section}
     {\centering\bfseries}
     {\thesection.	}
     {1em}
     {\MakeUppercase}
     
\hypersetup{
    bookmarks=true,         
    unicode=false,          
    pdftoolbar=true,        
    pdfmenubar=true,        
    pdffitwindow=false,     
    pdfstartview={FitH},    
    pdftitle={Disorder-induced restoration of massless Dirac fermions},    
    pdfauthor={Alex Thomson},     
    pdfsubject={Subject},   
    pdfcreator={Alex Thomson},   
    pdfproducer={}, 
    pdfkeywords={keyword1} {key2} {key3}, 
    pdfnewwindow=true,      
    colorlinks=true,       
    linkcolor=magenta, 
    citecolor=blue,        
    filecolor=magenta,      
    urlcolor=cyan           
}

\newcommand{\vkp}{{\boldsymbol{\kappa}}}
\renewcommand{\T}{{\mathcald{T}}}

\newcommand{\C}{{\mathcald{C}}}
\renewcommand{\M}{{\mathcald{M}}}
\newcommand{\aM}{a_{\hspace{-0.05em}M}}

\newcommand{\cCom}{\mathbin{\raisebox{0.5ex}{,}}}
\newcolumntype{C}[1]{>{\centering\arraybackslash}p{#1}}

\newcommand{\etal}{\emph{et~al.}}
\begin{document}

\title{Recovery of massless Dirac fermions at charge neutrality in strongly interacting twisted bilayer graphene with disorder}

\author{Alex Thomson}
 \affiliation{Institute for Quantum Information and Matter, California Institute of Technology, Pasadena, California 91125, USA}
 \affiliation{Walter Burke Institute for Theoretical Physics, California Institute of Technology, Pasadena, California 91125, USA}
 \affiliation{Department of Physics, California Institute of Technology, Pasadena, California 91125, USA}
 \author{Jason Alicea}
  \affiliation{Institute for Quantum Information and Matter, California Institute of Technology, Pasadena, California 91125, USA}
 \affiliation{Walter Burke Institute for Theoretical Physics, California Institute of Technology, Pasadena, California 91125, USA}
 \affiliation{Department of Physics, California Institute of Technology, Pasadena, California 91125, USA}
\date{\today}

\begin{abstract}
 
Stacking two graphene layers twisted by the `magic angle' $\theta \approx 1.1^\circ$ generates flat energy bands, which in turn catalyzes various strongly correlated phenomena depending on filling and sample details. At charge neutrality, transport measurements reveal superficially mundane semimetallicity (as expected when correlations are weak) in some samples yet robust insulation in others. We propose that the interplay between interactions and disorder admits either behavior, \emph{even when the system is strongly correlated and locally gapped}. Specifically, we argue that strong interactions  supplemented by weak, smooth disorder stabilize a network of gapped quantum valley Hall domains with spatially varying Chern numbers determined by the disorder landscape --- even when an entirely different order is favored in the clean limit. Within this scenario, sufficiently small samples that realize a single domain display insulating transport characteristics. Conversely, multi-domain samples exhibit re-emergent massless Dirac fermions formed by gapless domain-wall modes, yielding semimetallic behavior except on the ultra-long scales at which localization becomes visible. We discuss experimental tests of this proposal via local probes and transport.  Our results highlight the crucial role that randomness can play in ground-state selection of twisted heterostructures, an observation that we expect to have further ramifications at other fillings.
\end{abstract}

\maketitle

\section{Introduction}

The discovery of superconductivity and correlated insulators in magic-angle twisted bilayer graphene (mTBG) \cite{Cao18a,Cao18b} opened a fascinating new chapter in the field of strongly interacting quantum matter.  
The `magic' stems from the fact that upon twisting the two graphene layers by an angle $\theta \approx 1.1^\circ$ from one another, 
the bands immediately above and below the charge neutrality point become exceptionally flat  \cite{Lopes07,MacDonald11} 
---bringing interactions center stage.  
Accounting for spin and valley degrees of freedom, each of these two flat bands is essentially fourfold degenerate.  Correlated physics, including superconductivity, thus naturally arises when the number of charge carriers per moir\'{e} unit cell is between $\n=-4$ (four holes) and $\n=+4$ (four electrons).

The observed phenomenology of mTBG depends sensitively on sample details. Cao \emph{et al.}~\cite{Cao18a,Cao18b}  originally observed correlated insulating states at $\n=\pm2$ along with superconducting domes upon doping away from the $\nu = -2$ insulator. 
Near the charge neutrality point at $\nu = 0$, the conductance exhibited a V-shaped suppression indicative of semimetallicity.  Non-interacting band theory calculations \cite{MacDonald11} predict massless Dirac fermions at charge neutrality provided the system preserves $\C_2\T$ symmetry, with $\C_2$ a two-fold rotation and $\T$ time reversal \cite{Kim15,Po18a,deGail11,He13}; the latter observation thus at first sight suggests weak correlations at $\nu = 0$.  The magic-angle device examined by Yankowitz~\etal~\cite{Yankowitz18} additionally exhibited superconductivity adjacent to the $\nu = +2$ insulator and a resistive correlated state at $\nu = +3$.  Near charge neutrality, transport again appeared consistent with the semimetallic behavior expected from band theory.

A second class of mTBG systems arises upon aligning the hexagonal boron nitride (hBN) substrate with one of the graphene sheets \cite{Sharpe19,Serlin19}.
The alignment appears to underlie strikingly different correlated physics: an absence of superconductivity, removal of the $\nu = -2$ insulator, weak resistive peaks at $\nu = +2$ instead of robust insulation, and a quantum anomalous Hall state at $\nu = +3$.  
Furthermore, at charge neutrality the system becomes strongly insulating instead of semimetallic.
The behavior at charge neutrality is, however, yet again consistent with band theory.
Indeed, alignment-induced breaking of $\C_2$ symmetry renders the Dirac fermions massive, yielding a band gap at $\nu = 0$.
Explicit $\C_2$ breaking has also been proposed as a catalyst for the observed quantum anomalous Hall state \cite{Bultinck19,Zhang19}.  

\begin{figure}
\centering
\includegraphics[width=0.48\textwidth]{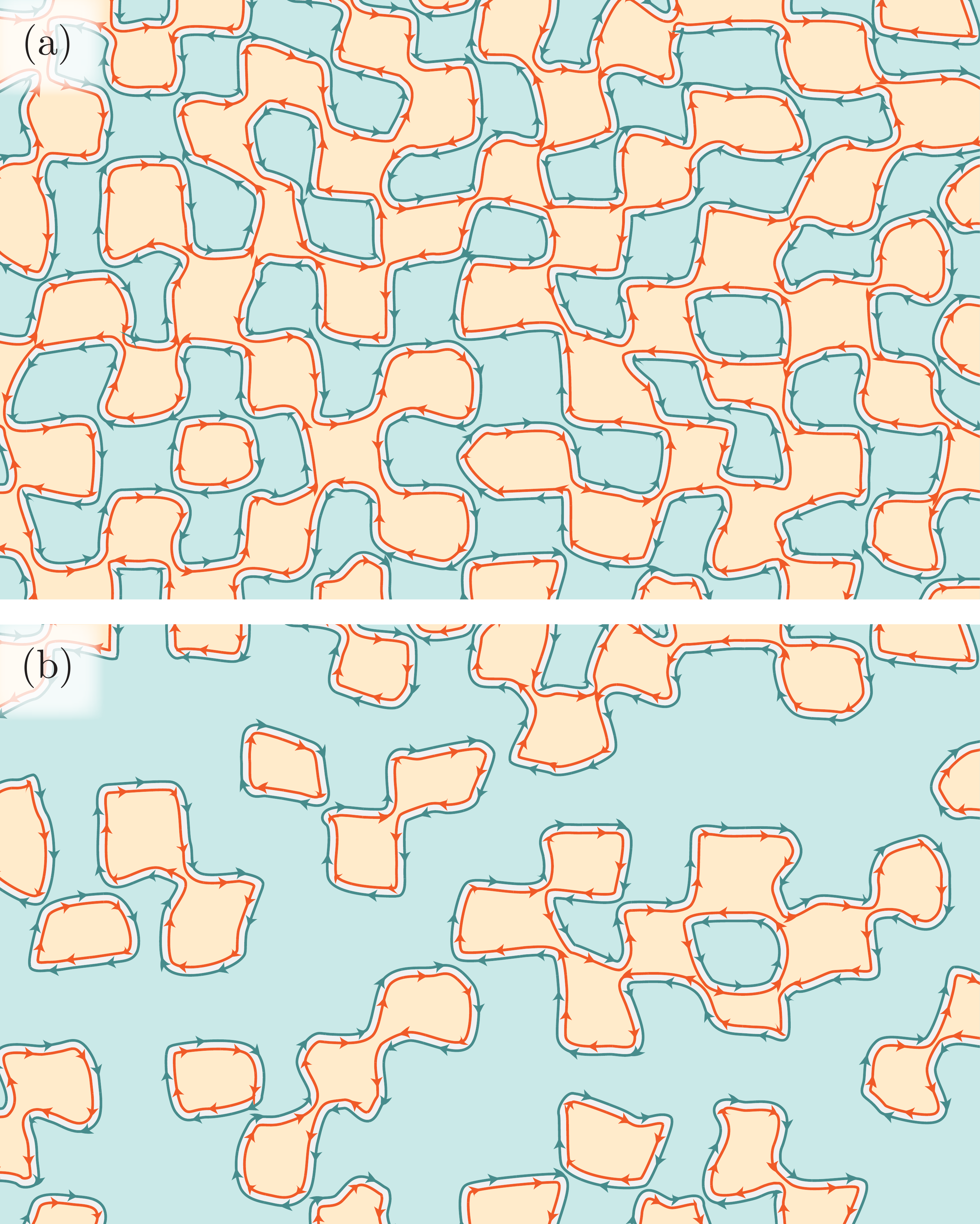}
\caption{Random tiling of quantum valley Hall states in a system where $\C_2$ is (a) preserved on average and (b)
explicitly broken (\emph{e.g.}, by hBN alignment).  For simplicity we show the domain structure only for a single valley.  Blue regions carry Chern number $C = +1$ for both spins, whereas orange regions carry Chern number $-1$. 
The arrows represent the two chiral edge modes (per spin) that traverse the domain boundaries.
The domain structure corresponding to the valley-sector not depicted here is simply obtained by exchanging the colours and reversing edge modes in (a) and (b).
}
\label{fig:InfSys}
\end{figure}

Still different phenomenology emerges in the ultra-homogeneous samples studied by Lu \etal~\cite{Lu19}.  These samples featured resistive peaks evincing either well-developed or incipient  insulators at \emph{all} integer fillings $\nu = 0, \pm 1, \pm 2, \pm 3$, as well as additional superconducting domes beyond those reported previously.
Notably, the strongest insulating state within the flat-band manifold occurred at charge neutrality, naively suggesting alignment with the hBN substrate as in Refs.~\onlinecite{Sharpe19,Serlin19}. 
Several factors challenge this interpretation, however.
First, Lu~\etal\;make no attempt to align the hBN, and it is unlikely to occur at random. 
Second, hBN-aligned samples and those of Lu~\etal~realize a largely disparate set of phenomena, suggesting against a common microscopic origin.    
Finally, the gap reported by Lu~\etal~\cite{Lu19} dwarfs by roughly an order of magnitude that measured in hBN-aligned mTBG \cite{Serlin19}, making its formation by explicit symmetry-breaking seem unlikely in comparison.
Thus the insulating behavior observed at all of the fillings indicated above---including $\nu = 0$---seems most naturally rooted in strong correlations.

A conservative interpretation of the available charge-neutrality transport data is that greater inhomogeneity in the samples from Refs.~\onlinecite{Cao18a,Cao18b,Yankowitz18} merely obliterates the strong correlations operative at $\nu = 0$ in the Lu \etal~samples.  Such a viewpoint is supported by the fact that significant ``twist-angle disorder" has been observed by multiple groups \cite{Cao18a,Cao18b,Yankowitz18,Kerelsky18,Choi19,Jiang19,Xie19,Uri19}; moreover, 
deviations from the magic angle locally enhance the flat-band dispersion \cite{Wilson19}, potentially diminishing correlation effects.  Scanning tunneling microscopy (STM) measurements from Refs.~\onlinecite{Kerelsky18,Choi19,Jiang19,Xie19}, however, do not simply fit this picture. All of these STM studies observed \emph{local} correlation effects at charge neutrality, manifested by a pronounced splitting of the flat-band van Hove peaks upon approaching $\nu = 0$ and, in Ref.~\onlinecite{Xie19}, evidence of a hard gap at charge neutrality\footnote{The issue of hBN alignment is subtle given the different nature of these experiments.  Nevertheless, alignment is  expected to enhance the van Hove peak splitting \emph{independent of filling}, whereas the splitting observed by STM is significantly larger at charge neutrality compared to when the bands are fully filled.}.  Much subtler signatures of correlated states were also seen at other integer fillings (typically most prominently at $\nu = +2$).
From a local perspective, it therefore appears that correlations in the STM samples are actually \emph{strongest} at $\nu = 0$.

\begin{figure*}[t]
	\centering
	\includegraphics[width=0.98\textwidth]{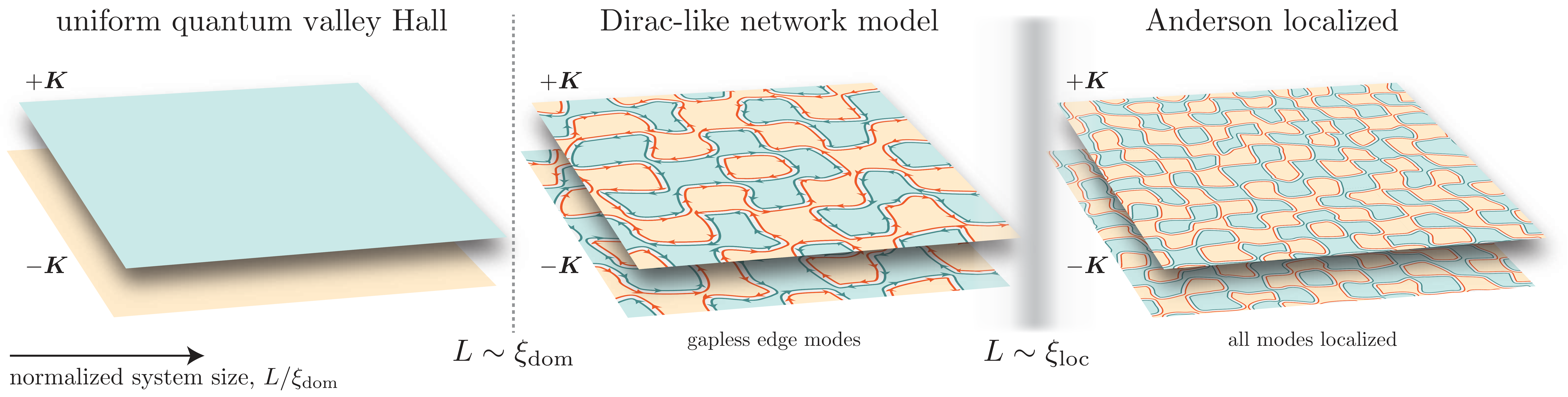}
	\caption{Phase diagram versus system size $L$ for a given disorder landscape.
	Left panel: With $L$ below the typical domain size $\xi_{\rm dom}$, a single domain is realized.  Here each valley exhibits Chern number $C=+1$ (depicted in blue) or $C=-1$ (depicted in orange) throughout the entire sample, yielding insulating transport as observed by Lu \etal~\cite{Lu19}.
	Central panel: When $L \gtrsim \xi_\mathrm{dom}$ multiple domains may be present within a single sample.  The domain structure results in a percolating network of gapless edge modes that underlies the Dirac-like conductance seen by Cao~\etal~\cite{Cao18a,Cao18b} and Yankowitz~\etal~\cite{Yankowitz18}.
	Right panel: When $L$ exceeds the localization length $\xi_{\rm loc}$ the sample localizes and ceases to conduct.
	}
	\label{fig:DisPhaseDiagram}
\end{figure*}

In this paper we propose a unifying explanation for the diverse phenomenology observed to date in mTBG at charge neutrality. Our scenario posits that strong correlations are ubiquitous---even in samples that observe semimetallic behavior expected from band theory---with disorder playing a secondary but still crucial role. We specifically assume that in a perfectly clean infinite system, interactions favor, or very nearly favor, correlated states that spontaneously break $\C_2\T$ symmetry in a way that yields Chern number $C = \pm 1$ for a given spin/valley sector.  This assumption is bolstered by existing numerical simulations \cite{Liu19,Choi19,Xie18,Lu19} and justified further below.  Among the many possible insulators, only two preserve translation symmetry, spin rotation symmetry, and time reversal: the pair of `quantum valley Hall' states 
\cite{ZhangFan11,Martin08,QiaoZH11,YinLJ16,JuL15,LiJ16}
with  $C = +1$ for both spins in one valley and $C = -1$ for both spins in the other valley, or vice versa.  Note that $\C_2$ transforms the quantum valley Hall states into one another; hence they are exactly degenerate provided $\C_2$ is not explicitly broken.   

Imagine now turning on smooth, non-magnetic disorder that explicitly violates the infinite system's $\C_2$ symmetry but preserves it in an average sense.  Within the manifold specified above, quantum valley Hall states are unique in that their order parameter directly couples to the disorder potential---allowing the system to efficiently gain energy by locally forming one of those two phases.  
We further assume that the energy gain outweighs any energy cost (should one exist) for forming quantum valley Hall order in the clean limit.  Under these circumstances the infinite system exhibits a random tiling of the two quantum valley Hall states, details of which are determined by the interplay between interactions and the disorder landscape; see Fig.~\ref{fig:InfSys}(a) for an illustration. 
Similar domain structures have been discussed in several other contexts, \emph{e.g.}, in systems with valley Hall nematic order \cite{Abanin10,Parameswaran18} or
as a source of non-Abelian `PH-Pfaffian' topological order \cite{Mross2018,WangC2018,Lian2018}.

Crucially, the infinite system is locally gapped within the quantum valley Hall domains but is not entirely electrically inert.  Each domain wall binds four `right-moving' and four `left-moving' charge-carrying modes, reflecting the fact that the Chern numbers for the spin/valley sectors change by $\pm 2$ upon passing between adjacent domains.
Smoothness of the disorder potential suppresses scattering among these modes and thus justifies treating the spin/valley sectors as decoupled (to a first approximation).  In this limit the
system realizes four copies of a Chalker-Coddington network model \cite{Chalker88} describing an integer-quantum-Hall plateau transition at which the Hall conductivity changes by $\delta \sigma_{xy} = \pm 2 e^2/h$ \cite{LeeChalker94a,LeeChalker94b}.  This set of plateau transitions can be described by eight massless Dirac fermions (two per sector) with disorder acting within each cone \cite{Ho96,Lee94}.  Hence essentially the same low-energy physics expected from band theory emerges from a strongly correlated framework!  
Residual scattering among the domain-wall modes generates inter-cone disorder that produces localization, but the localization lengths can be arbitrarily long.  

We emphasize that strong correlations form the bedrock of the scenario outlined above.
Without interactions and in the absence of explicit net $\C_2\T$-breaking, 
the fate of the system depends sensitively on the nonuniversal details of the disorder. 
The quantum valley Hall state is but one among many potential phases, both gapped and gapless, that disorder could locally favour.
Moreover, even if local quantum valley Hall order happened to develop, the gap would be set by disorder and could be exceedingly small.
In contrast,
by inducing the spontaneous breaking of $\C_2\T$, interactions select a small subset of energetically competitive states in our scenario.
Disorder then plays a subordinate role by favoring one of the two quantum valley Hall orders in that set, thereby generating the domain structure.
The local gap protecting the insulating domains is determined primarily by interactions rather than disorder.

Let us now revisit  experiments in light of our proposed picture.  Locally probing the quantum valley Hall domains should reveal signatures of a correlation-driven gapped spectrum (possibly dressed with disorder-induced subgap states), consistent with STM experiments \cite{Kerelsky18,Choi19,Jiang19,Xie19}.  
The outcome of global transport measurements depends on the ratio of sample size $L$ to the typical domain size $\xi_{\rm dom}$ that would occur in an infinite system.   
For homogeneous systems such that $L/\xi_{\rm dom} \lesssim 1$, transport probes essentially a single domain, yielding insulating behavior as observed by Lu \etal~\cite{Lu19}.  (Strong intervalley scattering induced by the sample boundary is expected to suppress edge conduction.)
Conversely, for more-disordered samples with $L/\xi_{\rm dom} \gg 1$, transport probes many domains; here the massless Dirac fermions emerging from the gapless domain walls underpin semimetallic conduction as measured in Refs.~\onlinecite{Cao18a,Cao18b,Yankowitz18}.  See Fig.~\ref{fig:DisPhaseDiagram} for a summary.  
We can also make contact with alignment-induced insulation observed in Refs.~\onlinecite{Sharpe19,Serlin19}.  
Turning on hBN alignment supplements the disorder landscape with a \emph{uniform} $\C_2$-breaking potential that shrinks the area occupied by one of the quantum valley Hall states and expands the area of the other, as shown in Fig.~\ref{fig:InfSys}(b).  Domain walls then no longer percolate, thereby gapping the re-emergent massless Dirac fermions and producing insulating transport when  $L/\xi_{\rm dom} \gg 1$.  

The arguments outlined above are justified through a Landau-Ginzburg theory describing the quantum valley Hall order parameter.
With the inclusion of disorder, we arrive at a classical 2$d$ random-field Ising model, which allows us to estimate the scaling of the typical domain size as a function of system parameters. 
Through a simple extension of this formulation, we can further study what occurs when a different phase that does \emph{not} couple directly to disorder is energetically favoured over the quantum valley Hall state in the clean limit.
As expected, when the (clean) ground state energy splitting between the two states is sufficiently small---in a sense that we quantify with our Ising formulation---quantum valley Hall order prevails throughout the majority of the sample. 

Our scenario for ubiquitous strong correlations at charge neutrality is not only compatible with existing charge-neutrality data, but further leads to falsifiable predictions both for STM and transport as described in Sec.~\ref{Discussion}.  
We also propose that two elements of this work may have broader applications in the study of mTBG.  First, disorder can play a key role in discriminating among nearly degenerate correlated states.  And second, disorder need not obliterate correlations, but can mask them as seen by global transport experiments.  

The rest of the paper is organized as follows. 
We begin by reviewing the low-energy theory and establishing our conventions for twisted bilayer graphene in Sec.~\ref{sec:LowETheory}.
Next, Sec.~\ref{sec:DisorderFree} describes the fate of non-interacting mTBG Dirac fermions at charge neutrality in the presence of disorder.
We then 
discuss the clean interacting theory in Sec.~\ref{sec:IntClean}. 
The interaction form is first outlined [Sec.~\ref{sec:CoulInt}], and then used to argue that the quantum valley Hall state is energetically competitive at charge neutrality [Secs.~\ref{sec:Groundstate} and~\ref{sec:SpinInteractions}].
Our main thesis is presented in detail in Sec.~\ref{sec:Int-Dis}, where each of the three regions illustrated in Fig.~\ref{fig:DisPhaseDiagram} is described in turn.  We conclude in Sec.~\ref{Discussion} by summarizing and highlighting future directions.  Supplemental details appear in numerous appendices.

\section{Review of low-energy theory}\label{sec:LowETheory}

In this section, we set the stage by reviewing the low-energy physics of mTBG at charge neutrality in the absence of interactions and disorder.  

\subsection{Continuum model}\label{sec:ContModel}

\begin{figure}[t]
	\centering
	\includegraphics[width=0.48\textwidth]{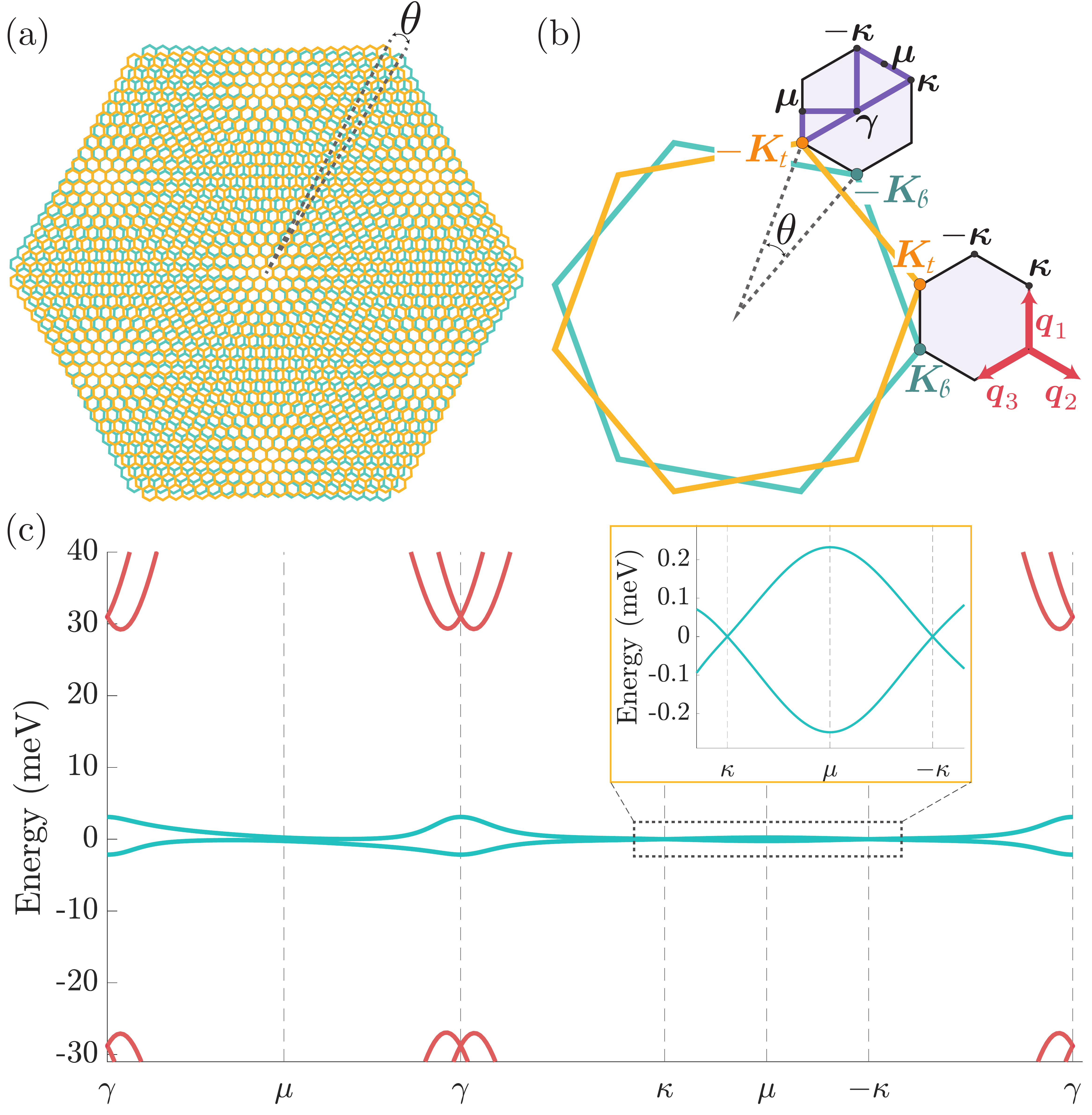}
	\caption{(a) Cartoon representation of twisted bilayer graphene. 
	The top and bottom graphene sheets are respectively represented by the orange and turquoise honeycomb lattices.
	The light, orange-tinted AA regions form a triangular superlattice, each of which is surrounded by a darker hexagonal rim whose vertices correspond to alternating AB and BA stacking regions.
	(b) Representation of the microscopic and moir\'{e} Brillouin zones. 
	The large orange and turquoise hexagons represent the microscopic Brillouin zones of the underlying graphene layers.
	The moir\'{e} Brillouin zones, shown in purple, are defined by the distance between the the $\vK$-points of the top and bottom layers.
	The size of the twist angles in both (a) and (b) has been exaggerated for clarity. 
	(c) Flat bands corresponding to one $\vK$-valley as calculated using the continuum model along the momentum line cut shown in purple in (b). 
	Here, we took $\th=1.05^\circ$. 
	Lattice relaxation is mimicked by decreasing the AA tunnelling amplitude $w_0$ relative to the AB tunnelling amplitude $w_1$. 
	In particular, we have $w_0=\unit[85]{\text{meV}}$ and $w_1=\unit[110]{\text{meV}}$ \cite{Nguyen17,Po18a,Koshino18}.
	The inset zooms in on the Dirac cones at $\vkp$ and $-\vkp$.
	}
	\label{fig:BZ}
\end{figure}

Consider two monolayer-graphene sheets stacked such that they are twisted relative to one another by an angle $\th$, as shown in Fig.~\ref{fig:BZ}(a) (for an arbitrary angle $\th$).
The twist dramatically reduces the system's translational symmetry.
While true translational symmetry requires special commensurate angles, when the twist angle is small, an effective moir\'{e} translational symmetry emerges;
the resulting triangular superlattice of orange AA regions, each surrounded by a hexagon of alternating AB and BA regions, is clearly visible in the cartoon of Fig.~\ref{fig:BZ}(a).
In this case,
the band structure at charge neutrality descends from the band structure of the individual graphene layers in a relatively straightforward manner 
when described in momentum space \cite{MacDonald11,Lopes07}.
Figure~\ref{fig:BZ}(b) shows the Brillouin zones (BZs) of the top and bottom graphene monolayers after applying a rotation by an angle $+\th/2$ and $-\th/2$, respectively.
The reciprocal lattice vectors of the resulting moir\'{e} pattern are given by $\v{G}_\ell =\mathcald{R}_{\th/2}\[\v{\mathcald{G}}_\ell\]-\mathcald{R}_{-\th/2}\[\v{\mathcald{G}}_\ell\]$ where $\v{\mathcald{G}}_{1,2}$ denote the reciprocal lattice vectors of the unrotated graphene sheets and $\mathcald{R}_\phi[\v{v}]$ rotates a vector $\v{v}$ by an angle $\phi$.
The length of the moir\'{e} reciprocal lattice vectors, $\abs{\v{G}_\ell}$, is therefore suppressed relative the graphene reciprocal lattice vectors by a factor of $2\sin\(\th/2\)\sim \th$, making it very small by assumption.
Equivalently, the moir\'{e} lattice constant is enlarged by $\sim1/\th$ relative to the graphene lattice constant.
We let $\pm\vK_\mathcald{t}=\mathcald{R}_{\th/2}[\pm\vK]$ and $\pm\vK_\mathcald{b}=\mathcald{R}_{-\th/2}[\pm\vK]$ denote the $\pm\vK$ points of the top and bottom layers, respectively.

As tunnelling between the two layers  turns on, states at momentum $\vK_\mathcald{t}+\vk$ on the top layer mix with those at momentum $\vK_\mathcald{b}+\vk+\vq_\ell+\v{G}$ on the top layer, where $\v{G}$ is a moir\'{e} reciprocal lattice vector and $\vq_\ell=\mathcald{R}_{2\pi(\ell-1)/3}\[\vK_\mathcald{t}-\vK_\mathcald{b}\]$, $\ell=1,2,3$ [see Fig.~\ref{fig:BZ}(b)].
Since the moir\'{e} BZ is much smaller than the BZ of monoloyer graphene, mixing states proximate to $\vK$ with those proximate to $-\vK$ is an extremely high-order tunnelling
process; the two valleys of the original graphene monolayers thus effectively decouple.
This decoupling is particularly convenient as it allows us to express the full band Hamiltonian $H_{\mathrm{cont}}$ as a sum of terms for the $\vK$ and $-\vK$ valleys:  $H_{\mathrm{cont}}=H_{+}+H_{-}$.
For convenience, we explicitly reproduce $H_+$ in Appendix~\ref{app:ContModel}.
Our ability to decompose the Hamiltonian into $\vK$-valley sectors is equivalent to the emergence of a U(1) ``valley" symmetry, which we denote U(1)$_v$.

In addition to moir\'{e} translations and U(1)$_v$, the continuum model  preserves the SU(2)$_s$ 
spin rotation symmetry (neglecting spin-orbit coupling), time reversal $\T$, $\C_2$ rotations by $\pi$, $\C_3$ rotations by $2\pi/3$, and a mirror symmetry $\M_y$ that takes $(x,y)\to(x,-y)$ and interchanges the two layers.
The latter three should be regarded as emergent symmetries similar to U(1)$_v$.
In our conventions the time-reversal operator  $\T$ does \emph{not} flip the electronic spins and accordingly obeys $\T^2=+1$.
Both $\T$ and $\C_2$ interchange the two valleys and hence are not symmetries of the individual single-valley Hamiltonians $H_{\pm}$.
Rather than keep track of these two symmetries separately, it is therefore convenient to consider $\T$ along with the composite operation $\C_2\T$---which commutes with U(1)$_v$.

The momenta $\vK_\mathcald{t}$ and $\vK_\mathcald{b}$ map to the corners of the moir\'{e} BZ.
In what follows, we denote these momenta by $\pm\vkp$ to distinguish them from the $\pm\vK$-valleys of the microscopic graphene layers.
Provided the $\C_2\T$ and $\C_3$ symmetries are present, the massless Dirac cones at $\vK_{\mathcald{t},\mathcald{b}}$ for the microscopic graphene layers evolve into massless Dirac cones at $\pm\vkp$ even once tunnelling is turned on.
This crucial property follows from the fact that the Berry phase enclosed within any loop is quantized to 0 or $\pi$ ($\mathrm{mod}\,2\pi$) when $\C_2\T$ is preserved.
Since a Dirac point necessarily exhibits Berry phase $\pi$, the Dirac cones at $+\vkp$ and $-\vkp$ are locally protected against a mass \cite{Kim15,Po18a}.
Breaking $\C_3$ can shift the location of the Dirac cones, but cannot gap them.
Importantly, since both cones in $H_+$ ($H_-$) descend from the Dirac cones at $\vK_\mathcald{t,b}$ ($-\vK_\mathcald{t,b}$) in a continuous fashion, they possess the same chirality \cite{deGail11,He13}---thereby obstructing the development of a two-band, single $\vK$-valley tight-binding model in which all symmetries are realized in a local fashion \cite{Po18a,Po18b,Kang18}. 

\subsection{Flat bands}\label{sec:FlBands}

The previous subsection highlighted generic features of small-angle twisted-bilayer graphene. 
At the \emph{magic} angle, the velocity of the massless Dirac fermions becomes very small, 
and the bands immediately above and below the charge neutrality point separate from the remaining bands by a finite energy (provided lattice relaxation is incorporated \cite{Nguyen17,Koshino18,Po18a}); see Fig.~\ref{fig:BZ}(c).
The resulting energetically isolated ``flat bands" are each (essentially) four-fold degenerate, reflecting spin and valley degrees of freedom.
We now describe the flat-band Hamiltonian by first focusing on the $+\vK$ valley and subsequently incorporating the $-\vK$ valley.

Let $c_{\alpha j}(\vk)$ denote momentum-space annihilation operators associated with the flat bands at valley $+\vK$; here $\alpha = {\uparrow,\downarrow}$ is a spin index and $j = 1,2$ is a band index. 
Reference~\onlinecite{Zou18} showed that these operators can be defined such that they transform under $\C_2\T$ via
\eq{
\C_2\T:\qquad
c(\vk)
&\to
\eta^xc(\vk),
&
i&\to-i
}
with Pauli matrices 
$\eta^{x,y,z}$ that act on the band indices.  (Here and below we often suppress indices for notational simplicity.) 
It follows that the $\C_2\T$-invariant flat-band Hamiltonian takes the form
\eq{\label{eqn:FlatBandH0}
H_{0}&=\int_{\vk\in BZ}c^\dag(\vk)\Big[h_0(\vk)+ h_x(\vk)\eta^x + h_y(\vk)\eta^y \Big]c(\vk).
}
Next we project onto the massless Dirac fermions at $\pm \vkp$ in the moire BZ by defining Dirac spinors
$\psi_{1\alpha j}(\vq) \sim c_{\alpha j}(+\vkp + \vq)$ and $\psi_{2\alpha j}(\vq) \sim c_{\alpha j}(-\vkp + \vq)$ and retaining only small $\vq$ modes.  
The fact that the massless Dirac cones exhibit the same chirality at $\pm \vkp$ implies that $h_x(\pm\vkp+\vq)\sim +q_x$ and $h_y(\pm\vkp+\vq)\sim +q_y$.
Upon shifting the energy such that $h_0(\pm\vkp)=0$ and reverting to real space, the low-energy Hamiltonian becomes 
\eq{\label{eqn:DirHam}
H_D&=-\int_\vr\,v_F\psi^\dag \(i\ptl_x\eta^x + i\ptl_y\eta^y\)\psi.
}
The Fermi velocity $v_F$ has been assumed isotropic and identical for both $\pm\vkp$ Dirac cones, which is guaranteed when all  symmetries outlined earlier are present.  

An insulating phase at charge neutrality may only be obtained by either breaking the $\C_2\T$ symmetry or by closing the gap separating the flat bands and the dispersing bands \cite{Kim15,Po18a}.
We focus entirely on the former scenario, which is straightforward to represent using the Dirac theory.
Let $\tau^{x,y,z}$ and $\sigma^{x,y,z}$ denote Pauli matrices that respectively act on $\vkp$-valley indices and spin indices.   Mass terms then take the form 
$\psi^\dag \eta^z M \psi$ with $M=\big\{ \id, \s^i,\t^i,\t^i\s^j\big\}$.
The Chern number for a given spin/valley sector depends on the relative sign of the masses gapping the $\vkp$ and $-\vkp$ Dirac cones. 
When both cones have the same-sign mass, the sector acquires Chern number $C=\pm1$, whereas opposite-sign masses yield $C = 0$.  
Consequently, mass terms with $M=\id,\,\s^i$ yield insulating bands with non-zero Chern number, while masses with $M=\t^i,\,\t^i\s^j$ yield trivial insulating bands\footnote{Given the Hamiltonian $H_D$, the above conclusions regarding Chern number hold true regardless of the details of the high-energy theory from which it was derived.
It is worth noting that for a single graphene sheet, expanding in small $\vq$ the functions analogous to $h_{x,y}(\pm\vkp+\vq)$ would not yield the low-energy Hamiltonian of Eq.~\eqref{eqn:DirHam}.
Instead, the two Dirac cones would possess opposite chirality.}.

We now restore the $-\vK$ valley. 
In terms of the low-energy Dirac Hamiltonian, the chirality of the massless Dirac fermions in the $-\vK$ valley is opposite that of the $+\vK$ valley.  
Defining the 
spinor $\Psi=(\psi_+,\psi_-)^T$, where $\psi_\pm$ describe Dirac fermions in valley $\pm \vK$, the full Dirac Hamiltonian may be written
\eq{\label{eqn:BothValleyDirHam}
H_{D,\mathrm{tot}}
&=
-\int_\vr\,v_F
\Psi^\dag\(i\ptl_x \,\m^z \eta^x + i\ptl_y\, \eta^y \)\Psi,
}
where we introduced Pauli matrices $\mu^{x,y,z}$ that act on $\vK$-valley indices.  
The presence of $\mu^z$ in the first term above implements the opposite-chirality requirement.  Our discussion of the mass terms and the associated Chern numbers extends straightforwardly to the $\Psi$ fermions. 
For details see Appendix~\ref{app:Bilinears}.
A notable consequence of the opposing chiralities of the $\pm\vK$ valleys is that a mass term $\Psi^\dag \eta^z \Psi = \psi_+^\dag\eta^z \psi_+ + \psi_-^\dagger \eta^z \psi_-$ generates an insulator with  $C=+1$ for the $+\vK$ valley and $C=-1$ for the $-\vK$ valley (or vice versa depending on the overall sign of the mass term).
These insulators correspond to the quantum valley Hall states that play a prominent role in this paper. 

In the following sections, we use the operator $\psi$ when restricting our discussion to a single $\vK$-valley. 
We suppress the ``$\pm$'' indices in such cases but assume for concreteness that the $+\vK$ valley is being considered (as in the beginning of this subsection).  
We reserve use of $\Psi$ for occasions when both valleys are discussed simultaneously.

\section{Free fermions with disorder}\label{sec:DisorderFree}

Next, we discuss the physics of non-interacting twisted bilayer graphene with disorder at charge neutrality.

\subsection{Sources of disorder in twisted bilayer graphene}

It is useful to review the specific types of disorder that are believed to be most relevant to experiments, though we attempt to keep the majority of our discussion as general as possible.
Charge disorder appears to be quite low: 
Refs.~\onlinecite{Cao18b,Yankowitz18,Uri19} estimate charge-carrier inhomogeneity in the range $\d n\sim {\unit[1-2\times10^{10}]{\text{cm}^{-2}}}$.
Yankowitz~\etal~\cite{Yankowitz18} further consider the observation of fractional quantum Hall states at magnetic fields as low as $\unit[4]{\text{T}}$ as additional proof of the high purity of their sample.  

Twist-angle disorder is perhaps the most prevalent type of inhomogeneity in mTBG systems.
Due to strain, different regions of a given sample may correspond to different twist angles, as directly imaged in STM \cite{Kerelsky18,Choi19,Jiang19}.
From topography, the AA regions of the moir\'{e} structure are very clear, allowing one to locally establish the moir\'{e} lattice constants and thus the twist angle.
Twist-angle variations were more recently characterized by Uri~\etal~\cite{Uri19} using a superconducting quantum interference device on a tip.
Under an applied magnetic field, these authors measured the electron density of the sample as a function of the tip location, which in turn allowed them to map out the twist angle throughout the entire sample.
Such measurements indicated local twist angles varying within a range $\d\th\sim0.1^\circ$. 
Both samples they studied developed correlated insulating states, but only the sample with a continuous magic-angle region percolating across the sample displayed clear signs of superconductivity. 

While transport measurements cannot access such local information, by comparing two-terminal conductance measurements between different pairs of contacts, Cao \emph{et al.} \cite{Cao18a,Cao18b} and Yankowitz \emph{et al.} \cite{Yankowitz18} nevertheless note that some regions require different electron densities to achieve the band insulator at full-filling, again implying that unit cells differ between regions.
Similar measurements by Lu~\etal~\cite{Lu19} returned a much more uniform signal across the sample. 
Disorder signatures are also observable from within the superconducting states.
Both Cao \emph{et al.} and Yankowitz \emph{et al.} observe phase-coherent Fraunhofer interference, indicating the coexistence of superconducting and normal regions.
Conversely, the interference patterns measured by Lu~\emph{et al.} are comparatively weak,
which they take as further indication of the high degree of sample homogeneity.

The hBN substrate may serve as yet another source of disorder.
When uniformly aligned with one of the graphene monolayers, $\C_2\T$ symmetry is explicitly broken and a gap at charge neutrality is opened \cite{Bultinck19,Zhang19}.
While the explicit gapping naturally explains the $\nu = 0$ insulator and anomalous Hall effect observed by Sharpe \emph{et al.} \cite{Sharpe19} and Serlin~\etal~\cite{Serlin19} at $\n=+3$, hBN-alignment is believed to be an otherwise small effect in the majority of samples studied. 
Nevertheless, it is possible that a \emph{local} alignment of the substrate, differing between regions, could weakly break the $\C_2\T$ symmetry---just as for twist-angle disorder---even though it may be present on average.

\subsection{Theoretical modelling of disorder}\label{sec:TypesOfDis}

Motivated by the preceding discussion, we now incorporate weak, smooth disorder that preserves time-reversal and spin-rotation symmetries.  
We model such disorder by coupling spatially varying (but static) fields to fermion bilinears of the non-interacting Dirac theory reviewed in Sec.~\ref{sec:FlBands}.
The most relevant forms of disorder couple to bilinears that do not contain derivatives, and so we focus our study on this subset\footnote{In particular, we neglect disorder-induced variation in the Fermi velocity.  
This omission is supported by the numerics of Ref.~\onlinecite{Wilson19}, which show that the velocity remains largely unaffected by the presence of twist-angle disorder.
}.
Time-reversal invariance and spin symmetry further reduce the number of bilinears capable of coupling to disorder; we enumerate all such symmetry-preserving terms in Appendix~\ref{app:Bilinears}.
Collectively denoting the set of symmetry-allowed operators by $\{\Psi^\dagger T^i\Psi\}$, the most general disorder Hamiltonian takes the form
\eq{
H_\mathrm{dis}
&=
\int_\vr\,\sum_i R_i(\vr) \Psi^\dag(\vr) T^i \Psi(\vr).
\label{Hdis}
}
We assume Gaussian-distributed $R_i(\vr)$ with zero mean and variance \eq{\label{eqn:DisorderCorr}
\overline{R_i(\vr)R_j(\vr')}
&=
\d_{ij}g_i^2\,K_i\big( (\vr-\vr')/\xi_i \big). 
}
Here, $g_i$ is the disorder strength with units of energy, $\xi_i$ is the disorder correlation length, and $K_i$ is a dimensionless function that characterizes the spatial correlations of the disorder and obeys $K_i(0) = 1$.
We frequently specialize to the case where the spatial correlations are Gaussian, \emph{i.e.}, 
\eq{\label{eqn:GaussianCorrFun}
K_i(\vr/\xi_i)&=e^{-\vr^2/(2\xi_i^2)}.
}
Weakness of the disorder implies that $g_i$ are small relative to the other scales of the theory, enabling a perturbative treatment.
Smoothness of disorder is imposed by requiring that $\xi_i\gtrsim \aM$, with $\aM$ the moir\'{e} lattice constant.
We assume that the correlation lengths corresponding to different forms of disorder do not differ substantially and simply set $\xi_i = \xi_{\rm dis}$ for all $i$.

The smoothness condition is physically very natural given that the existence of the moir\'{e} superlattice and the resulting band structure is predicated on the absence of fluctuations on the scale of the graphene lattice constant $a$.
In momentum space, smoothness implies the suppression of inhomogeneities mediating momentum exchanges of order $\sim\abs{\vK}$, \emph{i.e.}, disorder processes that couple to bilinears of the form $\Psi^\dag\mu^{x,y}M\Psi$.
In fact, we demonstrate in Appendix~\ref{app:InterKSuppression} that given Gaussian-correlated disorder [Eq.~\eqref{eqn:GaussianCorrFun}], the disorder strengths corresponding to inter-$\vK$-valley scattering are exponentially suppressed relative the intra-$\vK$-scattering disorder strengths: if $g$ is the magnitude of a typical intra-$\vK$-valley disorder field, then 
\eq{\label{eqn:KK'suppression}
g_{\!\mathit{KK'}}\sim g\, \smash{e^{-\vK^2\xi_\mathrm{dis}^2/4}}=g\,\smash{e^{-4\pi^2\xi_\mathrm{dis}^2/a^2}}
}
is the typical amplitude of an inter-$\vK$-valley scattering event.
 Neglecting such exponentially suppressed events for now, we focus on a single $\vK$-valley and couple disorder to the $\psi$ fermions described by $H_D$.

Since time-reversal interchanges $\vK$-valleys, it is \emph{not} a symmetry of the single-$\vK$-valley theory, implying that the system is described by the Wigner-Dyson class A \cite{Zirnbauer96,Altland97}.
Disorder can thus couple to all spin-rotation-invariant bilinears and takes the form
\eq{\label{eqn:HdisGeneric}
H_{\mathrm{dis}}^{\rm smooth}
&=
\int_\vr\,\psi^\dag(\vr)\Big\{
\mathcal{M}_0(\vr)\eta^z + \mathcal{M}_\ell(\vr)\eta^z\t^\ell
\nt&\quad
+
\sum_{i=x,y}\Big[\mathcal{A}_{i,0}(\vr)\eta^i + \sum_{\ell=x,y,z}\mathcal{A}_{i,\ell}(\vr)\eta^i\t^\ell\Big]
\nt
&\quad
+\mathcal{V}_0(\vr)+\sum_{\ell=x,y,z}\mathcal{V}_\ell(\vr)\t^\ell
\Big\}\psi(\vr),
}
where $\mathcal{M}$, $\mathcal{A}$, and $\mathcal{V}$ respectively represent various forms of mass, vector potential, and scalar potential disorder.

It is also useful to consider the limit where disorder is sufficiently smooth relative to the moir\'{e} lattice scale that inter-$\vkp$-valley scattering may also be neglected.  We can then further restrict our attention to one of the Dirac cones in the moir\'{e} BZ---say $+\vkp$.  
Denoting the spinor describing the Dirac cone at $+\vkp$ by $\chi(\vr)$, the disorder Hamiltonian becomes simply
\eq{\label{eqn:HDis1kapVal}
H_{\mathrm{dis}}^{{\rm ultra}{\text -}{\rm smooth}}
&=
\int_\vr\,\chi^\dag(\vr) \Big[
\mathcald{m}(\vr)\eta^z 
\nt
&\quad
+
\sum_{i=x,y}\mathcald{a}_i(\vr)\eta^i
+
\mathcald{v}(\vr)\Big]\chi(\vr),
}
where the random mass $\mathcald{m}$, vector potential $\mathcald{a}_{x,y}$ and scalar potential $\mathcald{v}$ satisfy Eq.~\eqref{eqn:DisorderCorr} with $R_i=\mathcald{m},\,\mathcald{a}_{x,y},\,\mathcald{v}$.
Since each moir\'{e} unit cell encompasses $\sim$10 000 carbon atoms, distilling the disorder Hamiltonian down to Eq.~\eqref{eqn:HDis1kapVal} is significantly more suspect than merely omitting inter-$\vK$-valley scattering terms.
Moreover, though it might naively appear that inter-$\vkp$-scattering should be suppressed in a manner analogous to Eq.~\eqref{eqn:KK'suppression}, we only expect such an effect to be manifest for extremely large correlation lengths $\xi_{\rm dis}$ relative to $\aM$ as discussed at the end of Appendix~\ref{app:InterKSuppression}.
We nevertheless argue in Sec.~\ref{sec:Int-Dis} that interactions greatly enhance the validity of Eq.~\eqref{eqn:HDis1kapVal} over a broader parameter regime.

\subsection{Free Dirac fermions coupled to disorder}\label{sec:DisFreeFerm}

While we are interested in the situation where the disorder strength is subleading relative to  interactions, it is instructive to review the expected fate of the free Dirac theory at charge neutrality in several limits.
Consider first the single-Dirac-cone theory with disorder described by $H_{\mathrm{dis}}^{{\rm ultra}{\text -}{\rm smooth}}$.  
Having restricted to this minimal theory, it is  convenient to abandon smooth disorder and instead take white-noise correlations such that
$\smash{K_i(\vr/\xi_\mathrm{dis})=\xi^2_\mathrm{dis}\d^2(\vr)}$.
Physically, this simplification implies that we are probing the system at long enough scales relative to $\xi_\mathrm{dis}$ that all correlations in $R_i$ are washed away.
The disorder correlation length is then encoded in the dimensionless (up to factors of $\hbar$ and $v_F$) disorder strength parametrized by ${g}_i^2\xi_\mathrm{dis}^2$.

Ludwig \emph{et~al.}~\cite{Ludwig94} analyzed the effect of each of the three remaining  disorder fields--- $\smash{\mathcald{m}(\vr)}$, $\smash{\mathcald{a}_{x,y}(\vr)}$, and $\smash{\mathcald{v}(\vr)}$. 
In the absence of all other types of disorder, the random mass, vector potential, and scalar potential fields were individually found to be marginally irrelevant, exactly marginal, and marginally relevant in turn. Ludwig \emph{et al.} further postulated that when all three disorder types are simultaneously present, the system flows to the integer quantum Hall (IQH) plateau transition fixed point. 

The correspondence between Landau-level physics and disordered Dirac theories may be understood from the perspective of a Chalker-Coddington network model \cite{Chalker88}.
This model can be employed to efficiently study the transition between a trivial insulator with Landau-level filling  $\tilde{\nu}=0$ an IQH state with $\tilde{\n}=1$ (the tilde distinguishes Landau-level filling from the mTBG filling).
The system is assumed to locally prefer either $\tilde{\n}=0$ or $\tilde{\n}=1$---thus forming domains of trivial and IQH states whose detailed structure depends on the total filling and the disorder potential. 
As the total filling varies, either the trivial state percolates, with small ``lakes" of $\tilde{\n}=1$, or vice versa.
At some critical value, the system transitions between these two limits and becomes gapless. 
The network model exploits the fact that each boundary between $\tilde{\n}=0$ and $\tilde{\n}=1$ regions binds a chiral edge mode, and maps the problem onto one of directed links scattering at different nodes.

The key observation for our purposes is that the network model may be directly mapped onto a massless Dirac cone coupled to random fields $\mathcald{m}(\vr)$, $\mathcald{a}_{x,y}(\vr)$, and $\mathcald{v}(\vr)$ \cite{Ho96,Lee94}.
The correspondence between a lone disordered Dirac cone and the IQH plateau transition has been studied more recently in the context of \emph{monolayer} graphene \cite{Ostrovsky07},
where reducing the problem to that of a single Dirac cone only requires that disorder correlations are smooth on the scale of the microscopic lattice.
When the effective time-reversal symmetry of the single Dirac cone is broken by strain \cite{Morpugo06},
the appropriate nonlinear $\s$-model was shown to possess a topological term with $\th=\pi$; consequently, the system exhibits universal conductivity, just as predicted for the IQH plateau transition by Pruisken \cite{Pruisken88,Pruisken90}.

Upon resurrecting inter-$\vkp$-valley scattering in mTBG, disorder is instead described by $H_{\rm dis}^{\rm smooth}$.  Here the theory localizes in the thermodynamic limit, and the conductivity accordingly approaches zero even at charge neutrality \cite{Aleiner06,Altland06}.
Nevertheless, the localization length is expected to be extremely long since   
the scaling theory of Anderson localization indicates a lower critical dimension of $d=2$
\cite{gangof4,Lee85}.
The conductance thus only vanishes logarithmically with system size, suggesting that the localization length may be exponentially long, at least for a typical metal. 
Fradkin studied the fate of a system featuring $N_f$ massless Dirac cones in the large-$N_f$ limit \cite{Fradkin86a,Fradkin86b}.
Denoting the disorder strength for all processes simply by $g$, he obtained an exponentially large mean free path,
\eq{
\ell_\mathrm{mfp}\sim \aM \,\exp\[{\pi\o2}\(\hbar v_F\o g\,\xi_\mathrm{dis}\)^{\!\!2}\],
}
and a still-larger localization length 
$\xi_\mathrm{loc}\sim\ell_\mathrm{mfp}\,\smash{\exp\big(64 N_f^2/9\big)}$.

\section{Interactions in the clean limit}
\label{sec:IntClean}

Turning away from the question of disorder, we now investigate the effect of interactions in a homogeneous sample (though we occasionally allude to disorder effects).
We begin with a discussion of the general form and magnitude of the interactions. 
 Drawing on numerical results, experimental observations, and symmetry considerations, we then argue that the quantum valley Hall  state
 is energetically competitive in interacting mTBG at charge neutrality.

\subsection{Coulomb interaction}\label{sec:CoulInt}

The Coulomb Hamiltonian $H_{C,\mathrm{tot}}={1\o2}\int_\vq V(\vq)\r(\vq)\r^\dag(\vq)$ encodes the leading interaction.  Here $V(\vq)$ is the Fourier-transform of the long-range Coulomb potential (which technically depends on both the layer and sublattice, but these microscopic corrections can be ignored for the purpose of our discussion).
The operator $\r(\vq)$ represents the Fourier transform of the full microscopic density. 
Specifically, we write $\r(\vq)=\smash{\sum_\ell\int_\vk \tilde{f}_\ell^\dag(\vk)\tilde{f}_\ell(\vk+\vq)}$, where $\tilde{f}_\ell(\vk)$ denotes the annihilation operator corresponding to one of the decoupled graphene monolayers, with $\ell$ a combined index labelling both layer and sublattice and $\vk$ taking values across the full microscopic BZ.
As explained in Sec.~\ref{sec:ContModel}, to a high degree of accuracy the flat-band wavefunctions are composed entirely of states originating proximate to the Dirac cones of the decoupled monolayers.
We focus on these important momenta by introducing operators $\smash{f_{\ell,n = \pm }(\vk)}\equiv\smash{\tilde{f}_\ell(\vk\pm \vK)}$ that are defined for $|\vk| \ll |\vK|$;
note that this ``small $\vk$'' condition does not necessarily imply that $\vk$ resides within the moir\'{e} BZ.
It follows that only the density operators $\r(\vq)$ and $\r(\vq\pm\vK)$ with $\vq$ small are physically relevant to the flat-band physics:
\eq{\label{eqn:FullDensityMicro}
\r(\vq)&\cong
\sum_{\ell,n}
\int_{\vk\;\mathrm{small}}f_{\ell,n}^\dag(\vk )f_{\ell,n}(\vk+\vq),
\nt
\r(\vq+\vK)
&\cong
\sum_\ell\int_{\vk\;\mathrm{small}}f_{\ell,+}^\dag(\vk)f_{\ell,-}(\vk+\vq)
\nonumber \\
&= \rho^\dagger(-\vq -\vK).
}
Inserting these definitions into our expression for $H_{C,\mathrm{tot}}$ we find $H_{C,\mathrm{tot}}\cong H_C+ H_C'$ where
\eq{\label{eqn:HcDef}
H_{C}
&=
{1\o2}\int_{\vq\;\mathrm{small}}V(\vq)\r(\vq)\r^\dag(\vq),
\nt
H_{C}'
&=
\int_{\vq\;\mathrm{small}}
V(\vq+\vK)
\r(\vq+\vK)\r^\dag(\vq+\vK). 
}

There is a vast separation of energy scales between $H_{C}$ and $H'_{C}$.
Since $V(\vq)\propto1/\abs{\vq}$, the largest contribution to $H_C$ comes from momenta $\vq$ within the moir\'{e} BZ, \emph{i.e.} $\abs{\vq}\lesssim\abs{\vkp}$.
On the other hand, in $H_C'$, the smallness of the internal momentum $\vq$ implies $V(\vq+\vK)\approx V(\vK)$.
It follows that the relative strength of $H_C$ and $H_C'$ is $V(\vK+\vq)/V(\vq)\lesssim V(\vK)/V(\vkp)\sim \abs{\vkp}/\abs{\vK}\sim\th\ll1$.  
Hereafter we focus our attention on the dominant term, $H_C$. 
Reverting to real space, the Coulomb potential is $V(\vr)=e^2/(4\pi \ep \abs{\vr})$.  For graphene on hBN, we estimate the dielectric constant to be $\ep\sim 8\ep_0$ with $\ep_0$ denoting the permittivity of free space. 
Using the moir\'{e} lattice spacing, $\aM=\smash{a/\big(2\sin(\th/2)\big)}$, where $a=\unit[0.246]{\text{nm}}$ is the lattice constant of monolayer graphene 
as a typical length scale,  
one finds a characteristic interaction energy $ V(\aM)\sim\unit[14]{\text{meV}}$ at the magic angle $\theta \sim 1.05^\circ$.  

Theory estimates the bandwidth of the flat bands to be about $\unit[10]{\text{meV}}$ and the splitting between van Hove peaks within those bands to be $\sim \unit[5]{\text{meV}}$ \cite{Carr19,Koshino18}.
The above Coulomb-interaction scale thus raises natural questions regarding the validity of our expansion about the Dirac cones at $\pm \vkp$ in Sec.~\ref{sec:FlBands}.
It appears that the entirety of the flat bands and perhaps even neighboring energy bands should be considered. Non-interacting simulations of mTBG systems with twist angle disorder, however, have been shown to increase the bandwidth with little change to the Dirac character at charge neutrality \cite{Wilson19}.
Moreover, STM measurements of the fully filled flat bands (\emph{i.e.}, in a regime where correlations are presumably less important) measure van Hove peak splittings of $\sim\unit[10-20]{\text{meV}}$ \cite{Kerelsky18,Choi19}---several times larger than the above theoretical estimate. 
The full bandwidth of the flat bands may therefore  significantly exceed $V(\aM)$, supporting our use of the Dirac theory.  

\subsection{Preferred ground state of single-flavour theory}\label{sec:Groundstate}

Before turning to the full theory, it is useful to examine interaction effects at charge neutrality in a minimal, single-flavour model that includes only one spin and one $\vK$-valley.  References~\onlinecite{Xie18,Lu19,Liu19,Choi19} addressed this problem numerically via self-consistent Hartree-Fock calculations.  
Liu~\etal~\cite{Liu19} incorporated Coulomb interactions in the continuum model while Choi~\etal~\cite{Choi19} studied a 10-band lattice model \cite{Po18b} with a simplified local interaction.  Both analyses find a  $\C_2\T$-breaking gapped state with Chern number $C = \pm1$ as the lowest-energy solution.

References~\onlinecite{Xie18,Lu19} also predict an interaction-induced gapped phase at charge neutrality. 
However, while certain parameter regimes again return a $\C_2\T$-breaking state with nonzero Chern number, other  regimes yield a $\C_2\T$-preserving, trivial insulator.  
The latter statement may seem at odds with our assertion in Sec.~\ref{sec:ContModel} that $\C_2\T$ protects the masslessness of the Dirac cones, but this protection only holds when the flat bands are energetically isolated.
In the calculations of Refs.~\onlinecite{Xie18,Lu19}, interactions close the gap separating the flat and dispersive bands, thus negating the protection conferred upon the Dirac cones by $\C_2\T$ symmetry.  It is worth noting that STM measurements show that the flat bands indeed remain isolated as a function of filling \cite{Kerelsky18,Choi19,Jiang19,Xie19}, and yet still resolve correlation effects.  We therefore view the formation of the $\C_2\T$-symmetric insulator as a less likely outcome.

Returning to the $\C_2\T$-breaking gapped states, we remark that from the perspective of the Dirac theory, it is natural to expect the phase with $C = \pm 1$ to be energetically favourable relative to the  $\C_2\T$-breaking trivial insulator with $C = 0$.
Recall from Sec.~\ref{sec:FlBands} that the mean-field order parameter for the $C = \pm 1$ state is $\psi^\dag \eta^z\psi$, which in principle can arise from a momentum-independent microscopic perturbation. The trivial $C = 0$ phase instead corresponds to an order parameter $\psi^\dag\eta^z\t^z\psi$ that yields opposite-sign masses for the Dirac cones at $\pm \vkp$---and hence cannot arise from a momentum-independent microscopic perturbation.  
In monolayer graphene the converse situation arises: the momentum-independent staggered sublattice potential generates a trivial insulator whereas the relatively baroque, momentum-dependent Haldane mass \cite{Haldane88} is instead required to enter a $C = \pm 1$ phase.  (This distinction reflects the fact that the Dirac cones at $\pm \vkp$ exhibit the same chirality in mTBG, while the Dirac cones at $\pm \vK$ in monolayer graphene have opposite chirality \cite{Zou18}.)  Spontaneously generating a Haldane mass in monolayer graphene is thus unnatural---see, \emph{e.g.}, Ref.~\onlinecite{Weeks2010}---and it is analogously difficult to spontaneously enter the $C=0$ phase in mTBG.

\subsection{\texorpdfstring{Inclusion of spin and $\vK$-valley flavours}{Inclusion of spin and K-valley flavours}}\label{sec:SpinInteractions}

\begin{figure}
	\centering
	\includegraphics[width=0.48\textwidth]{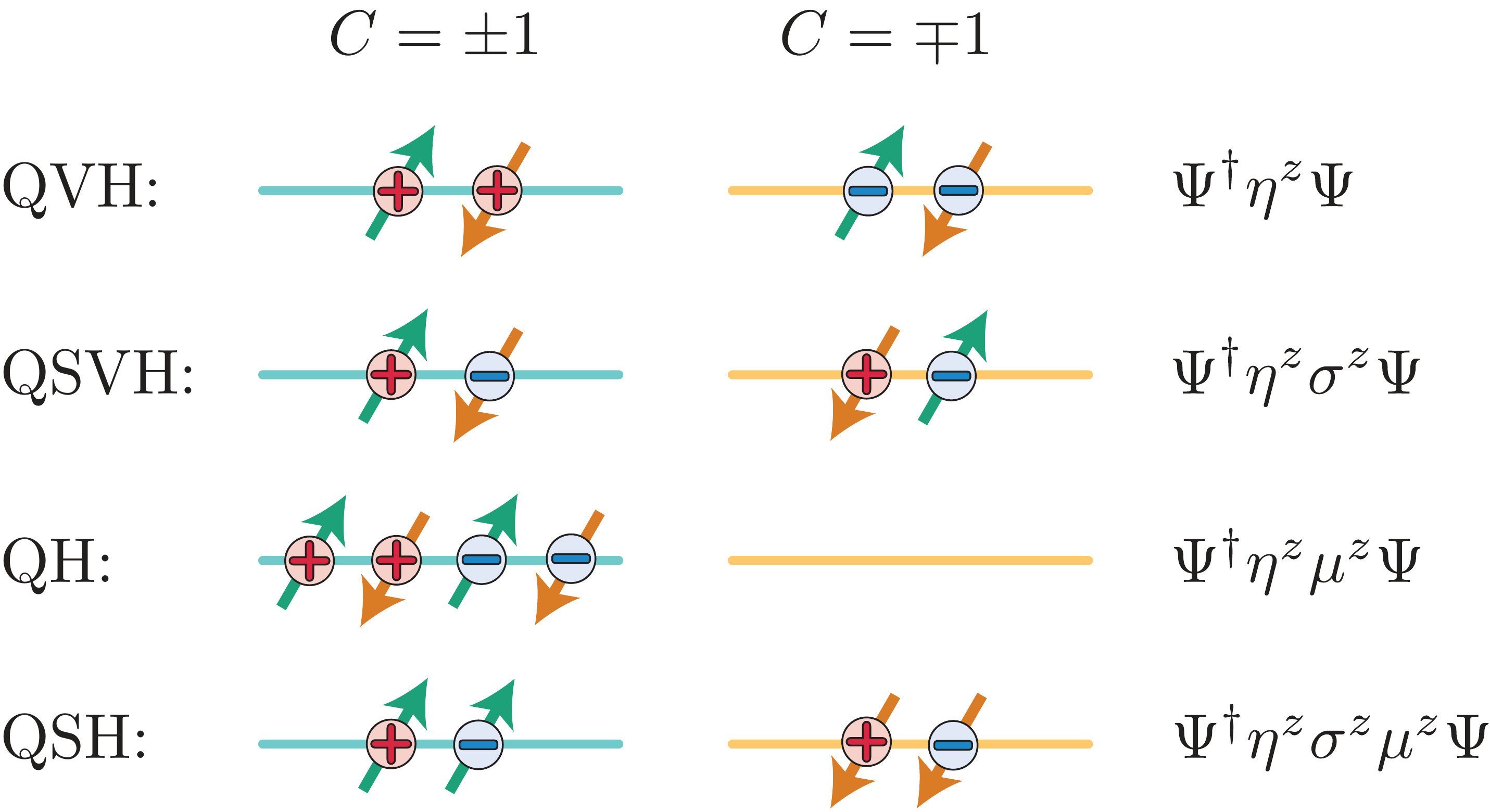}
	\caption{
	Four natural $\C_2\T$-breaking insulators at charge neutrality.
	In order from top to bottom: quantum valley Hall (QVH), quantum spin-valley Hall (QSVH), quantum Hall (QH), and quantum spin Hall (QSH).
	The direction of the arrow indicates spin, while the sign, `$+$' or `$-$,' labelling the arrow indicates the $\vK$-valley.
	}
	\label{fig:ChernNumberMass}
\end{figure}

We have so far argued that the single-flavour version of interacting, charge-neutral mTBG prefers to enter a gapped phase with Chern number $C = \pm 1$.  
Inclusion of spin and $\vK$-valley degrees of freedom not only allows for many distinct possible phases depending on the Chern numbers assigned to each sector, but further allows for additional phases that do not naturally descend from the single-flavour theory.  Let us begin by discussing the former.  

We specifically focus on four natural candidate insulators 
that we refer to as quantum valley Hall (QVH),  quantum spin-valley Hall (QSVH), quantum Hall (QH), and quantum spin Hall (QSH) phases.
Figure \ref{fig:ChernNumberMass} depicts these states along with their corresponding mass terms.  These insulators carry different symmetry properties as summarized in Table~\ref{tab:TopPhaseSymTrans}.
While all four  phases break  $\C_2\T$ symmetry, they do so in different ways: the QVH and QSVH states break $\C_2$ while preserving time reversal $\T$, whereas the converse is true of the QH and QSH states.
They are further distinguished by the action of the SU(2)$_s$ spin symmetry, which is preserved (broken) by the QVH and QH (QSVH and QSH) states.
Note that because of the additional $\vK$-valley flavour index, our QSH state differs from the 2$d$ topological insulator realized, \emph{e.g.}, in the Kane-Mele model \cite{KaneMele}.
We nevertheless adopt this nomenclature since the state breaks spin-rotation symmetry and preserves the `physical' electronic time reversal operation $\T_{\rm elec} \equiv i\s^y\T$ that obeys $\T_{\rm elec}^2 = -1$.

{\renewcommand{\arraystretch}{1.4}
\begin{table}[t]
	\centering
	\begin{tabular}{l c C{1.5cm} c C{1.5cm} c  C{1.cm} }
		&&	$\T$ &&	$\C_2$	&&	$\mathrm{SU}(2)_s$	\\\hline\hline
		QVH	&&	\cmark	&&	\xmark	&&	\cmark	\\
		QSVH	&&	\cmark	&&	\xmark	&&	\xmark	\\
		QH	&&	\xmark	&&	\cmark	&&	\cmark	\\
		QSH	&&	\xmark	&&	\cmark	&&	\xmark
	\end{tabular}
	\caption{
	Symmetry breaking pattern of the four topological states.
	Note that the QSH phase violates $\T$, but does preserve the `physical' electronic time reversal operation $\T_{\rm elec} = i\sigma^y \T$.
	}
	\label{tab:TopPhaseSymTrans}
\end{table}
}

It is useful to highlight some physical differences between these states and thus their compatibility with experimental observations.
Neither the QVH nor QSVH insulator is expected to possess gapless edge modes at a sample boundary. 
Our discussion has made significant use of the approximate U(1)$_v$ valley symmetry, but this symmetry is violently broken by the edge itself, which naturally occurs on the microscopic length scale $a$ of the underlying graphene monolayers. 
As a result, the edge modes from the two valleys scatter strongly, resulting in a purely insulating state.
By contrast, the QH state hosts robust gapless edge modes that are completely immune from scattering by virtue of their chirality.  Edge modes of the QSH insulator, while nonchiral, are nevertheless also robust since backscattering at a sample boundary must be accompanied by a spin flip.  
The sample studied by Lu~\etal \cite{Lu19} displayed  insulating transport with no signs of edge conduction. 
Among the four insulators, QH and QSH states thus appear unlikely, at least in that platform. 

As a result of the separation of scales between $\vK$-valleys and the SU(2)$_s$ symmetry, all four insulating states have very similar energies.
In Appendix~\ref{app:QSHvsSinglet}, we compare the QVH ground-state energy against the other three  insulators using a simple Hartree-Fock variational approach.
We show that all four states are exactly degenerate in the chiral model \cite{Grisha19}, a version of the continuum model that possesses an exact particle-hole symmetry that renders it exactly solvable.
Nevertheless, for more realistic versions of the continuum model (where particle-hole symmetry is absent), we find that the QVH state is actually disfavoured relative the other  insulators.
However, when computed numerically, we find the energy difference to be extremely small, less than $\sim\unit[10^{-5}]{\text{meV}}$ per electron, implying that the explicit breaking of particle-hole symmetry has little effect.

We turn now to alternative phases.
Polarized phases---for which the flat bands of two flavours are fully occupied---represent one class of competing ground states.
In general, both spin- and valley-polarized phases are degenerate at charge neutrality when $H_{C}'$ [Eq.~\eqref{eqn:HcDef}] is neglected  \cite{Zhang19b,LeeJY19}.
Liu~\etal~\cite{Liu19} find that, within the chiral model, these polarized states have identical Fock energies to the $\C_2\T$-breaking insulators with nontrivial Chern number in each flavour.
They also obtained self-consistent versions of these solutions numerically using a more realistic version of the continuum model; while no longer exactly degenerate, these states remained close in energy. 
Adding explicit $\C_3$-breaking strain---as observed in multiple STM experiments \cite{Kerelsky18,Choi19,Jiang19}---was, however, found to promote the $\C_2\T$-breaking insulators over the polarized states.  
Another proposed state is the inter-valley coherent phase (IVC) \cite{Po18a}, which spontaneously breaks  $\mathrm{U}(1)_v$ symmetry by coupling the $+\vK$ and $-\vK$ bands. 
General considerations \cite{Zhang19b} as well as calculations using the analytically tractable chiral model \cite{Liu19} indicate that IVC order is disfavoured at the Hartree-Fock level.
Other numerics nevertheless challenge these conclusions \cite{NickMURI}.

Importantly, among the gapped phases discussed here, only the QVH order parameter directly couples to disorder that is smooth and preserves $\T$ and SU(2)$_s$ spin symmetry. 
Time reversal and spin symmetry forbid coupling to the order parameters for QSVH, QH, QSH, and polarized phases, whereas  smoothness of disorder  prohibits coupling to an IVC order parameter. 
Hence, even if one of the latter states is energetically favourable in a perfectly clean system, the unavoidable presence of inhomogeneity in any physical sample may nevertheless stabilize the QVH phase, a possibility that we explore in Sec.~\ref{sec:CompPhase-Main}.

\section{Interplay of interactions and disorder}\label{sec:Int-Dis}

We are now in position to explore the fate of charge-neutral mTBG in the presence of interactions \emph{and} smooth disorder.  
Let us first recapitulate the expected behavior in the disordered, non-interacting limit (Sec.~\ref{sec:DisorderFree}) and in the clean but strongly interacting regime  (Sec.~\ref{sec:IntClean}):
\begin{enumerate}[topsep=3pt,parsep=0pt,itemsep=3pt]
\item  
In the absence of interactions, disorder localizes the massless Dirac fermions when any form of inter-$\vkp$-valley scattering is present in a manner that is formally analogous to physics of monolayer graphene. 
However, while monolayer graphene only requires that disorder be long-ranged on the scale of the microscopic lattice to avoid localization, disorder must be long-ranged on the scale of the moir\'{e} lattice to suppress localization in  twisted bilayer graphene.
\item
We have argued that in the strongly interacting, clean limit, the QVH phase that spontaneously breaks $\C_2$ symmetry constitutes (at the very least) an energetically competitive state that is compatible with experimental observations.  
Moreover, we observed that among various other candidate ground states, QVH order uniquely couples to smooth disorder respecting spin and time-reversal symmetries. 
\end{enumerate}
To simultaneously incorporate interactions and disorder below, we start with the assumption that the QVH state is the true ground state of the clean, interacting Hamiltonian. 
We construct an Ising formulation of the system in the presence of disorder, which allows us to systematically consider the crossover between the first and second panels of Fig.~\ref{fig:DisPhaseDiagram}.
We discuss the titular recovery of the massless Dirac cones before showing that even when the QVH insulator is not the true ground state in the clean theory, disorder may nevertheless tip the balance back in its favour.
We close with some comments on the eventual localization of the Dirac fermions, as illustrated in the final panel of Fig.~\ref{fig:DisPhaseDiagram}.

\subsection{Ising model formulation and domain formation}
\label{sec:DomainSize}

Suppose that the interaction energy scale dominates the physics, preferring to spontaneously break the $\C_2$ symmetry and form a QVH insulator.
Disorder terms that do not couple to the QVH order parameter can then be neglected, leaving only the random field $\mathcal{M}_0(\vr)$ that couples to $\psi^\dagger \eta^z\psi$ in Eq.~\eqref{eqn:HdisGeneric} (or, in the full theory, a random scalar field that couples to $\Psi^\dagger \eta^z\Psi$).
A random mass cannot produce localization 
but does compete against long-range order.
In fact, we show that even when disorder is weak and uncorrelated, the system always loses long-range order in the thermodynamic limit due to the formation of domains, as sketched in the central panel of Fig.~\ref{fig:DisPhaseDiagram}.
Destruction of long-range order only becomes observable, however, once the linear extent of the system, $L$, exceeds the typical domain size, $\xi_\mathrm{dom}$.
The goal of this subsection is to demonstrate that $\xi_\mathrm{dom}$ is finite and to determine its size as a function of the physical parameters of the theory.

We approach the problem in the standard fashion, via  the formulation of a Landau-Ginzburg theory.
The order parameter for the $\C_2$ symmetry breaking is simply an Ising field $\phi$ obtained by coarse graining the bilinear $\psi^\dagger\eta^z\psi$, \emph{i.e.}, 
\eq{\label{eqn:PhiDef}
\phi(\vr)
&\sim
\int_{\vr'\in R(\vr)} \psi^\dag \eta^z\psi(\vr')
\sim
\ell^2_\mathrm{UV}\psi^\dag \eta^z\psi(\vr),
}
where $R(\vr)$ is a spatial region centred at $\vr$ of typical size $\ell^2_\mathrm{UV}$ and $\ell_\mathrm{UV}$ is an ultraviolet cutoff quantified below.
Since we are interested in the physics deep within the ordered phase with $\Braket{\phi}\neq0$, a \emph{classical} Ising model suffices:
\eq{\label{eqn:IsingHamDef}
H_\mathrm{Ising}
&=
\int_\vr\Big[
\mathcald{K}\!\(\v{\nabla}\phi\)^2 + {r\o2}\phi^2 + {u\o4!}\phi^4
\Big].
}
The mass $r$ is clearly assumed to be negative.

The scales of the original fermionic Hamiltonian ultimately determine  parameters of the Ising model, though this assignment is not necessarily straightforward.  Consider first $\mathcald{K}$.  
Since $\phi$ is dimensionless, $\mathcald{K}$ has units of energy, and hence $\mathcald{K}\sim U$, with $U$ a characteristic energy scale of the system.
Both the rough estimate for the Coulomb potential, $V(\aM)\approx \unit[14]{\text{meV}}$, given at the end of Sec.~\ref{sec:CoulInt}, and the experimentally measured transport gap at charge neutrality, $\Delta_\mathrm{CNP}\approx\unit[1]{\text{meV}}$ \cite{Lu19}, provide natural candidates for $U$.
Given uncertainties in our calculation of $V(\aM)$ related to screening from other bands, we view the latter option as a more reasonable and conservative estimate.  We stress however that this choice has little direct bearing on the discussion that follows.

It is also important to assign a length scale to the interactions and hence the Ising theory.
Since our primary goal is to describe domain-wall physics, the most natural scale is
\eq{\label{eqn:XiIntDef}
\xi_\mathrm{int}\sim{\hbar v_F\o\Delta_\mathrm{CNP}}\cCom
}
which corresponds to the decay length of a Dirac fermion of mass $\Delta_\mathrm{CNP}/v_F^2$. 
In our context, these fermions are the chiral modes that bind to the domain walls at which the Chern numbers for each flavour change sign, identifying $\xi_\mathrm{int}$ as the domain boundary width.
Any physics occurring on scales smaller than $\xi_\mathrm{int}$  necessarily includes these fermionic degrees of freedom, and hence lies outside our Ising formulation's regime of validity.
The interactions length scale therefore defines a UV cutoff.\footnote{
The definition of $\xi_\mathrm{int}$ and $\ell_\mathrm{UV}$ is largely independent of our choice of $U$.
}
As a consistency check, we must verify that $\xi_\mathrm{int}$ exceeds the moir\'{e} lattice constant, $\aM\approx \unit[12.8]{\text{nm}}$.
Inserting $v_F\approx\unit[0.15\times10^6]{\text{m/s}}$ \cite{Cao18b} and $\Delta_\mathrm{CNP}\approx\unit[1]{\text{meV}}$ \cite{Lu19} into Eq.~\eqref{eqn:XiIntDef}, we indeed find $\xi_\mathrm{int}\approx \unit[100]{\text{nm}}\sim10\,\aM$.
We are therefore permitted to set $\ell_\mathrm{UV}\sim\xi_\mathrm{int}$.
In turn, dimensional analysis gives $r,\,u\sim U/\xi_\mathrm{int}^2$.  

Because disorder breaks $\C_2$, it should couple to the Ising field in a manner that breaks the $\Zt$ Ising symmetry. 
In other words, disorder appears as a random ``magnetic" field:
\eq{\label{eqn:PhiRandFieldDef}
H_{\phi,\mathrm{dis}}
&=
\int_{\vr}\mathcald{B}(\vr)\phi(\vr),
}
where $\smash{\mathcald{B}(\vr)\sim \int_{\vr'\in R(\vr)}\mathcal{M}_0(\vr')/\xi_\mathrm{int}^4}$.
The random field $\mathcal{M}_0$ is defined by the disorder strength $\d m$, correlation length $\xi_\mathrm{dis}$, and correlation function $K(\vr/\xi_\mathrm{dis})$ (in the notation of Sec.~\ref{sec:TypesOfDis}, these quantities correspond to $\smash{{g}_{\mathcal{M}_0}}$, $\smash{\xi_{\mathcal{M}_0}}$, and $\smash{K_{\mathcal{M}_0}}$, respectively).
We focus on the situation where the disorder is Gaussian correlated: $K(\vr/\xi_\mathrm{dis})=\smash{e^{-\vr^2/(2\xi_\mathrm{dis}^2)}}$.
Our assertion that the interaction energy scale dominates the disorder energy scale can now be more precisely stated as ${\d m/U}\ll1$.

In summary, the Hamiltonian controlling the ordering of $\phi$ is $H_\mathrm{RFIM}=H_\mathrm{Ising}+H_{\phi,\mathrm{dis}}$, which is none other than the much-studied random field Ising model (RFIM)\footnote{
This theory and its derivation should not be confused with the fact that a \emph{free} Dirac fermion with random mass disorder maps onto the random \emph{bond} Ising model \cite{Dotsenko83,Ludwig94}.
} \cite{Imry75,Natterman88,Natterman97}.
As claimed, the RFIM in 2$d$ is generically disordered \cite{Binder83,Aizenman89}, and so $\xi_\mathrm{dom}$ is finite.  
The mechanism of domain formation depends largely on the magnitude of the ratio
\eq{\label{eqn:xDef}
\a&={\d m\o U}{\xi_\mathrm{dis}\o\xi_\mathrm{int}}\cdot
}
This result and the scenarios we outline below are derived and further explained in Appendix~\ref{app:DisDomain}.

We first examine what occurs when $\a\gtrsim1$.
Since $U/\d m$ is already presumed large, in order for $\a$ to be larger than unity, this limit corresponds to that of extremely smooth disorder: $\xi_\mathrm{dis}/\xi_\mathrm{int}\gg1$.
In this scenario, the energy gained by having $\phi$ align in the direction preferred by $\mathcald{B}(\vr)$ is larger than the interaction energy cost associated with the misalignment of $\phi$ along the domain boundary.
The Ising field therefore directly tracks the disorder potential, implying that
\eq{\label{eqn:Ldom-LargeDisCorr}
\xi_\mathrm{dom}&\sim \xi_\mathrm{dis},
&
\a\gtrsim1.
}

The situation is more subtle when $\a\lesssim1$.
With stronger interactions, we naturally expect larger domains.
At some point, the domains are large enough that the correlated nature of the disorder is washed away, allowing us to treat it as white noise: $\overline{\mathcal{M}_0(\vr)\mathcal{M}_0(\vr')}\cong \d m^2\xi_\mathrm{dis}^2\d^2(\vr-\vr')$.
In this case, the destruction of long-range order occurs through the condensation of domain walls.
An evaluation of the domain-wall roughening yields a lower bound for their size of \cite{Binder83} 
\eq{\label{eqn:Ldom-SmDisCorr}
\xi_\mathrm{dom}
&\lesssim
\max\!\(\xi_\mathrm{int},\xi_\mathrm{dis}\)\,e^{c/\a^2},
&
\a&\lesssim1,
}
where $c\sim\mathcald{O}(1)$ is a non-universal constant.
We can verify that when $\a\ll1$ the domain length scale is indeed far greater than the disorder correlation length, \emph{i.e.}, $\xi_\mathrm{dom}\gg \xi_\mathrm{dis}$.

\subsection{Recovery of massless Dirac fermions}\label{sec:ResurgentDirac}

Next we discuss the physical consequences of the Ising model outlined above in the regime where the system size $L$ exceeds the typical domain size $\xi_\mathrm{dom}$.
For now we continue to assume suppression of both inter-$\vkp$- and inter-$\vK$-valley scattering. 
At least close to the crossover scale $\xi_\mathrm{dom}$, the Ising formulation should remain valid: the system is characterized by multiple domains of opposing Chern numbers with typical size $\xi_\mathrm{dom}$, as the central panel of Fig.~\ref{fig:DisPhaseDiagram} illustrates.
In this regime, the system can be described by eight independent Chalker-Coddington network models \cite{Chalker88}---one for each of the two Dirac cones within the four spin/valley sectors. 
As mentioned briefly in the introduction and  more fully in Sec.~\ref{sec:DisFreeFerm}, each network model may be mapped directly onto that of a single gapless Dirac cone \cite{Ho96,Lee94}, thus giving the promised restoration of massless Dirac fermions from a strongly correlated starting point.

We can alternatively motivate the recovery of massless Dirac cones without relying on network models.  Let us return to the full disordered Dirac theory described by Eqs.~\eqref{eqn:BothValleyDirHam} and \eqref{Hdis}, which includes all spin and valley degrees of freedom.  Notably, here we additionally allow for weak inter-valley scattering terms.  Upon including strong correlations at a mean-field level, interactions dramatically enhance the effective strength of the random field that couples to the QVH order parameter $\Psi^\dagger \eta^z \Psi$.  All other disorder fields, by contrast, remain weak and can be neglected to a first approximation.  The problem then reduces to a set of independent Dirac cones, each governed by the far simpler disorder Hamiltonian in Eq.~\eqref{eqn:HDis1kapVal} with \emph{only} random mass disorder.  As noted earlier, the random mass is a marginally irrelevant perturbation to the clean Dirac theory when it is the sole source of disorder \cite{Ludwig94}.  Massless Dirac fermions thus naturally re-emerge from this viewpoint as well.  

At sufficiently low energy scales, however, the additional disorder fields neglected above eventually kick in.  The dominant corrections are expected to arise from \emph{intra}-$\vkp$-valley scattering processes, encoded by the vector- and scalar-potential terms in Eq.~\eqref{eqn:HDis1kapVal}.  
When these terms are also present the theory is believed to flow to the IQH plateau transition, which is characterized by a finite a density of states with both universal longitudinal and Hall conductances (here, valley Hall).
At still lower energy scales, inter-$\vkp$-valley scattering is expected to produce localization, as Sec.~\ref{sec:Localization} discusses in more detail.  

Nevertheless, the perspective just outlined should be viewed as a consistency check and not a proof of concept. 
Crucially, it cannot account for the energy scales separating the Dirac fermions of the clean, non-interacting mTBG system  from the recovered Dirac cones of the interacting, disordered network model. 

\subsection{Competing phases}\label{sec:CompPhase-Main}

\begin{figure}
	\centering
	\includegraphics[width=0.47\textwidth]{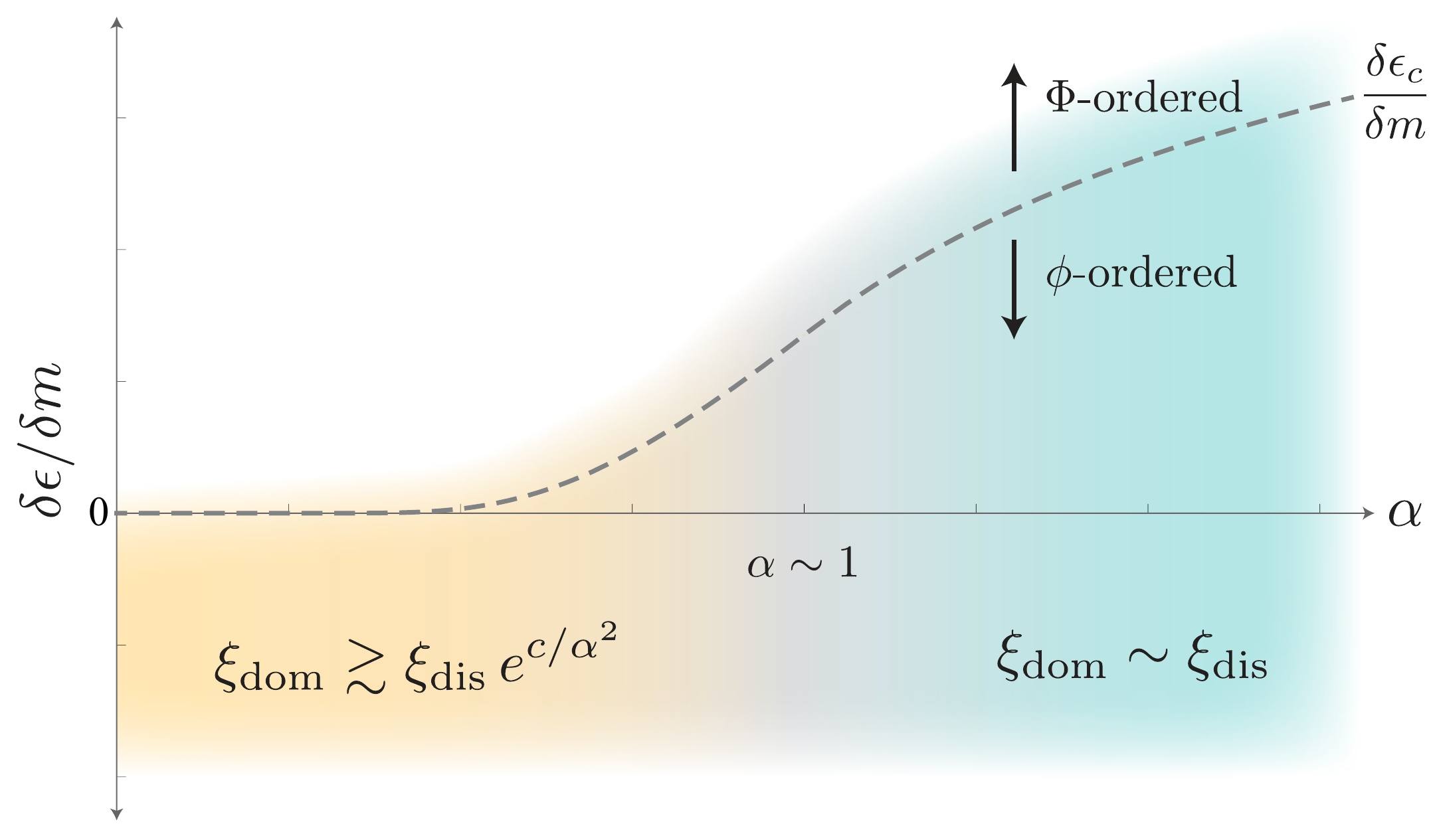}
	\caption{
	Schematic phase diagram as a function of disorder, $\a$, and ground state energy difference.
	The dashed line indicates the `critical' energy difference $\d\ep_c(\a)$ that characterizes the crossover from samples that are primarily $\phi$-ordered to  those that are primarily $\Phi$-ordered.
	There are two distinct $\phi$-ordered regimes.
	In the blue region, $\a\gtrsim1$, $\Braket{\phi}$ tracks the disorder so that domains are of the same size as the disorder correlation length, $\xi_\mathrm{dom}\sim\xi_\mathrm{dis}$.
	Conversely, in the orange region, $\a\lesssim1$, the correlated nature of the disorder is unimportant, and domains are exponentially large, $\smash{\xi_\mathrm{dom}\gtrsim \xi_\mathrm{dis}e^{c/\a^2}}$ [here, we assume that $\xi_\mathrm{dis}>\xi_\mathrm{int}$; see Eq.~\eqref{eqn:Ldom-SmDisCorr}].
	In the white region above the dashed line, the competing phase prevails, and $\Braket{\Phi}\neq0$ throughout most of the sample.
	}
	\label{fig:CompPhase}
\end{figure}

So far in this section, the QVH insulator has been taken as the true ground state of mTBG at charge neutrality, even in the absence of disorder.
We now address the possibility that interactions prefer a different state. 
To simplify the problem, we consider the situation in which a single competing phase is energetically favourable relative to the QVH insulator. 
In accordance with the conventions of Sec.~\ref{sec:DomainSize}, this competition can be quantified through the energy difference $\d\ep$ in an area of size $\ell_\mathrm{UV}^2=\xi_\mathrm{int}^2$:
\eq{
{\d\ep\o\xi_\mathrm{int}^2}\equiv\mathcald{E}_\mathrm{QVH}-\mathcald{E}_\mathrm{C}
\geq0,
}
where $\mathcald{E}_\mathrm{QVH}$ and $\mathcald{E}_\mathrm{C}$ respectively denote the ground-state energy densities of the QVH state and competing phase.
We further assume that the competing order may be described by an Ising field, $\Phi$, that does not linearly couple to any disorder field; recall the discussion at the end of Sec.~\ref{sec:SpinInteractions}.
Generalizing our arguments to include continuous order parameters (as needed for the QSH and QSVH insulators) is straightforward, and we therefore leave the competing phase's identity unspecified.

We again work in a regime where strong interactions obviate the need to include all disorder fields save for the random mass $\mathcal{M}_0$ that linearly couples to the QVH order parameter via Eq.~\eqref{eqn:PhiRandFieldDef}.  
Importantly, this type of disorder locally promotes the QVH state by lowering its energy relative to the competing phase, even though---as we saw earlier---it generally destroys true long-range order.  
When $\d\ep$ is small enough, we expect the majority of the sample to realize the QVH phase and the scenario outlined in the previous section to hold.
In terms of the Ising theory devised in the previous section, we can express this condition as
\eq{\label{eqn:phi-ordCond}
\[{1\o\mathrm{vol}}\int_\vr\,\Braket{\phi^2(\vr)}\]^{1/2}\gtrsim{1\o2}\cCom
}
where `$\mathrm{vol}$' denotes the sample volume.
When this equation holds, we say the system is `$\phi$-ordered'; otherwise, the system is `$\Phi$-ordered.'

Appendix~\ref{app:CompetingOrders} explores this problem in depth, ultimately deriving the schematic phase diagram shown in Fig.~\ref{fig:CompPhase}.
We again find that the primary control parameter is the ratio $\a$ [recall Eq.~\eqref{eqn:xDef}] corresponding to the horizontal axis.  
Motivated by the notion that $\Phi$-ordered regions may be viewed as annealed `vacancies,' we begin  
with a dilute Ising-model description.
At the lattice level, the theory is conveniently formulated by promoting the Ising variables $\sigma^z = \pm 1$ to three-state spin-1 variables $s$, where $s = \pm 1$ correspond to the two QVH phases and $s = 0$ corresponds to the competing phase.  We present a simple mean-field solution to the classical Blume-Capel model for these spin-1 degrees of freedom \cite{Blume66,Capel66,Kaufman90,Vasseur10} in Appendix~\ref{app:BlumeCapel}.
While the phase diagram we obtain resembles the one shown in Fig.~\ref{fig:CompPhase} in many respects, it erroneously predicts long-range $\phi$-order when $\d\ep<0$ and disorder is sufficiently small, $\a\lesssim1$; as discussed in Sec.~\ref{sec:DomainSize} and Appendix~\ref{app:DisDomain}, in reality, long-range order is unstable to the addition of any finite disorder.
This failure of mean-field theory is not unprecedented given the low dimensionality.

In Appendix~\ref{app:CompIsing-ImryMa}, we therefore return to the Imry-Ma type arguments of Sec.~\ref{sec:DomainSize} (see also Appendix~\ref{app:DisDomain}), which allow us to derive a `critical' energy difference $\d\ep_c(\a)$ that characterizes the crossover scale separating $\phi$- and $\Phi$-ordered regimes.
We plot $\d\ep_c(\a)$ with a dashed line in Fig.~\ref{fig:CompPhase}.
In the white region above the line, $\d\ep\gtrsim\d\ep_c(\a)$, the competing phase is realized throughout the majority of the system, and the network picture we propose is no longer relevant.
Conversely, Eq.~\eqref{eqn:phi-ordCond} holds in regions where $\d\ep\lesssim\d\ep_c(\a)$ (including the trivial case, $\d\ep<0$, where QVH states minimize the energy in the clean limit).
Just as we found above, depending on the strength of disorder, the destruction of long-range QVH order occurs in two fashions.
In Fig.~\ref{fig:CompPhase}, the parameter regime where $\Braket{\phi}$ tracks the disorder field is shown in turquoise.
The orange area indicates the opposite limit, where long-range order is eliminated by domain-wall condensation.
The intermediate regime where $\a\sim1$ is shown in neutral grey.

Notably, these considerations imply that disorder may not only be responsible for selecting which QVH order is locally realized, but that it may also determine whether or not QVH order is realized at all.
In particular, our proposal admits a scenario in which the clean samples of Lu~\emph{et al.} \cite{Lu19} are $\Phi$-ordered, while the less homogeneous samples of Cao~\emph{et al.} \cite{Cao18a,Cao18b} and Yankowitz~\emph{et al.} \cite{Yankowitz18} realize the QVH network picture displayed in the central panel of Fig.~\ref{fig:DisPhaseDiagram}---even supposing that the two sets of systems differ \emph{solely} in the amount of disorder they present.

\subsection{Localization}\label{sec:Localization}

\begin{figure}
\centering
\includegraphics[width=0.48\textwidth]{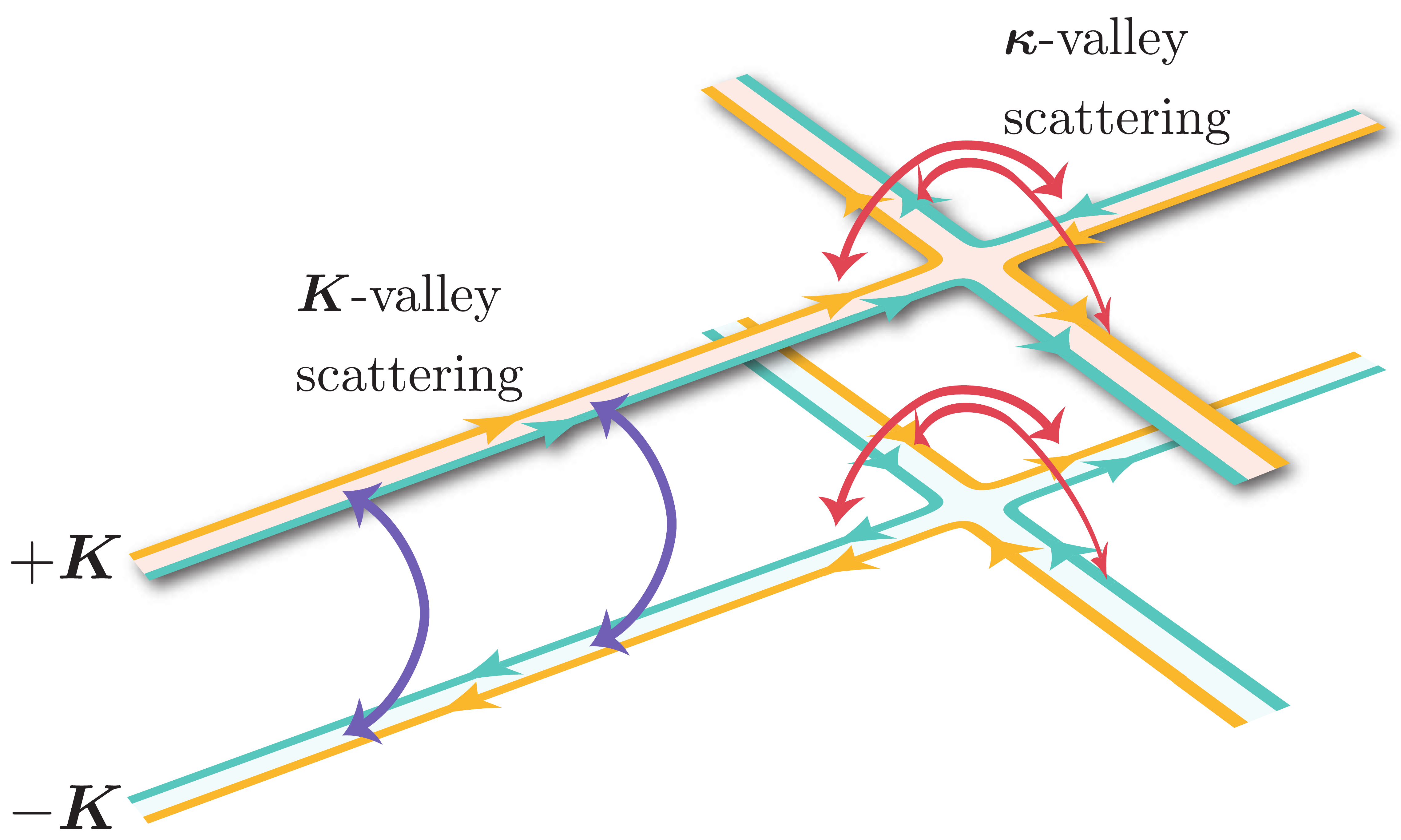}
\caption{Edge modes of the $+\vK$ and $-\vK$ valley sectors at a node connecting four domains.
The orange and turquoise arrows represent the chiral modes at the domain boundaries.
Red arrows at the node indicate $\mathrm{U}(1)_v$-preserving, inter-$\vkp$-valley scattering processes, which result from inhomogeneities at the moir\'{e} lattice scale, $\aM\approx\unit[12.8]{\text{nm}}$.
The $\mathrm{U}(1)_v$-breaking inter-$\vK$-valley scattering events are indicated by the purple arrows.
While this type of scattering is exponentially suppressed [see Eq.~\eqref{eqn:KK'suppression}], it can occur at any point along a domain boundary.
}
\label{fig:EdgeScattering}
\end{figure}

In the absence of any special symmetries, all two-dimensional systems are generically expected to localize in the thermodynamic limit, and our platform is no exception.
Localization is likely irrelevant for the previously studied mTBG samples, whose linear dimensions are $\sim\unit[2-8]{\mu\text{m}}\sim 150-600\aM$.  It is nevertheless instructive to briefly discuss  localization within our proposed scenario.  
The precise manner in which localization occurs in the presence of interactions poses a notoriously difficult and subtle problem that we will not wade into in detail.  Rather, our goal is to discuss some general features of the problem that can be deduced given some reasonable simplifying assumptions.

When discussing localization, one can imagine either increasing the system size or increasing the disorder strength.
In the latter case, the situation rapidly becomes unwieldy: as the disorder strength approaches the interaction energy ($\d m/U\to1$) or the disorder correlations become ultra-short-ranged ($\aM/\xi_\mathrm{dis}\to1$), our Ising formulation breaks down.
By contrast, the Ising-model perspective remains valid when we instead consider progressively larger samples with an otherwise identical set of parameters. 
Interactions can still of course pose complications; for instance, in the network-model picture localization involves a network of gapless domain-wall modes that generically form Luttinger liquids \cite{Chou19}. 
We do not address such subtleties, instead postulating that the primary effect of interactions is to catalyze the spontaneous breaking of $\C_2\T$.

The most straightforward manner by which the re-emergent Dirac fermions can localize is through inter-$\vkp$-valley scattering.
Such scattering events can also localize the original Dirac cones that appear in the free-fermion band structure for mTBG, but the physics is not quite identical: 
the network picture underlying the re-emergent Dirac cones effectively postpones localization by renormalizing the UV scale at which it occurs.
That is, if $\xi_{\mathrm{loc},\mathit{fr}}$ is the localization length in the free case, we have $\xi_\mathrm{loc}\sim \xi_\mathrm{dom}\xi_{\mathrm{loc},\mathit{fr}}/\aM$ with interactions.
One can intuitively understand this rescaling from the perspective of the gapless domain-wall modes in the network model.  As Fig.~\ref{fig:EdgeScattering} illustrates, in a given $\vK$-valley, the domain-wall modes corresponding to $\pm \vkp$ co-propagate, and hence non-forward-scattering processs can only occur at nodes where multiple domain walls intersect (see red arrows).  

Inter-$\vK$-valley scattering can also prompt localization \cite{QiaoZH11}.
Disorder coupling the two $\vK$-valleys has so far been completely ignored since it is exponentially suppressed relative to intra-$\vkp$-scattering [see Eq.~\eqref{eqn:KK'suppression}].
However, inter-$\vK$-valley scattering can occur at any point along the domain walls, as illustrated in  Fig.~\ref{fig:EdgeScattering}, making it a fundamentally one-dimensional process.
For very large domains, such intra-domain-wall scattering thus inevitably becomes the dominant localization mechanism. 
The localization length is then expected to be proportional to the mean free path of the domain-wall modes \cite{Evers08}, which is $\xi_\mathrm{loc}\sim \hbar v_F/g_{\mathit{KK'}}\sim\hbar v_F e^{4\pi^2\xi_\mathrm{dis}^2/a^2}/g$, and hence an exponentially large function of the disorder correlation length.

\section{Discussion}
\label{Discussion}

We have presented a theory that reconciles the seemingly conflicting experiments on charge-neutral mTBG by invoking a nontrivial interplay between strong interactions and weak disorder.
In our proposed picture, uniform order (QVH or otherwise) is realized throughout ultra-homogeneous samples, like those of Lu~\emph{et al.}~\cite{Lu19}, whereas QVH domains with opposite spin/valley Chern numbers appear in systems with more disorder, like the experiments of Cao~\emph{et al.}~\cite{Cao18a,Cao18b} and Yankowitz~\emph{et al.}~\cite{Yankowitz18}.
In the latter samples, gapless edge modes at domain boundaries form a network that may be mapped onto a theory of massless Dirac fermions, thereby explaining their semimetallic transport measurements.
By contrast, since a physical sample boundary strongly breaks the U(1)$_v$ symmetry protecting the edge modes, a uniformly ordered QVH state is an insulator at charge neutrality, in agreement with the observations of Lu~\emph{et al.} 
Both sample classes exhibit a local gap determined by the interaction strength---in harmony with STM experiments \cite{Kerelsky18,Choi19,Jiang19,Xie19}.  

The network model outlined in this paper is somewhat reminiscent of proposals aimed at describing `minimally' twisted bilayer graphene (minTBG) \cite{Efimkin18,WuXC19}.
When $\th\lesssim1^\circ$, it becomes energetically favourable for the microscopic lattices to distort such that the AB and BA regions occupying the moir\'{e} honeycomb sites enlarge at the expense of the AA regions situated at the centre of each moir\'{e} hexagon \cite{Woods14,vanWijk15,DaiSY16,Nguyen17,KimNY17,Gargiulo18,ZhangKuan18}; see Fig.~\ref{fig:BZ}(a) for an illustration of the undistorted case.
Under the application of a displacement field, AB and BA regions develop QVH order with opposing Chern numbers \cite{YinLJ16,JuL15,LiJ16}, yielding four edge modes per spin at the AB/BA boundaries.
While both our theory for mTBG and the theory proposed for minTBG are built on network models comprised of QVH domains, there are key qualitative distinctions that we wish to underscore.
The local QVH order in minTBG arises entirely as a single-particle effect, whereas the development of QVH order in our scenario relies principally on strong interactions. 
Moreover, the  shape and size of the AB and BA regions in minTBG are fixed; together, they comprise a single moir\'{e} unit cell.
The QVH domains discussed in this paper  instead result from the smooth disorder background and typically extend over many moir\'{e} unit cells.

\begin{figure*}[t]
	\centering
	\includegraphics[width=0.99\textwidth]{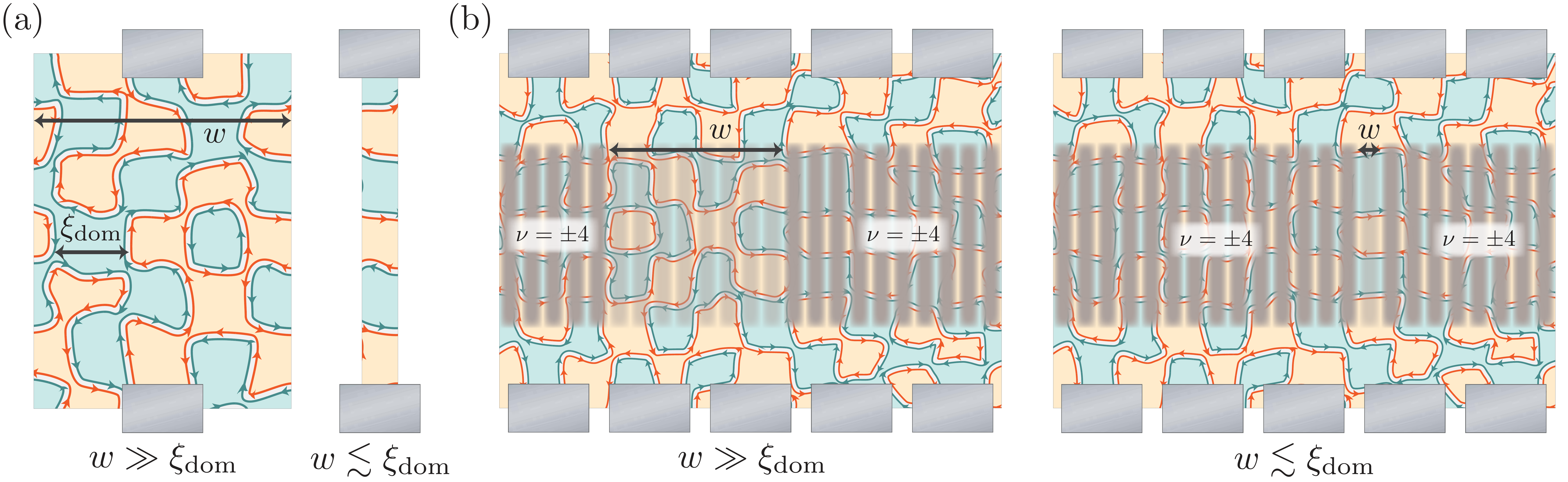}
	\caption{
	Schematic illustration of a proposed transport experiment. 
	The domain structure of one of the valley sectors is shown: regions carrying Chern number $C=+1$ are depicted in blue, whereas those carrying $C=-1$ are depicted in orange.
	The other valley sector is not shown explicitly.
	The grey rectangles above and below the sample indicate contacts through which the conductance is measured.
	(a) A single mTBG sample is sliced into multiple sub-systems of varying width $w$. 
	When $w\gg\xi_\mathrm{dom}$, as shown on the left, Dirac-like conductance is observed.
	Once the width is smaller than the typical domain size, $w\lesssim\xi_\mathrm{dom}$, the sample appears insulating, as shown on the right.
	(b) An alternate experiment in which the mTBG sample remains intact. 
	The taupe rectangles aligned in a row along the centre of the sample represent individually tunable gates through which the chemical potential may be locally varied.
	In regions where these gates are opaque, the chemical potential lies within the superlattice bandgap, \emph{i.e.} the flat bands are either completely empty or full ($\n=\pm4$).
	The system is tuned to charge neutrality in all other regions (either the gates are transparent or no gates are shown).
	On the left, $w\gg\xi_\mathrm{dom}$, and a semimetallic conductance should be observed.
	Conversely, since $w\lesssim\xi_\mathrm{dom}$, a large resistance is expected on the right.
	}
	\label{fig:TransportExp}
\end{figure*}

Our proposal is supported by available experimental data and crucially can be further tested in future experiments.  
One natural direction is to employ large-area STM scans to locally probe both gapped domains \emph{and} gapless domain-wall modes.  (To our knowledge evidence of the latter in mTBG has not yet been reported in the literature.)  
Samples that are simultaneously amenable to STM and transport would offer additional insight; for instance, the presence of gapless domain-wall modes should correlate with semimetallic transport, whereas such modes should be absent in homogeneous insulating samples.  Some caveats are warranted, however. 
First, discussions of local phenomena in STM measurements are often complicated, \emph{e.g.}, by disorder- or tip-induced localized states, and it may be difficult to unambiguously distinguish the domain physics we propose from such effects.
Additionally, the samples studied by Refs.~\onlinecite{Cao18a,Cao18b,Yankowitz18,Lu19} are enclosed on both sides by hBN, preventing STM study.  The nature of transport in mTBG with hBN only on the bottom side, as in the samples studied by STM in  Refs.~\onlinecite{Kerelsky18,Choi19,Jiang19,Xie19}, poses an interesting open question.  

One can also investigate our scenario entirely within transport \cite{DeanPrivateComm}. 
Consider a single mTBG sample etched into a series of strips of varying widths $w$, as shown in Fig.~\ref{fig:TransportExp}(a).
Transport through a given strip depends sensitively on the value of $w$ relative to the typical domain size, $\xi_\mathrm{dom}$. 
When $w\gg\xi_\mathrm{dom}$---the limit presumably relevant to the experiments of Refs.~\onlinecite{Cao18a,Cao18b,Yankowitz18}---semimetallic transport should occur.
In the opposite limit, $w\ll \xi_\mathrm{dom}$, no edge modes connect the contacts and the strip should appear insulating.
This experiment may be modified to preclude possible variations in the conductivity resulting from intrinsic variations between the strips, such as their local twist angle. 
Instead of physically cutting the sample, a `strip' can be electrostatically generated through  spatially varying gate voltages: Within a channel of width $w$, the system is locally tuned to charge neutrality, whereas elsewhere the Fermi energy is  tuned to lie within the gap separating the flat and dispersive bands. 
One could then study the conductivity as a function of width $w$ for all regions within the sample. 
 Figure~\ref{fig:TransportExp}(b) illustrates this refined version of the experiment.

Our proposal also spotlights various other avenues for future study.  
The fate of the network under an applied magnetic field poses a particularly interesting problem.
One possibility is that the magnetic field simply stabilizes a different competing phase, thereby destroying the network.
If this transition occurs at fields strengths close to or below \unit[1]{T}, where quantum oscillations are first clearly resolved, the re-emergent Dirac theory is unlikely to produce observable Landau-fan phenomena.
The occurrence of such a transition is neither necessary nor expected, however.
In the case where the QVH network survives a broader magnetic-field window, there are two limits to consider.
When the magnetic length $\ell_B$ far exceeds the typical domain size $\xi_\mathrm{dom}$, quantum oscillations are expected to be insensitive to the re-emergent nature of the fermions, implying that a Landau fan corresponding to massless Dirac fermions should be observed at low fields.
Given that the magnetic length is already quite large at \unit[1]{T}, $\ell_B\approx\unit[25]{\text{nm}}\approx 2\aM$, this regime may be difficult to access experimentally (recall that the UV cutoff for our network model was $\xi_\mathrm{int}\sim\unit[100]{\text{nm}}$).
The opposite limit, $\ell_B\ll\xi_\mathrm{dom}$, appears to be more subtle.
It is conceivable that the gapless edge modes do not affect the quantum oscillations,
resulting in a Landau level spectrum similar to that of  \emph{massive} Dirac cones associated with the gapped QVH domains.
Alternatively, the system could exhibit physics reminiscent of the Hofstadter butterfly \cite{butterfly}, though it seems likely that nonuniformity of the domain sizes may hinder any clear signal.
Quantifying these issues could shed additional light on the experimental relevance of our scenario.

The role of interactions at the edges of the domains is another topic that we have not touched on.
The edge modes may display interesting interacting phenomena that could be studied through the well-controlled bosonization formalism.
In fact, Wu~\emph{et al.}~\cite{WuXC19} have analyzed this problem in the context of the minimal twist angle samples described above.
Further, while disorder is generically expected to localize the edges, the inclusion of interactions may have nontrivial consequences \cite{Chou19}.

We have said little regarding transport away from the charge neutrality point.
While its semimetallic nature dictates that the conductivity $\s$ increase with doping, it can do so in different ways.
If transport is ballistic, far enough away from charge neutrality, the conductivity should essentially track the density of states: $\sigma\propto\smash{\sqrt{\abs{n}}}$, where $n$ is the electron density \cite{Tworzydlo06,Peres06}.
Provided inter-$\vkp$-scattering is the most important form of disorder, we expect the mean free path of the network model to be rescaled, implying that ballistric transport may not be unreasonable.
That is, letting $\ell_{\mathrm{mfp},\mathit{fr}}$ be the mean free path of the non-interacting Dirac fermions, we may postulate that the mean free path of the recovered network Dirac fermions is $\ell_\mathrm{mfp}\sim \xi_\mathrm{dom}\ell_{\mathrm{mfp},\mathit{fr}}/\aM$.
On the other hand, in monolayer graphene, the linear dependence of the conductivity on density away from charge neutrality,  $\sigma\propto\smash{\abs{n}}$, is largely ascribed to long-range Coulomb scattering \cite{GrapheneTransport11,Ando06,Cheianov06,Nomura06,Hwang07,Nomura07,Trushin08,Katnelson09}.
While it seems unlikely that a similar mechanism would play an important role in mTBG, it is possible that twist-angle disorder (which can also be long-range) could have a similar effect \cite{Katnelson08}.

Finally, exploring the interplay between interactions and disorder at other integer fillings constitutes perhaps the most interesting future direction.  The charge-neutrality regime that we examined here offers the virtue that the system is `almost' insulating even at the band structure level---thereby facilitating the study of (at least certain) correlated insulators.  Accessing correlated insulating states at other fillings requires a far more drastic modification of the band fillings.  Generalizing our analysis to such cases could provide valuable insight into the observed phenomenology of mTBG.

\section*{Acknowledgements}

We are grateful to Cory Dean, Arbel Haim, Eslam Khalaf, Stevan Nadj-Perge, Felix von Oppen, Seth Whitsitt, and Andrea Young for illuminating discussions.   This
work was supported by the Army Research Office under Grant Award W911NF-17-1-0323; the NSF through
grant DMR-1723367; the Caltech Institute
for Quantum Information and Matter, an NSF Physics
Frontiers Center with support of the Gordon and Betty
Moore Foundation through Grant GBMF1250; the Walter Burke Institute for Theoretical Physics at Caltech;
and the Gordon and Betty Moore Foundation’s EPiQS
Initiative, Grant GBMF8682 to JA.
This work was performed in part at the Aspen Center for Physics, which is supported by National Science Foundation grant PHY-1607611.

\addvspace{1.4\baselineskip}
{\onecolumngrid

\appendix 
\linespread{1.4}
\fontsize{11pt}{11.5pt}\selectfont

\titleformat{\section}[hang]
     {\scshape\bfseries\large}
     {\thesection.	}
     {1em}
     {\MakeUppercase}
\titleformat*{\subsection}{\bfseries}

\titleformat{\subsubsection}[hang]
     {\itshape}
     {\thesubsubsection.	}
     {1em}
     {}


\section{Continuum Model}\label{app:ContModel}

We briefly outline the continuum model in this section.
Spin indices are completely suppressed below.
We first decompose the microscopic graphene operators as
\eq{
\tilde{f}_\ell(\vr)=e^{i\vK\cdot\vr}f_{+,\ell}(\vr)+e^{-i\vK\cdot\vr}f_{-,\ell}(\vr),
}
where $\ell$ indicates both layer and sublattice.
As discussed in Sec.~\ref{sec:ContModel}, the continuum model Hamiltonian decouples into $\vK$-valley sectors $H_\mathrm{cont}=H_++H_-$, where $H_\pm$ act on $f_{\pm,\ell}$.
For the moment, we consider $H_+$.
We express $f_{+,\ell}$ as a vector $\(f_{\mathcald{t},\mathrm{A}}(\vr),f_{\mathcald{t},\mathrm{B}}(\vr),f_{\mathcald{b},\mathrm{A}}(\vr),f_{\mathcald{b},\mathrm{B}}(\vr)\)$, where the `$+$' has been dropped for convenience, $\mathcald{t}$, $\mathcald{b}$ denote layer, and $\mathrm{A}$, $\mathrm{B}$ denote sublattice.
In this basis, $H_+$ acts as
\eq{\label{eqn:HcontModelDef}
{H}_+
&=
\begin{pmatrix}
iv_0\v{\eta}_{\th/2}\cdot\v{\nabla}	&	T(\vr)	\\
T^\dag(\vr)	&	iv_0\v{\eta}_{-\th/2}\cdot\v{\nabla} 
\end{pmatrix},
}
where $\v{\eta}_\phi=e^{-i\phi\eta^z/2}\(\eta^x,\eta^y\)e^{i\phi\eta^z/2}$ act on the sublattice space and $\v{\nabla}=\(\ptl_x,\ptl_y\)$.
The tunnelling matrix $T(\vr)$ is given by 
\eq{\label{eqn:TunnellingOperator}
T(\vr)&=\sum_{\ell=1,2,3}t_\ell e^{-i\vq_\ell\cdot\vr},
&
\vq_\ell&=\mathcald{R}_{2\pi(\ell-1)/3}\[\vK_\mathcald{t}-\vK_\mathcald{b}\]
}
where $\mathcald{R}_\phi[\v{v}]$ rotates the vector $\v{v}$ by $\phi$ 
and the matrices $t_\ell$ are defined through
\eq{\label{eqn:TunnellingOperator-2}
t_\ell
&=
e^{2\pi i(\ell-1)\eta^z/3}\begin{pmatrix} 
w_0	&	w_1	\\	w_1	&	w_0	
\end{pmatrix}
e^{-2\pi i (\ell-1)\eta^z/3}.
}
The physical parameters of the model are the twist angle $\th$, the velocity of the microscopic graphene layers $v_0$, and the tunnelling amplitudes, $w_0$ and $w_1$.
We take the angle to be close to the magic angle, $\th=1.05^\circ$, and the graphene velocity to be $v_0=\unit[9.1\times10^5]{\text{m/s}}$.
The tunnelling amplitudes are typically taken to be $\(w_0,w_1\)=\(85,110\)$~meV \cite{MacDonald11}.
However, for the chiral version [see Sec.~\ref{app:ChiralModelCalc}] of the model, we set $w_0=0$, keeping $w_1=\unit[110]{\text{meV}}$ \cite{Grisha19}.

The Hamiltonian corresponding to the other valley, $\mathcal{H}_-$, may be obtained by acting time-reversal ($\T$) or by rotating by $180^\circ$ ($\C_2$).

The continuum Hamiltonians maybe also be expressed in momentum space.
Returning to second quantized notation, it may be written
\eq{
H_\m
&=
\sum_{\v{G},\v{G}',\ell,\ell'}\int_{\vk\in\mathrm{BZ}}
f^\dag_{\m,\ell}(\vk+\v{G})H^{(\m)}_{\v{G},\ell;\v{G}',\ell'}(\vk)f_{\m,\ell'}(\vk+\v{G}'),
}
where $\m=+,-$ labels the $\vK$-valley and the $\v{G}$s are moir\'{e} reciprocal lattice vectors.
Here, $H^{(\m)}(\vk)$ may be thought of as an infinite matrix taking values within the moir\'{e} BZ with indices $\(\v{G},\ell\)$.
It can be diagonalized through
the unitary rotation
\eq{\label{eqn:ContModWaveFun}
c_{\m,i}^\dag(\vk)
&=
\sum_{\v{G},\ell}u_{\m,i;\v{G},\ell}(\vk)f^\dag_{\m,\ell}(\vk+\v{G}),
&
f_{\m,\ell}^\dag(\vk+\v{G})
&=
\sum_i u^*_{\m,i;\v{G},\ell}(\vk)c_{\m,i}^\dag(\vk),
}
where $i$ indexes the band.
In terms of the $c_{\m,i}(\vk)$ operators $H_\m$ is
\eq{
H_\m&=\sum_i \int_{\vk\in\mathrm{BZ}} c_{\m,i}^\dag(\vk) \ep_i(\vk) c_{\m,i}(\vk).
}
We note that invariance of $c_{\m,i}(\vk)$ under shifts of $\vk$ by a reciprocal lattice vector, $\vk\to\vk+\v{G}$, implies $u_{\m,i;\v{G},\ell}(\vk+\v{G}')=u_{\m,i;\v{G}+\v{G}',\ell}(\vk)$.

\section{Spin and time-reversal symmetric bilinears}
\label{app:Bilinears}

Here, we enumerate some of the symmetries of the Dirac theory.
It is convenient to express them in terms of a large, unphysical, SU(8) symmetry generated by the $\vK$-valley, $\vkp$-valley, and spin symmetries.
The generators of these symmetries are
\eq{
\mathrm{SU}(2)_s&:\quad(\s^x,\s^y,\s^z),
&
\mathrm{SU}(2)_\kappa&:\quad(\t^x,\t^y,\t^z),
&
\mathrm{SU}(2)_K&:\quad(\m^x\eta^y,\m^y\eta^y,\m^z).
}
Since the SU(2)$_K$ triplet does not take a particularly simple form, we define $\bar{\m}^i=(\m^x\eta^x,\m^y\eta^x,\m^z)$.
Finally, the $\g$-matrices are $\g^\m=(\m^z\eta^z,i\eta^y,-i\m^z\eta^x)$.
By combining the $\g^\m$, $\s^i$, $\t^i$, $\bar{\m}^i$, we can generate all bilinears (pairing terms are not considered).

We are interested only in those bilinears that preserve the spin and time-reversal symmetries.
Clearly, spin-conservation requires that disorder not couple to any bilinear containing $\s^i$, so we ignore it completely, treating $\Psi$ as a spinless fermion.
Time reversal then acts as
\eq{
\T:\qquad\qquad
\Psi
&\to
\m^x\t^x\Psi,
&
i&\to-i.
}
It follows that the SU(2)$_\k$, SU(2)$_K$ triplets and the $\g$-matrices map as
\eq{
(\t^x,\t^y,\t^z)
&\to
(\t^x,\t^y,-\t^z),
\nt
\T:\qquad\qquad
(\m^x\eta^y,\m^y\eta^y,\m^z)
&\to
-(\m^x\eta^y,\m^y\eta^y,\m^z),
\nt
(\g^0,\g^x,\g^y)
&\to
(-\g^0,\g^x,\g^y)
}
These transformation properties result in the following time-reversal invariant bilinears:
\eq{\label{eqn:TRSbilinears}
\bar{\Psi}M\Psi,&\quad M\in\{\bar{\m}^i,\bar{\m}^i\t^{x,y},\t^z\},
\nt
\bar{\Psi}\g^0M\Psi,&\quad M\in\{\id,\t^{x,y},\t^z\bar{\m}^i\},
\nt
\bar{\Psi}\g^{x,y}M\Psi,&\quad M\in\{\bar{\m}^i,\bar{\m}^i\t^{x,y},\t^z\},
}
where $\bar{\Psi}=\Psi^\dagger\g^0$.
We are most concerned the mass bilinears, shown on the first line.
We note that $\bar{\Psi}\bar{\m}^i\Psi=\Psi^\dag(\m^y\eta^z,-\m^x\eta^x,\eta^z)\Psi$.
The last term, $\Psi^\dagger\eta^z\Psi$, is the order parameter for the QVH state.
For completeness we also list the bilinears that break time-reversal symmetry:
\eq{\label{eqn:TRSbreakBilinears}
\bar{\Psi}M\Psi,&\quad M\in\{\id,\t^{x,y},\t^z\bar{\m}^i\},
\nt
\bar{\Psi}\g^0M\Psi,&\quad M\in\{\bar{\m}^i,\bar{\m}^i\t^{x,y},\t^z\},
\nt
\bar{\Psi}\g^{x,y}M\Psi,&\quad M\in\{\id,\t^{x,y},\t^z\bar{\m}^i\}.
}

\section{\texorpdfstring{Suppression of inter-$\vK$-valley scattering}{Suppression of inter-K-valley scattering}}
\label{app:InterKSuppression}

We briefly outline a schematic argument for the exponential suppression of inter-$\vK$-valley scattering processes.
We begin by considering the operators on the microscopic graphene lattice.
Suppose disorder couples as
\eq{
H_\mathrm{micro}
&=
\sum_{\ell,\ell'}
\int_\vr\mathcal{R}(\vr)\tilde{f}^\dag_{\ell}(\vr)T_{\ell\ell'}\tilde{f}_{\ell'}(\vr)
}
Here, $\ell$ labels both the layer and sublattice of the fermion $\tilde{f}_{\ell}(\vr)$, $T_{\ell\ell'}$ is a matrix whose precise form is unimportant, and $\mathcal{R}(\vr)$ is the disorder field with values drawn from a Gaussian probability distribution:
\eq{
\overline{\mathcal{R}(\vr)}&=0,
&
\overline{\mathcal{R}(\vr)\mathcal{R}(\vr')}&=g^2\,e^{-(\vr-\vr')^2/(2\xi_\mathrm{dis}^2)}.
}
In momentum space, we find
\eq{
H_\mathrm{micro}
&=
\int_\vk\mathcal{R}(\vq)\tilde{f}^\dag_{\ell}(\vk)T_{\ell\ell'}\tilde{f}_{\ell'}(\vk+\vq),
}
where
\eq{
\overline{\mathcal{R}(\vq)\mathcal{R}^*(\vq')}=\d^2(\vq-\vq')\,g^2\xi_\mathrm{dis}^2\,e^{-\vq^2 \xi_\mathrm{dis}^2/2}.
}
We now wish to expand about the $+\vK$ and $-\vK$ points.
Letting $\tilde{f}_{n=\pm,\ell}(\vk)=f_{\ell}(\pm\vK+\vk)$, the Hamiltonian divides into two pieces
\eq{
H_\mathit{KK}
&=
\sum_{n=\pm}\int_{\vk,\vq}\mathcal{R}(\vq)f_{n}^\dag(\vk)Tf_{n}(\vk+\vq),
\nt
H_\mathit{KK'}
&=
\int_{\vk,\vq}\sum_{j=1}^3\mathcal{R}(\vq+\vQ_j)f^\dag_+(\vk)Tf_-(\vk+\vq)+h.c.
}
where $\vQ_j$ are the three (smallest) momenta such that $-\vK+\vQ_j=+\vK$, each of which has magnitude $\abs{\vK}=4\pi/3a$, where $a$ is the lattice constant of monolayer graphene.
We have also suppressed the summation over the $\ell$ indices of the fermions and matrix $T$.
Letting $\mathcal{R}_{(+-)}(\vq)=\sum_j \mathcal{R}(\vq+\vQ_j)$, we then see
\eq{
\overline{\mathcal{R}_{(+-)}(\vq)\mathcal{R}^*_{(+-)}(\vq')}
&=\d^2(\vq-\vq')\xi_\mathrm{dis}^2g^2\,\sum_j e^{-(\vq+\vQ_j)^2\xi_\mathrm{dis}^2/2}
\nt
&=
\d^2(\vq-\vq')\xi_\mathrm{dis}^2g^2e^{-\vK^2\xi^2/2}e^{-\vq^2\xi_\mathrm{dis}^2/2}\sum_j e^{-\vq\cdot\vQ_j\xi^2}.
}
Ignoring the anisotropic term on the right, the disorder field corresponding $\vK\to-\vK$ scattering has the same correlation length, $\xi_\mathrm{dis}$, but with an exponentially suppressed amplitude: $g_{KK'}\sim g \smash{e^{-4\pi^2 \xi_\mathrm{dis}^2/a^2}}.$

These arguments may appear to carry over directly to the case of inter-$\vkp$-scattering, \emph{i.e.}, we may wish to conclude that the typical inter-$\vkp$ valley scattering amplitude $g_{\k\k'}$ is exponentially suppressed relative to the typical intra-$\vkp$ scattering amplitude $g$: $g_{\k\k'}\sim g \smash{e^{-4\pi^2\xi_\mathrm{dis}^2/\aM^2}}$.
However, in this case, there are additional subtleties to take into account.
While the continuum Hamiltonian does not mix the $f$ fermions on the scale of the large BZ, $\sim 1/a$, they \emph{are} mixed on the scale of the moir\'{e} BZ, $\sim1/\aM$.
In particular, the flat band operator $c$ (or, equivalently, the Dirac operator $\psi$) at a momentum quantum number $\vk$ in the moir\'{e} BZ is composed of a superposition of $f$ fermions with momenta $\vk+\v{G}$ (in the microscopic BZ), where the $\v{G}$s are moir\'{e} reciprocal lattice vectors, as indicated in Eq.~\eqref{eqn:ContModWaveFun}.
With the exception of $\v{G}=0$, all such reciprocal lattice vectors are already of order $\abs{\vkp}$ or larger.
Importantly, this mixing is responsible for the very flatness of the bands and therefore constitutes a nonnegligible effect.
As a result, unless $\xi_\mathrm{dis}$ is much, much larger than $\aM$, these higher moments may nevertheless contribute substantially to the $\vkp\to-\vkp$ scattering processes.
We therefore emphasize that the analysis and proposal presented in this paper is \emph{not} predicated on the assumption that $g_{\k\k'}$ is small.

\section{Mean field analysis of insulating phases}\label{app:QSHvsSinglet}

Based on numerical results, we argued in Sec.~\ref{sec:Groundstate} that the ground state of a single flavour theory with interactions is a Chern insulator.
Upon including valley and flavour indices in Sec.~\ref{sec:SpinInteractions}, we identified four natural insulating states distinguished by their symmetry action, as summarized in Table~\ref{tab:TopPhaseSymTrans}.
Further, we noted that only the order parameter for the QVH insulator could couple to disorder, which is vital for the scenario we propose.

Here, we discuss the circumstances under which the QVH insulator is or is not energetically preferred compared to the QSH, QH, and QSVH.
We determine the band structure using the continuum model (see Appendix~\ref{app:ContModel}), and, in spite of 
the concerns raised at the end of Sec.~\ref{sec:CoulInt},
we model the interactions using $H_C$, as written in Eq.~\eqref{eqn:HcDef}.
Moreover, to further simplify the calculation, we project $H_C$ onto the flat bands, a simplification that may admittedly neglect relevant contributions from the dispersive bands.
We therefore view this exercise mainly as a guide intended to expose trends rather than provide rigorous quantitative energetics.
Nevertheless, we show that within a simple mean field analysis, the Fock terms are not expected to distinguish these phases.
While it appears that the Hartree terms favour the QSH, QSVH, and QH insulators over the QVH phase, we find that this preference is not the case for the chiral model \cite{Grisha19}---they remain degenerate.
We next calculate the energy difference between the QVH and other phases numerically for a more realistic set of parameters and demonstrate that while the energy difference is no longer zero, it remains negligibly small.

\subsection{Flat band projection}

The Hamiltonian $H_C$ of Eq.~\eqref{eqn:HcDef} is still quite complicated: it includes all bands of the model, whereas we are only interested in what happens to the flat bands.
Since these bands are separated from the dispersive bands by a gap $E_g$ by assumption, the latter states can be integrated out to give an effective Hamiltonian acting only on the flat band subspace. 
The leading order contribution is obtained simply by projecting $H_C$ to the flat bands:
\eq{\label{eqn:HcLeading}
H_{C,1}&=\int_{\vq\,\mathrm{small}}V(\vq)\rho_{\mathit{fl}}(\vq)\r_{\mathit{fl}}(-\vq),
}
where $\r_\mathit{fl}(\vq)$ is the density operator projected onto the flat bands.

We show that the mean field decoupling of $H_0+H_{C,1}$ [where $H_0$ is given in Eq.~\eqref{eqn:FlatBandH0}] are independent of the sign of the Dirac mass.
To do so, we define the variational Hamiltonian 
$H_\mathrm{MF}(\{M_\m\})=\sum_{\m}H_\mathrm{MF}^{(\m)}(M_\m)$ 
where $\m=(n,\a)$ sums over both $\vK$-valleys, $n=\pm$, and spin, $\a=\ua,\da$.
The individual mean field Hamiltonians are
\eq{\label{eqn:MFham}
H_{\mathrm{MF}}^{(\m)}(M_\m)
&=
\int_{\vk\in BZ}c^\dag_{\m}(\vk)
\underbrace{\Big[h_\m(\vk)+ M_\m\eta^z \Big]
}_{\bar{h}_\m(\vk;M_\m)}
c_{\m}(\vk),
\nt
h_\m(\vk)
&=
h_{\m,0}(\vk)+ h_{\m,x}(\vk)\eta^x + h_{\m,y}(\vk)\eta^y .
}
We study the dependence of $\Braket{\{M_\m\}|H_0+H_{C,1}|\{M_\m\}}$ on the signs of $M_{\m}$, where $\Ket{\{M_\m\}}$ denotes the groundstate of $H_\mathrm{MF}(\{M_\m\})$.

\subsection{Density operator and form factors}

One complication of this calculation is the presence of form factors in the definition of the densities and thus $H_{C,1}$ as well. 
In particular, we have 
\eq{
\r_\mathit{fl}(\vq)
&=
\sum_\m\r_\m(\vq),
\nt
\r_\m(\vq)
&=
\sum_{\m,\ell}\int_{\vk\;\mathrm{small}}f_{\m,\ell}^\dag(\vk)f_{\m,\ell}(\vk+\vq),
}
where $f_{\m,\ell}(\vk)=f_{n=\pm,\a,\ell}(\vk)$ denotes the electron operator with spin $\a=\ua,\da$ and total momentum $\pm\vK+\vk$.
As in Sec.~\ref{sec:IntClean} and Appendix~\ref{app:ContModel}, $\ell$ labels both layer and sublattice.
In what follows we omit the label ``$\mathit{fl}$.''
Recall that neither the momentum of the density operator, $\vq$, nor the momentum being summed over, $\vk$, is required to lie within the moir\'{e} Brillouin zone.
We therefore instead write
\eq{
\r_\m(\vq+\v{G}')
&=
\int_{\vk\in\mathrm{BZ}}\sum_{\v{G},\ell}f^\dag_{\m,\ell}(\vk+\v{G})f_{\m,\ell}(\vk+\vq+\v{G}+\v{G}'),
}
where $\v{G}$ and $\v{G}'$ are moir\'{e} reciprocal lattice vectors and both $\vk$ and $\vq$ lie within the moir\'{e} BZ.
Using the defintion of $c_{\m,i}$ in relation to $f_{\m,\ell}$ given in Eq.~\eqref{eqn:ContModWaveFun}, the density may now be expressed directly in terms of the flat band creation and annihilation operators:
\eq{
\r_\m(\vq+\v{G})
&=
\sum_{ij\in\mathit{fl}}c_{\m,i}^\dag(\vk)
\lam_{\m;ij}(\vk,\vk+\vq+\v{G}) c_{\m,j}(\vk+\vq),
\nt
\lam_{\m;ij}(\vk,\vk+\vq+\v{G})
&=
\sum_{\v{G}',\ell} u_{\m,i;\v{G}',\ell}^*(\vk)u_{\m,j;\v{G}',\ell}(\vk+\vq+\v{G}) .
}
We frequently refer to the functions $\lam_{\m,ij}$ as `form factors' in what follows.
We have used the fact that the band operators are invariant under reciprocal lattice translations up to a phase, $c_{\m,j}(\vp+\v{G})=e^{i\phi} c_{\m,j}(\vp)$. 
From the fact that $u_{\m,i;\v{G},\ell}(\vk+\v{G}')=u_{\m,i;\v{G}+\v{G}',\ell}(\vk)$, we also have $\lam_{\m;ij}(\vk,\vk'+\v{G})=\lam_{\m,ij}(\vk-\v{G},\vk')$.
Finally, with this notation, the flat-band Coulomb interaction is
\eq{
H_{C,1}
&=
\int_{\vq,\vk,\vk'}\sum_{\v{G}}\sum_{\m,\n}
c_\m^\dag(\vk)\lam_\m(\vk,\vk+\vq+\v{G})c_\m(\vk+\vq)
\cdot
c_\n^\dag(\vk'+\vq)\lam_\n(\vk'+\vq+\v{G},\vk')c_\n(\vk').
}

\subsection{Symmetry constraints}

We begin by discussing the symmetry properties of the mean-field kernel $\bar{h}_\m(\vk;M_\m)$.
We begin with the symmetry transformations
\eq{\label{eqn:FlBandSymAction}
\T&:\qquad
c(\vk)\to \m^x c(-\vk),
\nt
\C_2\T&:\qquad
c(\vk)\to \eta^xc(\vk),
}
where $\m^x$ acts on the $\vK$-valley indices and $\eta^x$ acts on the (flat) band indices.
Both are anti-Hermitian, taking $i\to-i$.
In terms of the mean field Hamiltonian, they imply
\eq{\label{eqn:MFhamTrans}
\bar{h}_{+,\a}(\vk;M)
&=
\bar{h}^*_{-,\a}(-\vk;M),
&
\bar{h}_{\m}(\vk;M)
&=
\eta^x\bar{h}^*_{\m}(\vk;-M)\eta^x.
}
Obviously, since $h_\m(\vk)=\bar{h}_\m(\vk;M=0)$, these relations also hold for the non-interacting part of the flat band Hamiltonian.

We now define the projector 
\eq{
P_{\m;ij}(\vk;M)
&=
\Braket{c_{\m,j}^\dag(\vk)c_{\m,i}(\vk)}_{\!M}.
}
The subscript $M$ is used as a shorthand to denote which mean field Hamiltonian the ground state begin used to compute the expectation value is associated with.
The equalities of Eq.~\eqref{eqn:MFhamTrans} then imply
\begin{subequations}
\eq{
\label{eqn:ProjConstraintTRS}
P_{+,\a}(\vk;M)
&=
P_{-,\a}^T(-\vk;M),
\\
\label{eqn:ProjConstraintC2T}
P_{\m}(\vk;M)
&=
\eta^x P_\m^T(\vk;-M)\eta^x.
}
\end{subequations}
Note that $P_\m^\dag(\vk;M)=P_\m(\vk;M)$.
Similarly, we find that the form factors must satisfy
\begin{subequations}
\eq{
\label{eqn:FormFactorConstraintTRS}
\lam_{+,\a}(\vk,\vk+\vq)
&=
\lam_{-,\a}^T(-\vk-\vq,-\vk),
\\
\label{eqn:FormFactorConstraintC2T}
\lam_\m(\vk,\vk+\vq)
&=
\eta^x \lam^T_\m(\vk+\vq,\vk)\eta^x.
}
\end{subequations}

\subsection{Evaluation of mean field Hamiltonian}

We wish to compute the expectation value $\Braket{\{M_\m\}|H_0+H_{C,1}|\{M_\m\}}$.
This function may be separated into three pieces:
\eq{
\Braket{\{M_\m\}|H_0+H_{C,1}|\{M_\m\}}
&=
\Braket{H_0}_{\{M_\m\}}
+
H_F(\{M_\m\})
+
H_H(\{M_\m\}),
}
where $H_F$ and $H_H$ are the Fock and Hartree decouplings of the Coulomb interaction.
These three terms are discussed in the following subsections.

\subsubsection{\texorpdfstring{Quadratic term: $\Braket{H_0}$}{Quadratic term: <H0>}}

We write the quadratic part of the Hamiltonian as a sum over the valleys and spins, $H_0=\sum_\m H^{(\m)}_0$, where
\eq{
H_0^{(\m)}
&=
\int_\vk c_\m^\dag(\vk) h_\m(\vk)c_\m(\vk).
}
The kernel $h_\m(\vk)$ is defined in Eq.~\ref{eqn:MFham}.
Taking the expectation value, we find
\eq{
\Braket{H_0^{(\m)}}_{M_\m}
&=\int_\vk 
\tr\big[P_\m(\vk;M_\m)h_\m(\vk)\big].
}
Inserting the relations given in Eqs.~\eqref{eqn:ProjConstraintC2T} and~\eqref{eqn:FormFactorConstraintC2T}, we arrive at
\eq{
\Braket{H_0^{(\m)}}_{M_\m}
&=
\int_\vk\tr\big[\eta^x P^T_\m(\vk;-M_\m)\eta^x \eta^x h_\m^T(\vk)\eta^x]
\nt
&
=
\Braket{H_0^{(\m)}}_{-M_\m}.
}
Hence, we have verified that $\Braket{H_0}$ is independent of the signs of the mass terms.

\subsubsection{\texorpdfstring{Fock term: $H_F$}{Fock term}}

The Fock term is 
\eq{
H_F(\{M_\m\})&=\sum_\m
H^{(\m)}_F(M_\m),
\nt
H_F^{(\m)}(M_\m)
&=
-\int_{\vk,\vp} \sum_{\v{G}}V(\vp-\vk+\v{G})
\tr\[\lam_\m(\vk,\vp+\v{G})P_\m(\vp;M_\m)\lam_\m(\vp+\v{G},\vk)P_\m(\vk;M_\m)\].
}
Inserting the relations from Eqs.~\eqref{eqn:ProjConstraintC2T} and~\eqref{eqn:FormFactorConstraintC2T}, we find
\eq{
H_F^{(\m)}(M_\m)
&=
-\int_{\vk,\vp}\sum_{\v{G}}V(\vp-\vk+\v{G})
\tr\big[
\lam_\m^T(\vp+\v{G},\vk)P_\m^T(\vp;-M_\m)\lam_\m^T(\vk,\vp+\v{G})P_\m^T(\vk;-M_\m)
\big]
\nt
&=
H_F^{(\m)}(-M_\m).
}
We again conclude that the Fock contribution is independent of the sign $M_\m$ takes.
 
\subsubsection{\texorpdfstring{Hartree term: $H_H$}{Hartree term}}\label{app:HartreeTerm}

The Hartree term can be written
\eq{
H_H(\{M_\m\})
&=
\sum_{\v{G}}V(\v{G})\sum_{\m,\n}\Braket{\r_\m(\v{G})}_{M_\m}\Braket{\r_\n(-\v{G})}_{M_\n}.
}
We therefore begin by calculating $\Braket{\r_\m(\v{G})}_M$:
\eq{
\Braket{\r_\m(\v{G})}_M
&=
\int_\vk
\tr\big[
P_\m(\vk;M)\lam_\m(\vk,\vk+\v{G})
\big].
}
We use the constraints imposed by time reversal [Eqs.~\eqref{eqn:ProjConstraintTRS} and~\eqref{eqn:FormFactorConstraintTRS}] to relate the expectation values of the densities of the two valleys to one another:
\eq{
\Braket{\r_{+,\a}(\v{G})}_M
&=
\int_\vk\tr\big[
P_{-,\a}^T(-\vk;M)
\lam_{-,\a}^T(-\vk-\v{G},-\vk)
\big]
=
\int_\vk\tr\big[
P_{-,\a}(\vk;M)\lam_{-,\a}(\vk,\vk+\v{G})
\big]
\nt
&=
\Braket{\r_{-,\a}(\v{G})}_{M}.
}
We see that the expectation value of the density operator is independent of the valley and spin degree of freedom, motivating us to define the function
\eq{
R(M;\v{G})\equiv\Braket{\r_\m(\v{G})}_M.
}
Note that the identity $\r_\m(\v{G})=\r_\m^\dag(-\v{G})$ implies $R(M;\v{G})=R^*(M;-\v{G})$.
The $\C_2\T$ symmetry [Eqs.~\eqref{eqn:ProjConstraintC2T} and~\eqref{eqn:FormFactorConstraintC2T}] then gives,
\eq{
\Braket{\r_\m(\v{G})}_M
&=
\int_\vk
\tr\big[
P_\m^T(\vk;-M)\lam_\m^T(\vk+\v{G},\vk)
\big]
=
\int_\vk 
\tr\big[
P_\m(\vk;-M)\lam_\m(\vk,\vk-\v{G})
\big]
\nt
&=
\Braket{\r_\m(-\v{G})}_{-M}
=
\Braket{\r_\m(\v{G})}^*_{-M}.
}
We conclude that $R(-M;\v{G})=R^{*}(M;\v{G})$.

The Hartree term is therefore
\eq{
H_H(\{M_\m\})
&=
\sum_{\v{G}}V(\v{G})\Big|\sum_\m R(M_\m;\v{G})\Big|^2.
}
The relative signs of the mass terms of the four states under consideration are shown in Tab.~\ref{tab:MassTerms}.
Separating $R(M;\v{G})$ into real and imaginary parts, $R(M;\v{G})=R'(M;\v{G})+iR''(M;\v{G})$, we conclude that 
\eq{\label{eqn:HartreeEnergies}
H_H^\mathrm{QVH}&
=16\sum_{\v{G}}V(\v{G})\[R'(M;\v{G})^2+R''(M;\v{G})^2\],
\nt
H_H^\mathrm{QSVH}
&=
H_H^\mathrm{QH}
=
H_H^\mathrm{QSH}
=
16\sum_{\v{G}}V(\v{G})R'(M;\v{G})^2.
}
It follows that the QVH state is \emph{higher} in energy than the other three insulating states by $16\sum_{\v{G}}V(\v{G})R''(M;\v{G})^2$.

We note that since $\lam(\vk,\vk)=\id$, for $\v{G}=0$ we necessarily have $R''(M;\v{0})=0$, implying that for this term at least, there is no difference in energy between the QVH insulator and the other three.
In a typical tight-binding model, the $\v{G}=0$ term accounts for the entirety of the Hartree energy.
For the continuum model, however, the internal spatial structure of the wavefunctions also affects the Hartree energy.
Nevertheless, the form factors $\lam_\m(\vk,\vk+\v{G})$ decay quite quickly as a function of $\v{G}$ \cite{Liu19}---implying that the spatial variation of the density within the unit cell is not too large.
As we discuss in the next two sections, the contribution from $R''(M;\v{G})$ is essentially negligible.

{\renewcommand{\arraystretch}{1.4}
\begin{table}[t]
	\centering
\begin{tabular}{lcccccccc}
		&&	$\quad M_{+,\ua}$ 	&& $\quad M_{+,\da}$	&&	$\quad M_{-,\ua}$	&&	$\quad M_{-,\da}$
		\\\hline\hline
	QVH	&&	$\ph{-}1$	&&	$\ph{-}1$	&&	$\ph{-}1$	&&	$\ph{-}1$	\\
	QSVH	&&	$\ph{-}1$	&&	$-1$	&&	$\ph{-}1$	&&	$-1$	\\
	QH	&&	$\ph{-}1$	&&	$\ph{-}1$	&&	$-1$	&&	$-1$	\\
	QSH	&&	$\ph{-}1$	&&	$-1$	&&	$-1$	&&	$\ph{-}1$	\\
\end{tabular}
\caption{Relative signs of the mass terms corresponding to the four phases depicted in Fig.~\ref{fig:ChernNumberMass}.}
\label{tab:MassTerms}
\end{table}
}

\subsection{Chiral model}\label{app:ChiralModelCalc}

We show that in the chiral model \cite{Grisha19}, the functions $R(M;\v{G})$ are purely real, implying that the Hartree terms are all degenerate.
The chiral model is a particular case of the continuum model in which hopping only occurs between A and B sites both within and between graphene layers.  
This constraint is implemented by setting $w_0$ in Eq.~\eqref{eqn:TunnellingOperator-2} to zero.
The result is an exact particle-hole (chiral) symmetry $\Gamma$ that interchanges positive and negative energy states.
We follow the discussion in the Appendix of Ref.~\onlinecite{Liu19}.
$\Gamma$ may be assumed to act as
\eq{
\Gamma&:\qquad
c(\vk)\to \eta^zc(\vk).
}
In fact, in this basis, the sublattice index of the $c(\vk)$'s can be identified with the A and B sublattices of the two layers.
It's then convenient to reinterpret the wavefunctions written in Eq.~\eqref{eqn:ContModWaveFun}, $u_{\m,i;\v{G},\ell}(\vk)$. 
We explicitly identify the index $i=\mathrm{A},\mathrm{B}$ with the sublattice, leaving $\ell$ to denote the layer.
It then follows that the form factor may be written
\eq{
\lam_{\m,ij}(\vk,\vk'+\v{G})
&=
\Big[
\lam^{(0)}_{\m}(\vk,\vk'+\v{G})\id_{2\times2} 
+
i\lam^{(z)}_{\m}(\vk,\vk'+\v{G})\eta^z
\Big]_{ij},
}
where both $\lam^{(0)}_\m$ and $\lam^{(z)}_\m$ are real functions.

An additonal symmetry allows one to rotate the two layers in opposite directions.
The authors of Ref.~\onlinecite{Grisha19} use this observation to simplify the problem substantially, resulting in an exact expression for the ground state wavefunction at the magic angle.
For any angle, however, it implies that the Hamiltonian of Eq.~\eqref{eqn:HcontModelDef} satisfies
\eq{
\begin{pmatrix}
iv_0\v{\eta}_{\th/2}\cdot\v{\nabla}	&	T(\vr)	\\
T^\dag(\vr)	&	iv_0\v{\eta}_{-\th/2}\cdot\v{\nabla} 
\end{pmatrix}
&=
\eta^z\tau^z
\begin{pmatrix}
-iv_0\v{\eta}_{\th/2}\cdot\v{\nabla}	&	T(\vr)	\\
T^\dag(\vr)	&	-iv_0\v{\eta}_{-\th/2}\cdot\v{\nabla} 
\end{pmatrix}
\t^z\eta^z,
}
where Pauli operators $\eta^z$ and $\tau^z$ act on the sublattice (A,B) and layer $(\mathcald{t},\mathcald{b})$ indices respectively.
The continuum representation of the wavefunction given in Eq.~\eqref{eqn:ContModWaveFun} therefore satisfies
\eq{
u_{\m,i;\v{G},\ell}(\vk) = e^{i\vphi_\vk}\sum_{i',\ell'}\eta^z_{ii'}\t_{\ell\ell'}^zu_{\m,i';-\v{G}\ell'}(-\vk),
}
which in turn implies
\eq{
\lam_\m(\vk,\vk'+\v{G})=\lam_\m(-\vk,-\vk'-\v{G}).
}
Similarly, the mean field Hamiltonian must give $\bar{h}_\m(\vk;M)=\bar{h}_\m(-\vk;M)$ and therefore
\eq{
P_\m(\vk;M)=P_\m(-\vk;M).
}
These relations provide an additional constraint on the form of $\Braket{\r(\v{G})}_M$:
\eq{
\Braket{\r_\m(\v{G})}_M
&=
\int_\vk
\tr\[ P_\m(\vk;M)\lam_\m(\vk,\vk+\v{G})\]
=
\int_\vk
\tr\[ P_\m(-\vk;M)\lam_\m(-\vk,-\vk-\v{G})\]
\nt
&=
\int_\vk
\tr\[ P_\m(\vk;M)\lam_\m(\vk,\vk-\v{G})\]
\nt
&=
\Braket{\r_\m(-\v{G})}_M
=
\Braket{\r_\m(\v{G})}_M^*.
}
That is, $R(M;\v{G})$ is \emph{real}: $R''(M;\v{G})=0$. 
From Eq.~\eqref{eqn:HartreeEnergies}, we conclude that the Hartree energies corresponding to all four insulating states are fully degenerate in the chiral limit:
\eq{
H_H^{\mathrm{QVH}}
=
H_H^\mathrm{QSVH}
&=
H_H^\mathrm{QH}
=
H_H^\mathrm{QSH}.
}

\subsection{Numerical evaluation of Hartree term}

\begin{figure}[t]
\includegraphics[width=0.95\textwidth]{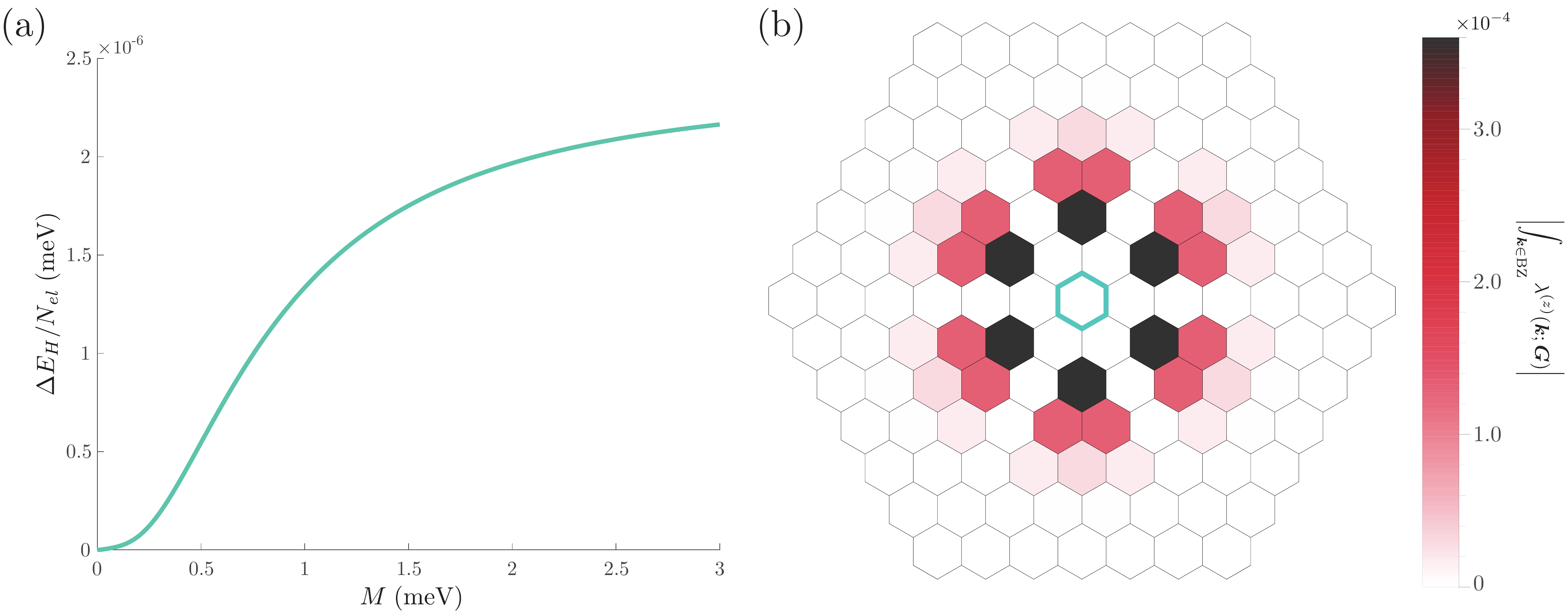}
\caption{(a) Energy difference as a function of the variational mass $M$ between the QVH phase, $H_H^\mathrm{QVH}$, and the other three phases, $H_H^\mathrm{other}=H_H^\mathrm{QSH}=H_H^\mathrm{QH}=H_H^\mathrm{QSVH}$, per electron at charge neutrality: $\Delta E_H/N_\mathit{el}=(H_H^\mathrm{QVH}-H_H^\mathrm{other})/N_\mathit{el}$.
(b)
Colour plot of $\big|\int_\vk \lam^{(z)}_\m(\vk;\v{G})\big|$ as a function of the moir\'{e} reciprocal lattice vector $\v{G}$.
Each hexagon represents a different $\v{G}$, with the central hexagon outlined in turquoise corresponding to $\v{G}=0$.
Noticeably, $\smash{\big|\int_\vk\lam^{(z)}(\vk;\v{G})\big|}=0$ along all mirror axes, as we showed in the main text.
}
\label{fig:HartreeNumerics}
\end{figure}

We now return to the non-chiral version of the model.
In Fig.~\ref{fig:HartreeNumerics}(a) we plot the energy difference per electron of the Hartree term for the model using the parameters given in Appendix~\ref{app:ContModel} as a function of the Dirac mass $M$.
Even for a mass $M=\unit[3]{\text{meV}}$, the energy difference is as small as $\unit[2.5\times10^{-6}]{\text{meV}}$ --- certainly our rough model is not expected to be reliable for such small energy differences.

We can understand the smallness in several ways.
As mentioned at the end of Appendix~\ref{app:HartreeTerm}, the form factors $\lam(\vk,\vk+\v{G})$ decay quite quickly as a function of $\v{G}$.
We can further show that $R''(M;\v{G})=0$ for all $\v{G}$ such that $\v{G}=\M_{y}[\v{G}]$, $\v{G}=\C_3\M_{y}[\v{G}]$, or $\v{G}=\C_3^2\M_{y}[\v{G}]$.
To do so, we start by using the fact that a basis exists in which Eq.~\eqref{eqn:FlBandSymAction} holds and the mirror symmetry acts as \cite{Zou18}
\eq{
\M_{y}:\qquad
c(\vk)\to \eta^xc(\M_{y}[\vk]).
}
Since $h_\m(\vk)$ satisfies the symmetry whereas the mass term $M\eta^z$ does not (\emph{e.g.} $\bar{h}_\m(\vk;M)=\eta^x\bar{h}_\m(\M_y[\vk];-M)\eta^x$), we must have
\eq{
P_\m(\vk;M)&=\eta^xP_\m(\M_{y}[\vk];-M)\eta^x,
&
\lam_\m(\vk,\vk+\v{G})
&=
\eta^x\lam_\m(\M_{y}[\vk],\M_{y}[\vk+\v{G}])\eta^x.
}
We therefore find
\eq{
\Braket{\r_\m(\v{G})}_M
&=
\int_{\vk\in\mathrm{BZ}}
\tr\[\eta^xP_\m(\M_{y}[\vk];-M)\eta^x\eta^x\lam_\m(\M_{y}[\vk],\M_{y}[\vk+\v{G}])\eta^x\]
\nt
&=
\int_{\vk\in\mathrm{BZ}}
\tr\[P_\m(\vk;-M)\lam_\m(\vk,\vk+\M_{y}[\v{G}])\]
\nt
&=
\Braket{\r_\m\(\M_{y}[\v{G}]\)}_{-M}
=
\Braket{\r_\m\(\M_{y}[\v{G}]\)}_{M}^*.
}
It follows that $\Braket{\r_\m(\v{G})}_M$ is \emph{real} for all moir\'{e} reciprocal lattice vectors such that $\v{G}=\M_{y}[\v{G}]$: $R''(M;\v{G}=\M_{y}[\v{G}])=0$.
The reflection axis chosen for $\M_{y}$ was actually arbitrary---by $\C_3$ rotational symmetry, the same should hold for the two equivalent axes given by $\C_3\M_{y}$ and $\C_3^2\M_{y}$.
Notably, this means that $R''(M;\v{G})=0$ for the shortest set reciprocal lattice vectors.

We can quantify the size of $R''(M;\v{G})$ for arbitrary $\v{G}$ through the follow set of observations.
First, we note that the energies of the flat bands may be written as $E_{\m,\pm}(\vk)=h_{\m,0}(\vk)\pm\ep_\m(\vk)$, where $\ep^2_\m(\vk)=h_{\m,x}^2(\vk)+h_{\m,y}^2(\vk)$.
This allows us to express the projection matrix as
\eq{
P_\m(\vk;M)
&=
{1\o2}\( \id - {1\o \sqrt{\ep_\m^2(\vk)+M^2}}\( h_{\m,x}(\vk)\eta^x + h_{\m,y}(\vk)\eta^y + M\eta^z\)\).
}
It then follows that
\eq{\label{eqn:imRdef}
{R''(M;\v{G})}&={1\o2}
\({\Braket{\r_\m(\v{G})}_M-\Braket{\r_\m(\v{G})}_{-M}}\)
=
-{1\o2}\int_{\vk\in\mathrm{BZ}}{M\o \sqrt{\ep_\m^2(\vk)+M^2}}\tr\[\eta^z\lam_\m(\vk,\vk+\v{G})\]
\nt
&=
-\int_{\vk\in\mathrm{BZ}}
{M\o \sqrt{\ep_\m^2(\vk)+M^2}}\lam_\m^{(z)}(\vk;\v{G})\cCom
}
where we've defined
\eq{
\lam^{(z)}_\m(\vk;\v{G})
&=
-{i\o2}\tr\[\eta^z\lam_\m(\vk,\vk+\v{G})\].
}
We can verify through Eq.~\eqref{eqn:FormFactorConstraintC2T} and the identity $\lam_\m(\vk,\vk+\v{G})=\lam_\m^\dag(\vk,\vk-\v{G})$ that $\tr[\eta^z\lam_\m(\vk,\vk+\v{G})]$ must be imaginary.
In limit that $M$ is large, Eq.~\eqref{eqn:imRdef} implies
\eq{
R''(M;\v{G})
\to
-\int_{\vk\in\mathrm{BZ}}\lam^{(z)}_\m(\vk;\v{G}).
}
Assuming that $R''(M;\v{G})$ is a monotonically increasing function of $M$ (which Fig.~\ref{fig:HartreeNumerics}(a) verifies at least for the parameters considered), we expect $\lam^{(z)}_\m$ to supply an upper bound on $R''$:
\eq{
\abs{R''(M;\v{G})}
\leq
\abs{\int_{\vk\in\mathrm{BZ}}\lam^{(z)}_\m(\vk;\v{G})}.
}
In Fig.~\ref{fig:HartreeNumerics}(b) we plot the right hand side of the above equation as a function of $\v{G}$. 
The fact that $\int_\vk \lam^{(z)}_\m(\vk;\v{G})$ vanishes for all $\v{G}$ such that $\v{G}=\M_{y}[\v{G}]$, $\v{G}=\C_3\M_{y}[\v{G}]$, and $\v{G}=\C_3^2\M_{y}[\v{G}]$ follows from the symmetry analysis given at the beginning of this section---as we see, the reciprocal lattice vectors with the smallest amplitudes do not contribute to $R''(M;\v{G})$.

More importantly, the largest value of $\int_\vk \lam^{(z)}_\m(\vk,\v{G})$ is already incredibly small--its maximum value is $\sim3.6\times10^{-4}$.
Even when multiplied by the relatively large interaction scale $V(\aM)$, the energy difference between the QVH and the other insulating phases remains small, as evinced by Fig.~\ref{fig:HartreeNumerics}(a).
We conclude that, at least within the approximation considered here, the QVH insulator is indistinguishable from its cousins, the QSVH, QH, and QSH states.

\section{Random field Ising model domain estimates}
\label{app:DisDomain}

In this appendix, we discuss the Imry-Ma \cite{Imry75} arguments used in Sec.~\ref{sec:DomainSize} to obtain the estimates given in Eqs.~\eqref{eqn:Ldom-LargeDisCorr} and~\eqref{eqn:Ldom-SmDisCorr} for the minimal domain size $\xi_\mathrm{dom}$.
We consider the regime where the homogeneous system would like to order---in this sense, we are assuming that disorder is weak compared to the interaction energy: $\d m\ll U$.
We next estimate the energy cost $E_\mathrm{dom}(L)$ of changing the sign of $\phi$ within a domain $D$ of linear extent $\sim L$, as depicted in Fig.~\ref{fig:Domain}(a).
There are two contributions to $E_\mathrm{dom}$: one from the interaction energy, $E_\mathrm{int}(L)$, and another from the disorder potential, $E_\mathrm{dis}(L)$.
As reasoned in the main text, we assume that $\abs{\Braket{\phi}}\sim\mathcald{O}(1)$.
Since we are primarily interested in the relative scaling of the two energy terms, coefficients of $\mathcald{O}(1)$ are not be included.

\begin{figure}[t]
\includegraphics[width=0.95\textwidth]{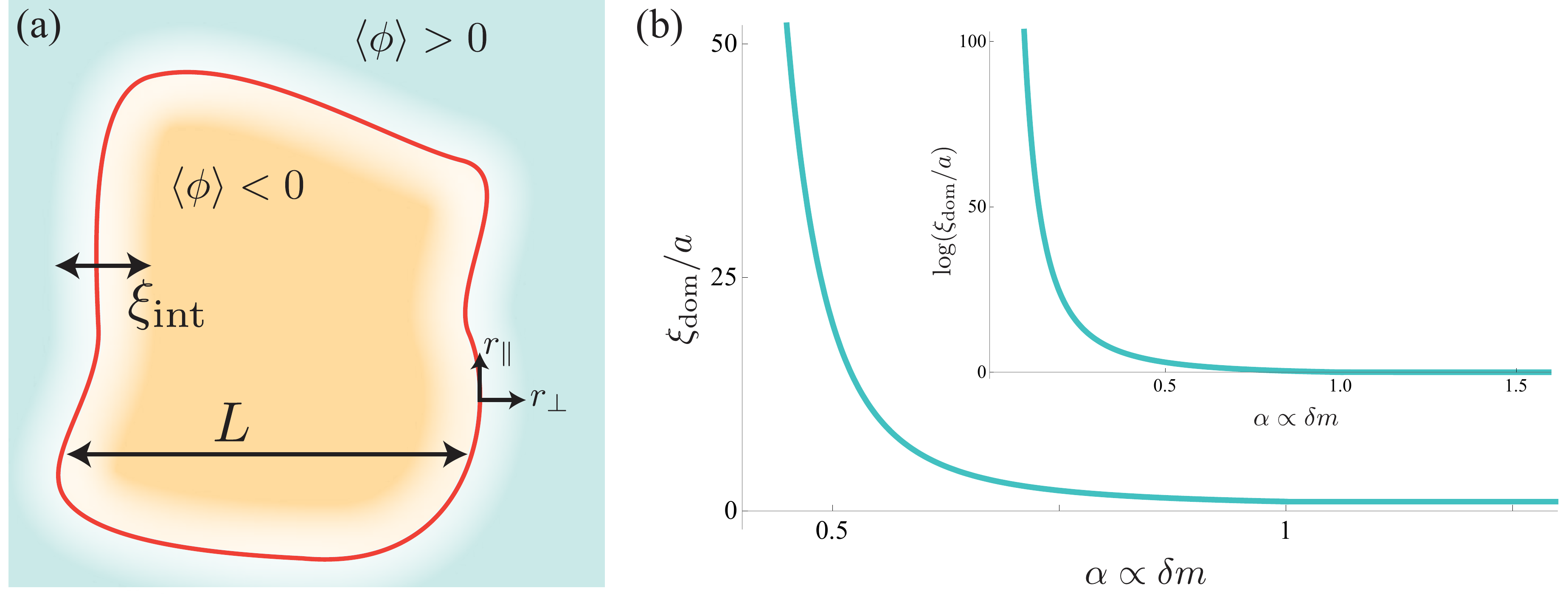}
\caption{
	(a) Illustration of a domain $D$ of linear size $\sim L$ with $\Braket{\phi}<0$ (orange region) immersed within a region of $\Braket{\phi}>0$ (blue region).
	The boundary region of the domain, $\ptl D$, is indicated in white.
	Its width, $\sim\xi_\mathrm{int}$, is shown with an arrow.
	The coordinates $(r_\perp,r_\parallel)$ used to estimate $\smash{\int_\vr \mathcald{K}\(\v{\nabla}\phi\)^2}$ are shown to the right of the domain.
	(b) Schematic plot of domain size, $\xi_\mathrm{dom}$, as a function of $\a$ [Eq.~\eqref{eqn:xDef}] for Gaussian-correlated disorder, Appendix~\ref{app:GaussCorrDisDomain}.
	The inset plots the logarithm of the domain size.
	In both, $a=\max(\xi_\mathrm{dis},\xi_\mathrm{int})$. 
	When $\a\lesssim1$, the disorder is effectively local and the domains are exponentially large, as per Eq.~\eqref{eqn:xiDom-WNregime}.
	On the other hand, for $\a\gtrsim1$, the domain size is set by the disorder correlation length $\xi_\mathrm{dis}$.
	Coefficients of $\mathcald{O}(1)$ have been chosen by hand to smoothen the crossover between these two regimes.
	Since we assume that $\d m\ll U$, $\a\gtrsim1$ implies that $\xi_\mathrm{dis}\gg\xi_\mathrm{int}$.
}
\label{fig:Domain}
\end{figure}

The interaction energy of the domain is determined by the kinetic term of the Ising model:
\eq{
E_\mathrm{int}(L)\sim\int d^2\vr\,\mathcald{K}\(\v{\nabla}\phi\)^2.
}
The coefficient $\mathcald{K}$ should have units of energy, and so we naturally set $\mathcald{K}\sim U$, as discussed in the main text.
The Ising field $\phi$ changes only within the boundary region $\ptl D$ of the flipped domain $D$.
Given our initial definition of $\phi$ [Eq.~\eqref{eqn:PhiDef}], this change can only occur on the scale of $\xi_\mathrm{int}$ [Eq.~\eqref{eqn:XiIntDef}], implying that $\(\v{\nabla}\phi\)^2\sim U\phi/\xi_\mathrm{int}^{2}\sim 1/\xi_\mathrm{int}^{2}$.
Integrating over $\ptl D$, including its width, contributes a factor of $\xi_\mathrm{int}L$ so that the total cost is
\eq{\label{eqn:EintCost}
E_\mathrm{int}(L)= U {L\o\xi_\mathrm{int}}\cdot
}
More concretely, this estimate can be obtained through the {\it ansatz} $\phi(\vr)\sim\tanh\(r_\perp/\xi_\mathrm{int}\)$, where $r_\perp$ is the direction perpendicular to the domain boundary, with the boundary itself occurring at $r_\perp=0$ [see Fig.~\ref{fig:Domain}(a)].
Ignoring the effect of curvature, we again find
\eq{
E_\mathrm{int}(L)\sim
U \int dr_\parallel \int dr_\perp {1\o\xi_\mathrm{int}^2}\sech^4\(r_\perp-r_0\o\xi_\mathrm{int}\)
\sim
U\cdot{1\o\xi_\mathrm{int}^2}\cdot L \cdot {4\o3}\xi_\mathrm{int}
\sim U {L\o \xi_\mathrm{int}}\cdot
}

We now consider the contribution to the energy cost of the domain due to the random field $\mathcald{B}(\vr)$ [as defined in and below Eq.~\eqref{eqn:PhiRandFieldDef}].
For a given realization of disorder, we have
\eq{
E_\mathrm{dis}(L)&\sim\int_{\vr \in D} \mathcald{B}(\vr).
}
Depending on where the domain is placed, disorder can either increase or decrease the domain energy.
For an arbitrarily chosen $D$, $E_\mathrm{dis}$ will average to zero, with a standard deviation given by
\eq{
E_\mathrm{rms}^2
&\sim
\overline{ \[\int_{\vr\in D}\mathcald{B}(\vr)\]^2}
=
{\d m^2\o\xi_\mathrm{int}^4}\int_{\vr,\vr'\in D}K\(\vr-\vr'\o\xi_\mathrm{dis}\).
}
Importantly, however, the location of the domain is not arbitrary.
We can choose to place our domain in a region where this contribution is negative, taking the typical value
\eq{\label{eqn:EdisCos}
E_\mathrm{dis}
&\sim
-\sqrt{E_\mathrm{rms}^2}
\sim
-{\d m\o \xi_\mathrm{int}^2}\[\int_{\vr,\vr'\in D}K\(\vr-\vr'\o\xi_\mathrm{dis}\)\]^{1/2}.
}
The total cost of the domain is therefore
\eq{\label{eqn:EdomGen}
E_\mathrm{dom}(L)
&\sim
U{L\o\xi_\mathrm{int}}
-{\d m\o\xi_\mathrm{int}^2}\[\int_{\vr,\vr'\in D}K\(\vr-\vr'\o\xi_\mathrm{dis}\)\]^{1/2}.
}
If $L_*$ exists such that $E_\mathrm{dis}(L_*)=0$, the formation of the domain is energetically favourable and long-range order is destroyed. 
This destruction occurs in all of the examples we consider.

\subsection{Long-range disorder}

The simplest example actually turns out to be the case of long-range disorder \cite{Nattermann83}:
\eq{
K\(\vr\o\xi_\mathrm{dis}\)={\xi_\mathrm{dis}\o\abs{\vr}}\cdot
}
We do not discuss this form of $K$ in the main text since it is unlikely to describe the physical system; it nevertheless serves as a convenient example.
We note that while $\xi_\mathrm{dis}$ is a lengthscale, it does not truly represent a correlation length in this context. 
Instead, it simply enters into the disorder strength as a multiplicative factor:
\eq{
\overline{\mathcald{B}(\vr)\mathcald{B}(0)}
&=
\({\d m^2 \xi_\mathrm{dis}}\){1\o\xi_\mathrm{int}^4}{1\o\abs{\vr}}\cdot
}
Inserting this definition into Eq.~\eqref{eqn:EdomGen}, we find that the change in energy expected for a (judiciously-chosen) domain of size $L$ is
\eq{
E_\mathrm{dom}(L)
&\sim
{L\o\xi_\mathrm{int}}\( U - \d m {\sqrt{\xi_\mathrm{dis}L}\o\xi_\mathrm{int}}\).
}
For large $L$, it's clear that the domain energy eventually becomes negative, destabilizing the ordered phase.
This destruction first occurs at the emergent length scale
\eq{
L_*
&\sim
\({U\o\d m}{ \xi_\mathrm{int}\o \xi_\mathrm{dis}}\)^{\!2}\xi_\mathrm{dis}.
}
We conclude that when the disorder is long-range, domains are expected to form once the system size is larger than $L_*$.

\subsection{White noise (short-range) disorder}\label{app:WNdisDom}

We now consider local, white noise disorder:
\eq{\label{eqn:WNdis}
K\(\vr\o\xi_\mathrm{dis}\)
&=
\xi^2_\mathrm{dis}\d^2(\vr).
}
As in the long-range case, the parameter $\xi_\mathrm{dis}$ enters only as a multiplicative factor.
Together with the disorder strength $\d m$ and the Fermi velocity $v_F$, they form a dimensionless parameter 
$\d m\,\xi_\mathrm{dis}/\hbar v_F$ discussed in Sec.~\ref{sec:TypesOfDis}. 

Following the arguments above, an appropriately chosen domain therefore contributes an energy
\eq{\label{eqn:EdisWN0}
E_\mathrm{dis}(L)
\sim
-\d m\,{\xi_\mathrm{dis}L\o\xi_\mathrm{int}^2}\cdot
}
The total energy cost of the domain is
\eq{\label{eqn:WNlinEdom}
E_\mathrm{dom}(L)
&\sim
U{L\o\xi_\mathrm{int}}
-\d m\,{\xi_\mathrm{dis}L\o\xi_\mathrm{int}^2}
=
U{L\o\xi_\mathrm{int}}\Big(1-\a\Big),
}
where we have defined
\eq{\label{eqn:xDef-App}
\a
&\equiv
{\d m\o U}{\xi_\mathrm{dis}\o\xi_\mathrm{int}}\cCom
}
as given in Eq.~\eqref{eqn:xDef} of the main text.
Notably, it is not $\d m/U$ that controls the domain energy cost, but instead the ratio $\a$.
This feature is related to our remark that the true disorder strength is actually $g=\d m\,\xi_\mathrm{dis}$.
The correct energy scale is therefore obtained in units of the UV cutoff, giving $g/\xi_\mathrm{int}=\a\,U$, from which it follows that $\a$ is the appropriate tuning parameter, \emph{not} $\d m/U$.
Equation~\eqref{eqn:WNlinEdom} simply tells us that when disorder is larger than the interaction scale, $\a\gtrsim1$, there is no reason for the system to order. 
In this limit, the domain structure and fate of the theory is complicated and will not be relevant for us \cite{Seppala98,Seppala01}. 

Conversely, for $\a\lesssim1$, Eq.~\eqref{eqn:WNlinEdom} may appear to imply that that the system should order.
However, while Eq.~\eqref{eqn:xDef-App} is sufficient for large $\a$, the analysis above omits the effect of domain roughening.
This effect should be included in general, and it completely alters our conclusions when $\a$ is small.

Roughening in the context of the RFIM was first discussed in Ref.~\onlinecite{Binder83}, and we now summarize the reasoning made there.
We begin by considering a portion of a domain wall of linear extent $y$, displacing it by a (small) length $w$, and determining the change in energy, $\d E(w,y)$.
First, the displacement increases the length of the boundary by $\d E_\mathrm{int}\sim U w/\xi_\mathrm{int}$. 
With regards to the disorder field, we can choose to displace the boundary to either the left or the right direction, each of which has a 50\% likelihood of decreasing the energy.
There is therefore a 75\% probability that the displacement lowers the energy by a typical amount $\d E_\mathrm{dis}\sim -\d m \xi_\mathrm{dis} \sqrt{wy}/\xi_\mathrm{int}^2$.
In total, the displacement results in a typical energy change
\eq{\label{eqn:dE_disp}
\d E(w,y)
&\sim
U {w\o \xi_\mathrm{int}} - \d m{\xi_\mathrm{dis}\o \xi_\mathrm{int}^2}\sqrt{ w y}.
}
We now minimize $\d E$ with respect to $w$, to obtain
\eq{
w_*
&\sim 
\({\d m\o U}{\xi_\mathrm{dis}\o\xi_\mathrm{int}}\)^{\!2}y
= 
\a^2\,y,
\nt
\d E_*(y)&\equiv \d E(w_*,y)
\sim 
-{\a^2\,U}{y\o\xi_\mathrm{int}}\cdot
}
Next, we note that this procedure may be performed for segments of all sizes along the domain boundary.
In particular, there are $N(y_\ell)=L/y_\ell$ segments of size $y_\ell=e^{-\ell}L$, 
each of which contributes an energy $\d E_*(y_\ell)$.
Summing over all scales returns the total energy contribution from domain wall roughening:
\eq{\label{eqn:EdisWNrough}
\d E_\mathrm{tot}(L,a)
&=
\int_0^{\log\(L/a\)}d\ell \,N(y_\ell)\d E_*(y_\ell)
\sim
-\int_a^L {dy\o y}{L\o y}\a^2{U} {y\o\xi_\mathrm{int}}
\nt
&\sim-\a^2{U}{L\o\xi_\mathrm{int}}\log\(L\o a\).
}
Here, $a$ is the smallest scale at which roughening may occur; in this context, $a\sim \xi_\mathrm{int}$, though we will find otherwise in the next section.
(Note that this `$a$' should \emph{not} be confused with the microscopic lattice constant of monolayer graphene.)
Throughout this derivation, we have assumed that $a$ is significantly smaller than $L$.
Finally, the total energy cost of the domain is
\eq{
E_\mathrm{dom}(L)
&\sim
{L\o\xi_\mathrm{int}}\[
U
-
\a^2{U} \log\(L\o \xi_\mathrm{int}\)
\].
}
Solving for $E_\mathrm{dom}(L_*)=0$, we find
\eq{\label{eqn:LdomMinWN}
L_*
&\sim
\xi_\mathrm{int}\,e^{c/\a^2},
}
where we have introduced the non-universal constant ${c}\sim\mathcald{O}(1)$ to account for the imprecise nature of our scaling arguments.
Once more, for systems larger than $L_*$, multiple domains should be apparent.

As we mentioned below Eq.~\eqref{eqn:EdisWNrough}, our integration was predicated on the assumption that the domain size $L$ was much larger than $\xi_\mathrm{int}$. 
It is clear that this is only satisfied provided the disorder is weak: $\a\ll 1$.
When the disorder is stronger, the situation is more complicated.

\subsection{Gaussian-correlated disorder}\label{app:GaussCorrDisDomain}

We now consider the situation considered in the main text, that of Gaussian correlated disorder:
\eq{\label{eqn:GaussDis}
K\(\vr\o\xi_\mathrm{dis}\)
&=
e^{-{\vr^2\o 2\xi_\mathrm{dis}^2}}.
}
Unlike the previous two cases, the scale $\xi_\mathrm{dis}$ is a true correlation length in this scenario, 
as is clear from the form of the disorder-induced energy reduction:
\eq{\label{eqn:EdisGaussCorr}
E_\mathrm{dis}(L)
&\sim
-{\d m}{\xi_\mathrm{dis} L\o\xi_\mathrm{int}^2}\sqrt{1-e^{-L^2/2\xi_\mathrm{dis}^2}}.
}
While the domain size appeared as a ratio of the UV cuttoff $\ell_\mathrm{UV}=\xi_\mathrm{int}$ in the previous two examples, here $E_\mathrm{dis}(L)$ is also a function of $L/\xi_\mathrm{dis}$.

There are two natural limits to consider.
In the first, we take the domain size to be small enough relative to $\xi_\mathrm{dis}$ that the smoothness of the disorder is still important, \emph{i.e.} we cannot simply ignore the exponential in Eq.~\eqref{eqn:EdisGaussCorr}.
As an extreme example, when $L\ll\xi_\mathrm{dis}$, 
\eq{
E_\mathrm{dis}(L)
&\sim
-\d m {L^2\o\xi_\mathrm{int}^2}\cdot
}
That is, the change in energy is proportional to the \emph{volume} of the domain.
This observation makes sense given that $\mathcald{B}(\vr)$ should be essentially constant for two points within a distance $\xi_\mathrm{dis}$ of one another.
In fact, it seems clear that an energetically favourable domain should be at least $\xi_\mathrm{dis}$ in extent: $L_*\gtrsim\xi_\mathrm{dis}$.
We therefore examine the theshold scenario given by $L=\xi_\mathrm{dis}$.
We conclude that domain formation is favourable when 
\eq{\label{eqn:GaussCorrDom}
E_\mathrm{dom}(\xi_\mathrm{dis})
\sim
U{\xi_\mathrm{dis}\o\xi_\mathrm{int}}\(
1-{\d m\o U}{\xi_\mathrm{dis}\o\xi_\mathrm{int}}\)
=
U{\xi_\mathrm{dis}\o\xi_\mathrm{int}}\big(1-\a\big)
\lesssim1.
}
The parameter $\a$ that appeared in the white noise case, Eq.~\eqref{eqn:xDef-App}, has showed up again.
When it is greater than unity, $\a\gtrsim1$, the disorder destroys long-range order, resulting in domains of typical size $\xi_\mathrm{dom}\sim\xi_\mathrm{dis}$.

When $\a\lesssim1$, the interaction energy cost associated with the boundary of a domain of linear extent $\xi_\mathrm{dis}$ is greater than the gain associated with aligning with the random field.
For domains larger than $\xi_\mathrm{dis}$, the random field within the domain is only weakly correlated.
The exponential under the square root may therefore be neglected, resulting in an expression identical to our original estimate for the domain energy with white noise disorder in Eq.~\eqref{eqn:WNlinEdom}.
As we discussed there, this expression was not complete: the roughening of the domain walls must also be taken into account, resulting in the contribution given in Eq.~\eqref{eqn:EdisWNrough}.
The arguments made in Sec.~\ref{app:WNdisDom} follow through for weak, Gaussian-correlated disorder in all respects save for one minor caveat.
Unlike the white noise disorder case, the roughening cutoff for Gaussian-correlated disorder is not necessarily $\xi_\mathrm{int}$. 
Instead, only scales down to \emph{at most} $\xi_\mathrm{dis}$ should be included, since this is where our omission of the exponential ceases to be valid, \emph{i.e.} $a=\max(\xi_\mathrm{int},\xi_\mathrm{dis})$.
Setting the domain energy to zero, we find
\eq{\label{eqn:xiDom-WNregime}
\xi_\mathrm{dom}
&\lesssim
\max(\xi_\mathrm{int},\xi_\mathrm{dis})\,e^{c/\a^2},
}
where $c\sim\mathcald{O}(1)$ is again a non-universal constant.
In Fig.~\ref{fig:Domain}(b), we show $\xi_\mathrm{dom}$ for Gaussian-correlated disorder for both regimes, $\a\lesssim1$ and $\a\gtrsim1$.

\section{Competing orders}\label{app:CompetingOrders}

We now address the possibility considered in Sec.~\ref{sec:CompPhase-Main} that the QVH state is not the ground state of the clean theory at charge neutrality---either one of the other three $\C_2\T$-breaking insulators (QSH, QH, or QSVH) or a completely different order minimizes the energy of the homogeneous theory.

We are interested in studying the conditions under which the QVH phase is realized.
To simplify the analysis, we assume that there is a single competing phase whose order parameter does not couple to disorder, but whose ground state energy density, $\mathcald{E}_\mathrm{comp}$, is lower than the energy density of the QVH phase, $\mathcald{E}_\mathrm{QVH}$, by a small amount.
We measure this distinction in terms of the energy difference $\d\ep$ within a region of area $\ell_\mathrm{UV}^2=\xi_\mathrm{int}^2$:
\eq{\label{eqn:depDef}
{\d\ep\o\xi_\mathrm{int}^2}
&=
\mathcald{E}_\mathrm{QVH}-\mathcald{E}_\mathrm{C}
\geq0
}
Thoughout this section, we assume that $\d\ep\ll U$.
While this ground state energy difference implies that the competing phase is realized in a perfectly clean sample, disorder exclusively favours the local realization of the QVH phase. 
We therefore expect the majority of the sample to be in the QVH phase when $\d\ep$ is sufficiently small.
Using the Ising notation of Sec.~\ref{sec:DomainSize} and Appendix~\ref{app:DisDomain}, we quantify this expectation as
\eq{
\label{eqn:MostlyQVH}
\[{1\o\mathrm{vol}}\int_\vr\Braket{\phi^2(\vr)}\]^{1/2}\gtrsim{1\o2}\cCom
}
where `$\mathrm{vol}$' denotes the sample volume.

We approach the problem in two complementary fashions.
The question of an Ising order parameter competing with another phase may bring to mind dilute Ising physics, where here `vacancies' represent regions where the Ising $\phi$ field is not ordered.
In Appendix~\ref{app:BlumeCapel}, we describe a mean field solution of a classical 2$d$ lattice model formulated to tackle this type of question.

While useful, because of the low-dimensionality of the problem, mean field theory is not particularly reliable.
In particular, we are free to take the limit $\d\ep\to-\infty$, effectively removing the `competing' phase from the problem.
In this limit, our results should agree with those of Sec.~\ref{sec:DomainSize} and Appendix~\ref{app:DisDomain}.
There, we found that any disorder was sufficient to destroy long-range order.
In contrast, the mean field calculation falsely finds long-range order in this limit.
We therefore devise an Ising formulation of the problem in Appendix~\ref{app:CompIsing-ImryMa}, which allows us to make Imry-Ma arguments similar to those of Appendix~\ref{app:DisDomain}.

\subsection{Blume-Capel description}\label{app:BlumeCapel}

In keeping with the Ising description of the QVH insulator, we view the ordering of the competing phase as the presence of an annealed `vacancy.' 
At finite temperature, this physics is known to give rise to the tricritical Ising fixed point, though this observation is not relevant for our discussion.
While continuum descriptions do exist, for our purposes, it is most convenient to employ a lattice model.
We therefore consider the Blume-Capel model \cite{Blume66,Capel66} on an (unspecified) lattice of coordination number $z$ with quenched random-field disorder:
\eq{
H_\mathrm{BC}&=-{J\o z}\sum_{\Braket{\vr,\vr'}}s_\vr s_{\vr'}
+
\m \sum_\vr s_\vr^2
+
\sum_\vr h_\vr s_\vr,
} 
where the classical spins may take three values: $s_\vr\in\{+1,\,-1,\,0\}$.
As above, the quenched disorder is represented through a random `magnetic field' $h_\vr$.
For simplicity, we assume that $h_\vr$ satisfies Gaussian white noise disorder.
The corresponding probability distribution reads
\eq{\label{eqn:h0ProbDis}
\mathcald{P}(h_\vr)
&=
{e^{-{h_\vr^2\o2h_0^2}}\o\sqrt{2\pi h_0^2}}\cdot
}
The use of this distribution is equivalent to our previous definitions of the disorder distribution, entirely in terms of moments:
\eq{
\overline{h_\vr}&=0,
&
\overline{h_\vr h_{\vr'}}&=h_0^2 \d_{\vr,\vr'}.
}
We associate $s_\vr=\pm1$ with the realization of the QVH phase, \emph{i.e.} $\Braket{\phi}\sim\pm1$, and `vacancies' $s_\vr=0$ with competing phase.
The exchange energy $J$ corresponds to the Coulomb interaction strength, $J\sim U$, while the random field strength $h_0$, should be mapped to the disorder strength ${h_0\sim \d m\, \xi_\mathrm{dis}/\xi_\mathrm{int}}$ in units of the UV cutoff $\xi_\mathrm{int}$ [see the discussion below Eq.~\eqref{eqn:xDef-App}].
Finally, the 
so-called `crystal field,' $\m$, can be related to the energy splitting $\d\ep$ by establishing when the competing phase (all $s_\vr=0$) and QVH phase (all $s_\vr=+1$ or $-1$) are degenerate, indicating that $\m=\d\ep+J/2\sim\d\ep+U/2$.

As discussed, we analyze this model in mean field theory \cite{Kaufman90,Vasseur10}.
Letting $m\equiv\Braket{s_\vr}$ be the average magnetization, the mean-field free energy is
\eq{
f_\mathrm{BC}(m)
&=
{1\o2}J m^2
-
\overline{
\log
\!\(1+e^{-\b\m} 2\cosh\[\b(Jm+h)\]\)}
\nt
&=
{1\o2}J m^2
-
\int {dh\o\sqrt{2\pi}h_0}e^{-h^2/2h_0^2}
\log\!\(1+e^{-\b\m} 2\cosh\[\b(Jm+h)\]\)\cCom
}
where $\b$ is the inverse temperature and we explicitly average over the Gaussian distribution of Eq.~\eqref{eqn:h0ProbDis} in the second line.
Taking the zero temperature limit, $\b\to\infty$, the integral can be evaluated exactly, giving
\eq{\label{eqn:FrEnBC}
f_\mathrm{BC}(m)
&=
{1\o2}Jm^2
+
{1\o2}
\[ (\m-Jm)\mathrm{Erfc}\!\(\m-Jm\o\sqrt{2}h_0\)
+
(\m+Jm)\mathrm{Erfc}\!\(\m+Jm\o\sqrt{2}h_0\)\]
\nt
&\quad
-
{h_0\o\sqrt{2\pi}}\( e^{-(\m-Jm)^2/2h_0^2}+e^{-(\m+Jm)^2/2h_0^2}\),
}
where $\mathrm{Erfc}(x)$ is the complementary error function.
The magnetization is determined by extremizing $f_\mathrm{BC}$, resulting in the self-consistency equation
\eq{
m
=
{1\o2}\[
\mathrm{Erfc}\({\m- Jm\o\sqrt{2}h_0}\)
-
\mathrm{Erfc}\({\m+ Jm\o\sqrt{2}h_0}\)
\].
}
The expectation value of the spin squared, $q\equiv\sqrt{\braket{s_\vr^2}}$, is directly analogous to the expression on the right-hand side of Eq.~\eqref{eqn:MostlyQVH}, \emph{i.e.}, when $q\gtrsim1/2$, QVH order prevails.
It is calculated by taking the derivative of $f_\mathrm{BC}$ with respect to $\m$:
\eq{\label{eqn:q2Def}
q^2
&=
{\ptl\o\ptl\m}f_\mathrm{BC}
=
{1\o2}\[
\mathrm{Erfc}\({\m- Jm\o\sqrt{2}h_0}\)
+
\mathrm{Erfc}\({\m+ Jm\o\sqrt{2}h_0}\)
\].
}
In Figs.~\ref{fig:BlumeCapelMF}(a) and~(c), we plot $m$ and $q$ as functions of $\d\ep/U$ and $\a$, respectively.
To make contact with the phase diagram in the main text, Fig.~\ref{fig:CompPhase}, we also plot $m$ and $q$ with the $y$-axis given by $\g \,\d\ep/\d m$, where $\g=\xi_\mathrm{int}/\xi_\mathrm{dis}$, in Figs.~\ref{fig:BlumeCapelMF}(b) and~(d).

\begin{figure}
	\centering
	\includegraphics[width=0.99\textwidth]{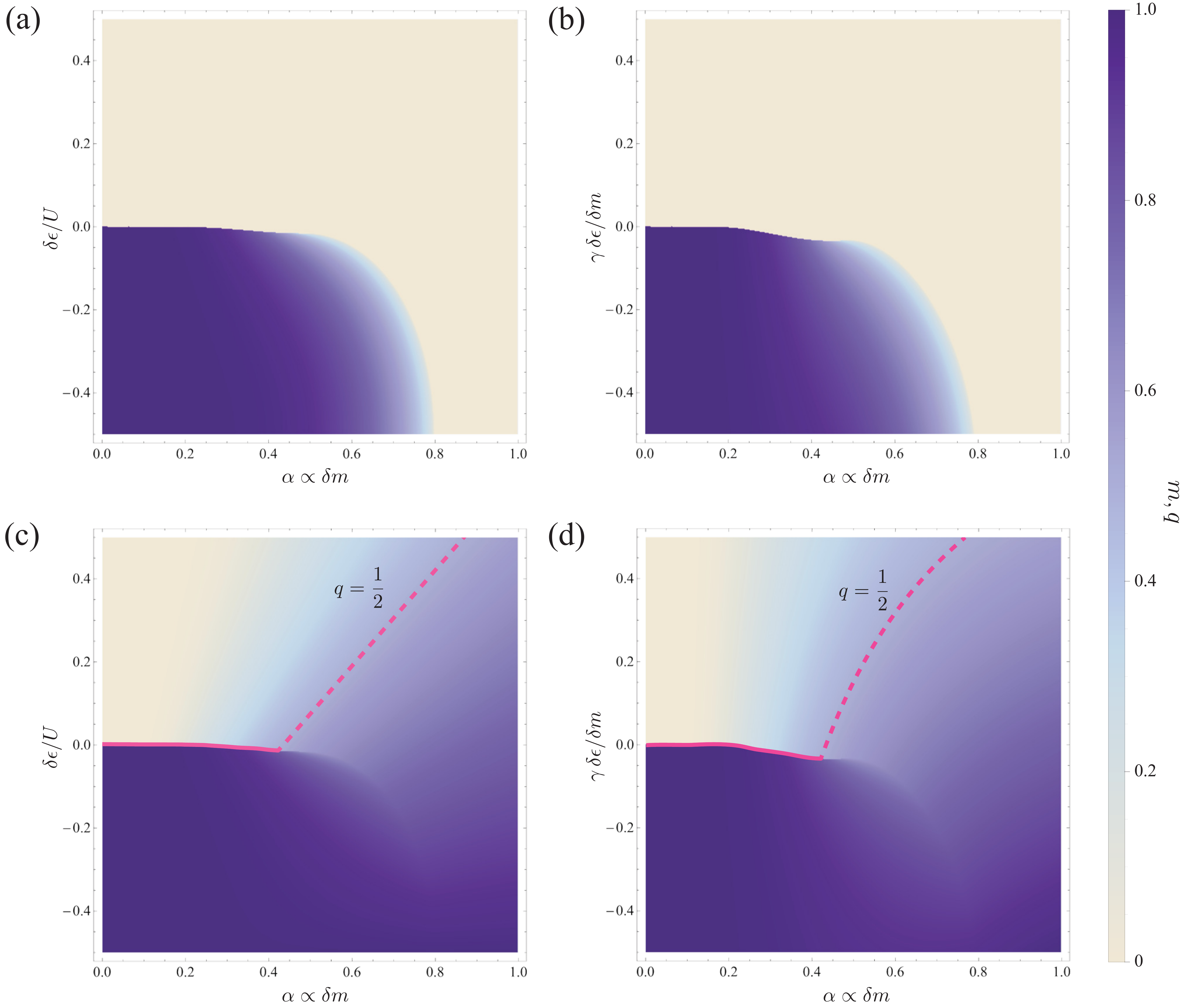}
	\caption{
	(a), (b) Density plot of the (absolute value of the) magnetization, obtained by minimizing $f_\mathrm{BC}(m)$ in Eq.~\eqref{eqn:FrEnBC}.
	(c), (d) Density plots of $q$, as given in Eq.~\eqref{eqn:q2Def}.
	The colour scheme for all plots, (a)-(d), is shown on the right, and, in (b) and (d), $\g=\xi_\mathrm{int}/\xi_\mathrm{dis}$.
	The solid pink line in (c) and (d) indicates the first order phase transition between regions with $q$ small and regions with $q\sim1$ (as follows from having $m\sim\pm1$ in that region).
	The dashed pink line, on the other hand, is the contour along which $q=1/2$ and $m=0$; we view it as demarcating a crossover between regions where the competing phase percolates and regions where the QVH insulator percolates.
	It follows that for both (c) and (d), the network scenario we propose should be valid in the regions below and to the right of the pink lines.
	}
	\label{fig:BlumeCapelMF}
\end{figure}

Figures~\ref{fig:BlumeCapelMF}(a) and~(b) indicate that $m$ orders for $\d\ep\lesssim0$ when disorder is sufficiently small.
While these calculations agree with our expectations when $\d m=0$, we showed in Appendix~\ref{app:DisDomain} that any nonzero disorder destroys long-range order. 
The presence of regions with $m\neq0$ is therefore an artifact of the mean field theory; given the low dimension, the failure of mean field theory in this regard is not surprising.
Nevertheless, we take it as a good sign that $m$ approaches zero close to $\a\sim0.8\sim1$ for $\d\ep<0$ since this condition defines the crossover regime identified in Appendix~\ref{app:DisDomain}.
We therefore optimistically associate mean field ordered regions with those that in reality possess exponentially large domains.

The density plots in Figs.~\ref{fig:BlumeCapelMF}(c) and~(d) display $q$.
Obviously, when our mean field prescription indicates that $m$ is ordered, $q$ is non-zero as well, as a quick comparison with (a) and (b) clearly shows.
Outside of these regions, however, we find that $q$  only vanishes exactly when $\d m\to0$ (equivalently, $h_0\to0$) as well.
From Eq.~\eqref{eqn:q2Def}, we verify that when $m=0$, 
\eq{
q(m=0)=\sqrt{\mathrm{Erfc}\!\(\m\o\sqrt{2}h_0\)}\cCom
}
implying that contours of constant $q$ are represented by straight lines extending from the $\m=0$ origin (not to be confused with $\d\ep=0$ origin), as shown in Fig.~\ref{fig:BlumeCapelMF}(c).
More precisely, we can numerically solve for the line along which $q=1/2$:
\eq{
{1\o2}=\sqrt{\mathrm{Erfc}\!\(\eta_{1/2}\o\sqrt{2}\)}\cCom
}
to obtain $\eta_{1/2}\cong1.15$.
Then, provided ${\d\ep}/{U}$ and $\a$ are such that $m=0$, we find that $q=1/2$ along the line
\eq{
{\d\ep\o U}&=\eta_{1/2}\a-{1\o2}\cdot
}
We plot this contour with a pink dashed line in Fig.~\ref{fig:BlumeCapelMF}(c).
It follows that the system is primarily in the QVH phase when either
\eq{
\d\ep&\lesssim U\(\eta_{1/2}\a-{1\o2}\)
\qquad
\text{or}
\qquad
\d\ep\lesssim0.
}
The modification needed to obtain the crossover lines drawn in (d) is straightforward:
\eq{\label{eqn:qHalfContour-2}
{\d\ep}
&\lesssim
{\d m}{\xi_\mathrm{dis}\o\xi_\mathrm{int}}\(\eta_{1/2}-{1\o2\a}\)
\qquad
\text{or}
\qquad
\d\ep\lesssim0.
}

\subsection{Competing Ising field description}\label{app:CompIsing-ImryMa}

The mean field theory discussed above had the advantage of simplicity, but did not correctly capture the absence of long-range order.
We therefore employ an Imry-Ma description, similar to the analysis of Appendix~\ref{app:DisDomain}.
The ordering of both phases is now modelled by two distinct Ising fields. 
As above, we associate $\phi$ with the QVH insulator (\emph{i.e.}, $\C_2$ symmetry breaking) and $\Phi$ with the competing phase.
The total energy is given by $H_\mathrm{Ising}+H_\mathrm{Ising}'+H_{\phi\Phi}+H_{\phi,\mathrm{dis}}$ where
\eq{\label{eqn:CompetingIsingHams}
H_\mathrm{Ising}
&=
\int d^2\vr \[ \mathcald{K}\(\v{\nabla}\phi\)^2 - {\abs{r}\o2}\phi^2 + {u\o4!}\phi^4 \],
\nt
H_\mathrm{Ising}'
&=
\int d^2\vr \[ \mathcald{K}'\(\v{\nabla}\Phi\)^2 - {\abs{r'}\o2}\Phi^2 + {u'\o4!}\Phi^4 \],
\nt
H_{\phi\Phi}
&=
\int d^2\vr\, \lam\,\phi^2\Phi^2,
\nt
H_\mathrm{dis}
&=
\int d^2\vr\,\mathcald{B}(\vr)\phi(\vr).
}
Since both $\phi$ and $\Phi$ are dimensionless, $\mathcald{K}$, $\mathcald{K}'$ have dimensions of energy.
We assume that the interaction scales of the QVH and competing phases are similar, prompting us to set both to $\sim U$.
Similarly, the remaining parameters describing $H_\mathrm{Ising}$ and $H_\mathrm{Ising}'$, $r$, $r'$, $u$, and $u'$, have units of energy over length squared.
Their natural scale is therefore $U/\ell_\mathrm{UV}^2$ where $\ell_\mathrm{UV}$ is the UV cutoff, which should in turn be approximately given by $\xi_\mathrm{int}=\hbar v_F/\Delta_\mathrm{CNP}$, as discussed in Sec.~\ref{sec:DomainSize}.
However, this assignment of energy scales cannot be the entire story since the difference in ground state energies, Eq.~\eqref{eqn:depDef}, has not yet been included. 
Because $\d\ep$ is assumed to be much smaller than $U$, and we ignore coefficients of $\mathcald{O}(1)$, the exact implementation is unimportant.
Nevertheless, to be concrete, we note that if one wishes to ensure that Eq.~\eqref{eqn:depDef} holds while also requiring the magnitudes of $\phi$ and $\Phi$ to be identical in their respective ordered phases, the following choice is sufficient:
\eq{\label{eqn:ShiftedPhiConsts}
\abs{r'}
&\sim
\abs{r}+{2\abs{r}\o3u}{\d\ep\o\xi_\mathrm{int}^2}\cCom
&
u'
&\sim
u+{2\o3}{\d\ep\o\xi_\mathrm{int}^2}\cdot
} 
The parameter $\lam$ in $H_{\phi\Phi}$ is assumed to be larger than the other scales of the theory in order to guarantee that $\Braket{\phi}\neq0$ and $\Braket{\Phi}\neq0$ do not occur within the same region.
Finally, the last term, $H_\mathrm{dis}$, describes the behaviour of disorder.
We will consider both white noise and Gaussian-correlated, as defined in Eqs.~\eqref{eqn:WNdis} and~\eqref{eqn:GaussDis} respectively.

We examine this system in several steps. 
Using Imry-Ma type arguments similar to those of Appendix~\ref{app:DisDomain}, we begin by studying the formation of a $\phi$-ordered domain within a uniformly $\Phi$-ordered system for both white noise and Gaussian-correlated disorder.
As we did in Appendix~\ref{app:DisDomain}, coefficients of $\mathcald{O}(1)$ are ignored.
Next, we argue that if the physical parameters favour the formation of a single $\phi$-ordered domain, a macroscopically large fraction of the system should also $\phi$-order.
Our final result is a function of the ratio $\a$ [see Eq.~\eqref{eqn:xDef-App}], $\d\ep_c(\a)$, that parametrizes a crossover between the two regimes of interest: when $\d\ep\lesssim\d\ep_c(\a)$, the system is primarily $\phi$-ordered, whereas when $\d\ep\gtrsim\d\ep_c(\a)$, the system is primarily $\Phi$-ordered.
Figure~\ref{fig:CompPhase} shows the resulting phase diagram.

\subsubsection{\texorpdfstring{Single $\phi$-domain formation: white noise disorder}{Single phi-domain formation: white noise disorder}}\label{app:CompCord-WNsingleDom}

To make contact with the mean field theory of Appendix~\ref{app:BlumeCapel}, we begin by considering white noise disorder.
We assume that the competing phase is realized, $\Braket{\Phi}\neq0$, and examine the energy cost associated with the formation of a $\phi$-ordered domain.
As in Appendix~\ref{app:DisDomain}, there are energy contributions from interactions along the domain boundary and from the random field $\mathcald{B}(\vr)$.
Since we assume that $\mathcald{K}\sim\mathcald{K}'\sim U$, the interaction energy cost $E_\mathrm{int}$ is identical to the expression given in Eq.~\eqref{eqn:EintCost}\footnote{
One might argue that it is more honest to define $\mathcald{K}'\sim\mathcald{K}+\d\ep\sim U+\d\ep$ in analogy with the definitions of Eq.~\eqref{eqn:ShiftedPhiConsts}.
However, since $\d\ep\ll U$ by assumption, this difference is negligible.}.
Similarly, the contribution from disorder, $E_\mathrm{dis}$, follows from the expression in Eq.~\eqref{eqn:EdisCos}, giving the same result as in Eq.~\eqref{eqn:EdisWN0}.
Unlike Appendix~\ref{app:DisDomain}, there is an important additional cost associated with the difference in ground state energy.
On general grounds, the cost must increase with the domain \emph{area}:
\eq{
E_\mathrm{comp}(L)
&\sim
\d\ep{L^2\o\xi_\mathrm{int}^2}\cdot
}
We could also have obtained this result from the Hamiltonian defined in Eq.~\eqref{eqn:CompetingIsingHams} with the coefficients defined in Eq.~\eqref{eqn:ShiftedPhiConsts}.
The total energy cost of a $\phi$-ordered domain is given by the sum of this expression with $E_\mathrm{int}$ and $E_\mathrm{dis}$:
\eq{
E_{\phi\text{-}\mathrm{dom}}(L)
&\sim
\d\ep{L^2\o\xi_\mathrm{int}^2}
+
U{L\o\xi_\mathrm{int}}
-
\d m{\xi_\mathrm{dis}L\o\xi_\mathrm{int}^2}
=
\d m{\xi_\mathrm{dis}L\o\xi_\mathrm{int}^2}
\( {\d\ep\o\d m}{L\o\xi_\mathrm{dis}} + {1\o\a} -1 \)
\cdot
}
This result is the analogue of Eq.~\eqref{eqn:WNlinEdom}. 
There, we concluded that when $\a\gtrsim1$, disorder was ``large" and the system would not order.
While this expression also indicates that $\a\gtrsim1$ is necessary to destroy the local order (here, $\Phi$-order instead a different type of $\phi$-order), the energy cost of the $\phi$-domain is also dependent on its size, $L$:
the smaller the domain size, the more favourable it is.
A threshold value of $\d\ep$ can therefore be defined by the condition $E_{\phi\text{-}\mathrm{dom}}(a)<0$, where $a$ is the smallest possible domain size. (Again, `$a$' should not be confused with the microscopic lattice constant of monolayer graphene here or below.)
For the current situation, clearly $a\sim\xi_\mathrm{int}$; nevertheless, with an eye to the subsequent section, it is convenient to leave $a$ unspecified.
That is, $E_{\phi\text{-}\mathrm{dom}}(a)<0$ provided
\eq{\label{eqn:depCrit-WN}
{\d\ep}
&\lesssim 
\d\ep_c(\a),
&
\d\ep_c(\a)
&\equiv
\d m{\xi_\mathrm{dis}\o a}\bigg(1-{1\o\a}\bigg),
\quad\text{when}\quad
\a\gtrsim1.
}
Here, we have defined the `critical' energy difference $\d\ep_c(\a)$ in the region where $\a\gtrsim1$ for white noise disorder with a minimal domain size $a=\xi_\mathrm{int}$.
We generalize this definition to smaller values of $\a$ below.

We note that up to coefficients of $\mathcald{O}(1)$, this inequality has the same dependence on $\a$ as our mean field result in Eq.~\eqref{eqn:qHalfContour-2}!
At least in the simple regime, the Blume-Capel and Imry-Ma descriptions are in agreement.

As we saw in Appendix~\ref{app:WNdisDom}, once $\a\lesssim1$, the effects domain wall roughening become important and must be included.
Because roughening does not change the domain area significantly, the roughening contribution Eq.~\eqref{eqn:EdisWNrough} remains valid\footnote{Alternatively, we can argue that since the displacement is equally likely to increase or decrease the domain area, Eq.~\eqref{eqn:dE_disp} remains valid on average}.
We note that this situation is similar to what occurs in the absence of a competing order when a small, uniform magnetic field is applied \cite{Binder83,Seppala01}.
The resulting cost of a $\phi$ domain is
\eq{\label{eqn:Ephidom-WallRough}
E_{\phi\text{-}\mathrm{dom}}(L)
&\sim
\d\ep\(L\o\xi_\mathrm{int}\)^{\!2}
+
U{L\o\xi_\mathrm{int}}
-
U\({\d m\o U}{\xi_\mathrm{dis}\o\xi_\mathrm{int}}\)^{\!2}{L\o\xi_\mathrm{int}}\log \(L\o a\)
\nt
&=
U {L\o\xi_\mathrm{int}}\[
{\d\ep\o U}{L\o\xi_\mathrm{int}}+1-\a^2\log\(L\o a\)\].
}
Again, $a$ is the minimal domain size, which is equivalent to $\xi_\mathrm{int}$ in this case.
We can now define a critical energy difference in the small $\a$ regime.
We find that there exists a solution $E_\mathrm{dom}(L)=0$ provided $\d\ep$ satisfies
\eq{
\d\ep&\lesssim\d\ep_c(\a),
&
\d\ep_c(\a)
&\equiv
{\xi_\mathrm{dis}\o a}\d m\,\a\,{e^{-c\({1\o\a^2}+1\)}}\cCom
\quad
\text{when}
\quad
\a\lesssim1.
}
In Fig.~\ref{fig:depC_WN}, we plot $E_{\phi\text{-}\mathrm{dom}}(L)$ as a function $L$ for several values of $\d\ep$.
As indicated in the figure, when $\d\ep<\d\ep_c$, there is an entire region where $E_{\phi\text{-}\mathrm{dom}}<0$ for $L_-<L<L_+$.
Naturally, as $\d\ep\to0$, $L_-\to L_*$ [as defined in Eq.~\eqref{eqn:LdomMinWN}] while $L_+\to\infty$.

\begin{figure}
	\centering
	\includegraphics[width=0.6\textwidth]{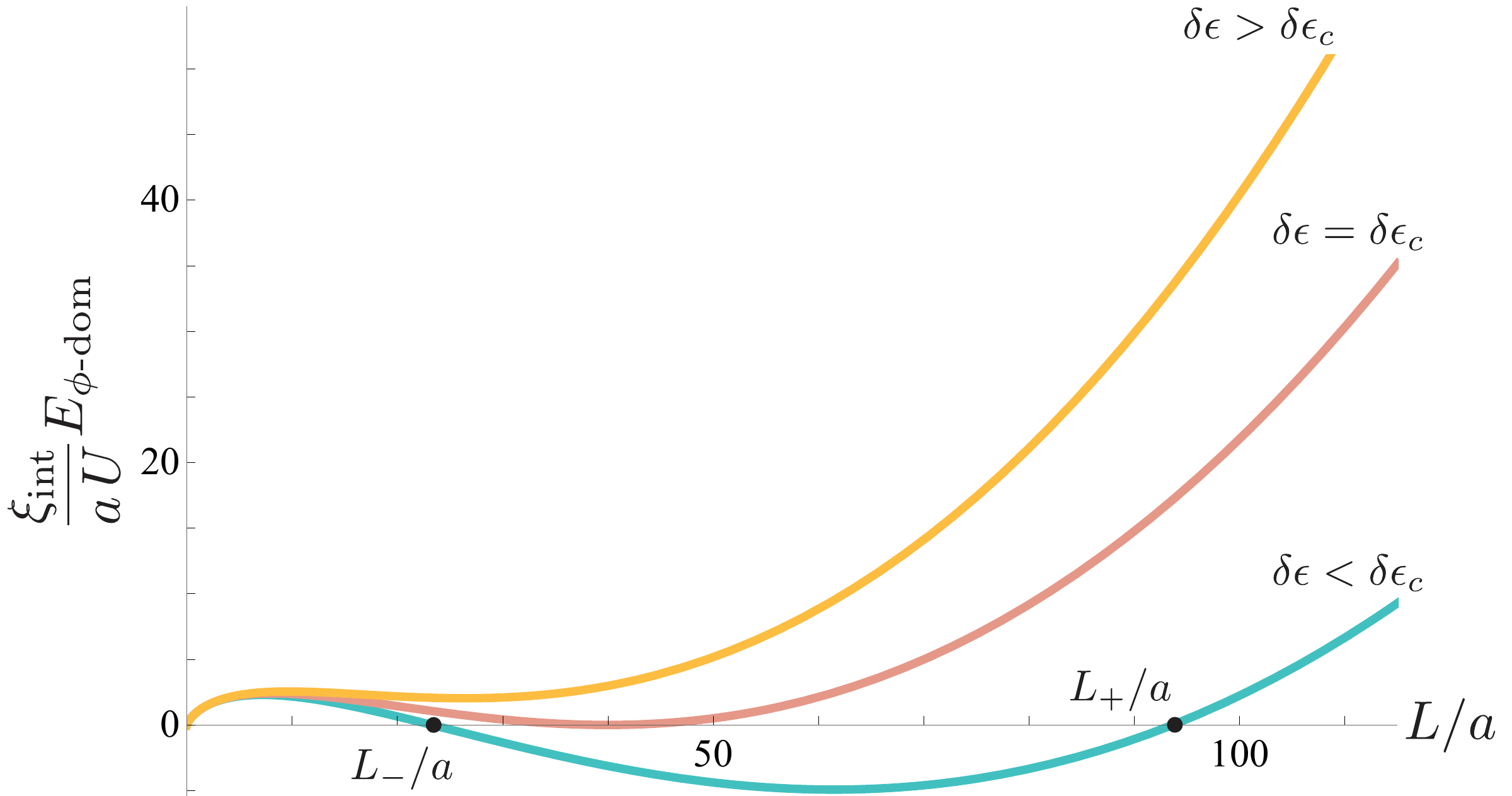}
	\caption{
	Plot of the energy cost associated with adding a $\phi$-ordered domain to a uniformly $\Phi$-ordered system when $\d\ep>\d\ep_c$ (orange), $\d\ep=\d\ep_c$ (pink), and $\d\ep<\d\ep_c$ (turquoise).
	For $\d\ep<\d\ep_c$, we see that domain formation is energetically favourable, $E_{\phi\text{-}\mathrm{dom}}<0$, for domains with linear extent $L$ satisfying $L_-<L<L_+$. 
	Here, we have set $\a\sim0.6$, for which $\d\ep_c\sim0.015\,\d m\, \xi_\mathrm{dis}/a$.
	}
	\label{fig:depC_WN}
\end{figure}

\subsubsection{\texorpdfstring{Single $\phi$-domain formation: Gaussian correlated disorder}{Single phi-domain formation: Gaussian correlated disorder}}

We now repeat the exercise above for Gaussian-correlated disorder.
The energy cost of inserting a $\phi$-ordered domain into a uniformly $\Phi$-ordered system is on average
\eq{
E_{\phi\text{-}\mathrm{dom}}(L)
&\sim
\d\ep\(L\o\xi_\mathrm{int}\)^{\!2}
+
U{L\o\xi_\mathrm{int}}
-
\d m{\xi_\mathrm{dis}L\o\xi_\mathrm{int}^2}\sqrt{1-e^{-L^2/2\xi_\mathrm{dis}^2}}.
}
We first study the regime where the smoothness of the disorder is important, \emph{i.e.} the exponential under the square root is important. 
In this case, we expect the $\phi$-domains to track the disorder potential and therefore be of the same size as the disorder correlation length $\xi_\mathrm{dis}$.
In order for this to be energetically favourable, we must have
\eq{
0>
E_{\phi\text{-}\mathrm{dom}}(\xi_\mathrm{dis})
&\sim
\d\ep\,{\xi_\mathrm{dis}^2\o\xi_\mathrm{int}^2}
+
U{\xi_\mathrm{dis}\o\xi_\mathrm{int}}
-
\d m{\xi_\mathrm{dis}^2\o\xi_\mathrm{int}^2}
=
U\(\xi_\mathrm{dis}\o\xi_\mathrm{int}\)^{\!2}
\[
{\d\ep\o\d m}+{1\o\a}-1\],
} 
It follows that $\phi$-ordered domains of linear extent $\xi_\mathrm{dis}$ should form once
\eq{
{\d\ep}
&\lesssim 
\d\ep_c(\a),
&
\d\ep_c(\a)
&\equiv
\d m\(1-{1\o\a}\),
\quad\text{when}\quad
\a\gtrsim1.
}
This critical energy difference is nearly identical to the analogous expression obtained for white noise disorder in Eqs.~\eqref{eqn:qHalfContour-2} and~\eqref{eqn:depCrit-WN}.
The most notable difference between the two inequalities is the prefactor $\xi_\mathrm{dis}/\xi_\mathrm{int}$ multiplying the right-hand side.
Going back to the previous section, we see that this coefficient originates from setting the minimal domain size to $\xi_\mathrm{int}$.
In contrast, for Gaussian-correlated disorder, the smallest allowed domains are expected to be $\xi_\mathrm{dis}$, and so $\d\ep_c(\a)$ contains no such prefactor.

As we saw in Appendix~\ref{app:DisDomain}, once $\a\lesssim1$, Gaussian-correlated disorder can be treated as local white-noise disorder, which necessitates a treatment that includes the effects of domain wall roughening.
The relevant expression for $E_{\phi\text{-}\mathrm{dom}}(L)$ is therefore identical to the one given in Eq.~\eqref{eqn:Ephidom-WallRough}, save that the smallest domain size is given by $a=\max(\xi_\mathrm{int},\xi_\mathrm{dis})$.
The inequality describing the favourability of domain formation  is now
\eq{
\d\ep&\lesssim\d\ep_c(\a),
&
\d\ep_c(\a)
&\equiv
{\xi_\mathrm{dis}\o a}\d m\,\a\,{e^{-c\({1\o\a^2}+1\)}}\cCom
\quad
\text{when}
\quad
\a\lesssim1.
}

\subsubsection{\texorpdfstring{Multiple $\phi$-domains}{Multiple phi-domains}}

The formation of a single domain does not necessarily imply the network model we propose as a description for mTBG at charge neutrality.
Instead, we want the $\phi$-ordered regions to percolate throughout the sample, as implied by the condition given in Eq.~\eqref{eqn:MostlyQVH}.
We argue that $\phi$-order should start dominating at a crossover set by the scale $\d\ep_c(\a)$.
As discussed in Appendix~\ref{app:CompCord-WNsingleDom}, within our approximation, 
domain boundaries between different $\phi$ orientations have the same cost as domains between $\Phi$- and $\phi$-ordered regions.
As a result, we can imagine `tiling' the $\phi$-ordered regions into domains of some size $\xi_*$.
For instance, when $\a\lesssim1$, the energy difference between a uniformly $\Phi$-ordered system and a (non-uniformly) $\phi$-ordered system is
\eq{
\Delta E&\sim
\d\ep \(L\o\xi_\mathrm{int}\)^{\!2}
+
\(L\o\xi_*\)^2\[U {\xi_*\o\xi_\mathrm{int}}
-
U\({\d m\o U}{\xi_\mathrm{dis}\o\xi_\mathrm{int}}\)^{\!2}{\xi_*\o\xi_\mathrm{int}}\log \!\(\xi_*\o a\)
\]
\nt
&=
U{L^2\o\xi_\mathrm{int}\xi_*}
\[
{\d\ep\o U}{\xi_*\o\xi_\mathrm{int}}+1-\a^2\log\!\(\xi_*\o a\)
\]
\nt
&=
{L^2\o\xi_*^2}
E_{\phi\text{-}\mathrm{dom}}(\xi_*),
}
where this expression is the same for both white noise and Gaussian-correlated disorder provided we recall that $a=\xi_\mathrm{int}$ in the former case while $a=\max(\xi_\mathrm{dis},\xi_\mathrm{int})$ in the latter.
It follows that when the typical domain size $\xi_*$ is such that $E_{\phi\text{-}\mathrm{dom}}(\xi_*)<0$ (\emph{i.e.} $L_-<\xi_*<L_+$), a wholly (but non-uniformly) $\phi$-ordered sample may be considered energetically favourable.
An identical argument holds for $\a\gtrsim1$ with $\xi_*=\xi_\mathrm{dis}$.
If we now imagine fixing $\a$ and increasing $\d\ep$, we expect Eq.~\eqref{eqn:MostlyQVH} to hold up to some value, $\d\tilde{\ep}_c(\a)$, of the same order as $\d\ep_c(\a)$.
Given the general lack of precision throughout this appendix, we assume that $\d\tilde{\ep}_c(\a)\sim\d\ep_c(\a)$.
This identity sets the dashed line in Fig.~\ref{fig:CompPhase}.

}
\twocolumngrid
\linespread{1.}

\bibliographystyle{apsrev4-2}
\bibliography{DisorderTBGRefs}
\end{document}